%% file: phi2-draft.tex
\documentclass[prd,nofootinbib,eqsecnum,superscriptaddress,citorder]{revtex4-2}

\PassOptionsToPackage{breaklinks=true,colorlinks=true,linkcolor=blue,urlcolor=blue,citecolor=blue}{hyperref}

\usepackage{hyperref}
\usepackage{color}
\usepackage{xcolor}
\usepackage{amsmath}
\usepackage{amssymb}
\usepackage{graphicx}
\usepackage{standalone}
\usepackage{tikz}
\usetikzlibrary{decorations.pathmorphing,calc}
\usepackage{subcaption}
\usepackage{setspace}
\usepackage[T1]{fontenc}
\usepackage[utf8]{inputenc}

\usepackage[normalem]{ulem}

\usepackage{epsfig}

\usepackage{dcolumn}

\usepackage{bm}

\newcommand{\w}{\omega}

\newcommand{\tort}{r_*}
\newcommand{\RinrefInt}{B^{\text{ref}}_-}
\newcommand{\RintraInt}{B^{\text{tra}}_-}

\newcommand{\Rin}{R^{\text{in}}_{\ell m}}

\newcommand{\Rinhat}{\hat{R}^{\text{in}}_{\ell m}}
\newcommand{\Bref}{B^{\text{ref}}}
\newcommand{\Binc}{B^{\text{inc}}}
\newcommand{\Btra}{B^{\text{tra}}}

\newcommand{\nn}{\nonumber}
\newcommand{\an}[1]{a_{#1}}
\newcommand{\rsST}{\tort^\text{ST}}
\newcommand{\RinrefIntNorm}{\hat{B}^{\text{ref}}_-}
\newcommand{\RintraIntNorm}{\hat{B}^{\text{tra}}_-}
\newcommand{\A}[1]{{}_{#1}A^{\text{ST}}_{\omega\ell m}}
\newcommand{\B}[1]{{}_{#1}B^{\text{ST}}_{\omega\ell m}}
\newcommand{\Rup}{R^{\text{up}}_{\ell m}}
\newcommand{\Ctra}{C^{\text{tra}}}
\newcommand{\Cinc}{C^{\text{inc}}}
\newcommand{\Cref}{C^{\text{ref}}}
\newcommand{\rhoup}[1]{{}_{#1}\rho^{\text{up,ST}}_{\omega\ell m}}
\newcommand{\tauup}[1]{{}_{#1}\tau^{\text{up,ST}}_{\omega\ell m}}

\newcommand{\psiup}[1]{{}_{#1}\psi^{\text{up}}_{\omega \ell m}}
\newcommand{\Btracc}{B^{\text{tra}^*}}
\newcommand{\Brefcc}{B^{\text{ref}^*}}

\begin{document}

\title{Quantum fluxes and $\langle\hat{\Phi}^2\rangle$ for a non-minimally coupled scalar field: ringdown and tail on approaching the polar Kerr inner horizon}
\author{Maria Alberti}
\email{mariaalberti@campus.technion.ac.il}
\affiliation{Department of Physics, Technion, Haifa 32000, Israel}
\affiliation{Max Planck Institute for Mathematics in Sciences (MiS), Inselstraße 22, 04103 Leipzig, Germany}
\author{Noa Zilberman}
\email{nz3745@princeton.edu}
\affiliation{Princeton Gravity Initiative, Princeton University, Princeton NJ 08544, USA}
\author{Marc Casals}
\email{marc.casals@uni-leipzig.de}
\affiliation{Institut f\"ur Theoretische Physik, Universit\"at Leipzig,\\ Br\"uderstra{\ss}e 16, 04103 Leipzig, Germany}
\affiliation{School of Mathematics and Statistics, University College Dublin, Belfield, Dublin 4, D04 V1W8, Ireland}
\affiliation{Centro Brasileiro de Pesquisas F\'isicas (CBPF), Rio de Janeiro, CEP 22290-180, Brazil}
\author{Adrian C. Ottewill}
\email{adrian.ottewill@ucd.ie}
\affiliation{School of Mathematics and Statistics, University College Dublin, Belfield, Dublin 4, D04 V1W8, Ireland}

\date{\today}
\begin{abstract}
We compute $\langle\hat{\Phi}^{2}\rangle_\text{ren}$ as well as the energy fluxes
$\langle \hat{T}_{uu}\rangle_\text{ren}$ and $\langle \hat{T}_{vv}\rangle_\text{ren}$ (where $u$
and $v$ are the standard Eddington-Finkelstein coordinates) associated with a
quantum massless real scalar field $\hat{\Phi}$, with a general curvature coupling
constant $\xi$, near the inner horizon (IH) of a Kerr black hole, along the axis of rotation. The quantum field is in the Unruh
state, corresponding to an evaporating black hole. We renormalize these quantities
by the state-subtraction method. We drop the assumption of minimal
coupling to the curvature, thereby generalizing the results of \cite{2022PhRvL.129z1102Z}  for the fluxes {\it at} the IH. This requires understanding the asymptotic behavior of  $\langle\hat{\Phi}^{2}\rangle_\text{ren}$ {\it near} the IH. State subtraction allows us to push the computation of $\langle\hat{\Phi}^{2}\rangle_\text{ren}$ along the axis of rotation in the Kerr interior in \cite{Zilberman:2024jns} deeper into the near-IH region, exposing their final asymptotic behavior on approaching the IH.
For $\langle\hat{\Phi}^{2}\rangle_\text{ren}$ (a $\xi$-independent quantity in the
Kerr case), we find that the approach to its finite asymptotic IH
value is given, per $\ell$-mode, by a ringdown phase
(namely exponentially damped oscillations), followed by an inverse-power
tail, both in the tortoise coordinate $r_{*}$ (which diverges at
the IH). Interestingly, in the regime where the ringing dominates, the ringing's complex frequencies are (numerically) found to match twice the well-known classical quasinormal-mode frequencies
in Kerr, and the inverse-power tails are found to be $r_{*}^{-2\ell-3}$
(resembling Price's law in the classical black hole exterior, upon replacement $t\to r_*$). In particular, the sum over $\ell$ behaves as $r_*^{-3}$ (with a prefactor obtained here analytically for the first time), resembling the known behavior of classical (axially symmetric) scalar perturbations on approaching the IH \cite{Ori:1998gp}. 
These results allow computing the flux components $\langle \hat{T}_{uu}\rangle_\text{ren}$
and $\langle \hat{T}_{vv}\rangle_\text{ren}$ for general $\xi$ at the IH vicinity. We find that the limiting IH values of $\langle \hat{T}_{uu}\rangle_\text{ren}$
and $\langle \hat{T}_{vv}\rangle_\text{ren}$ are \emph{independent} of the coupling constant
$\xi$. When translated to the Kruskal coordinate $V$ (regular and vanishing at
the Cauchy horizon), it implies that the prefactor $C$ characterizing the
divergence of the renormalized quantum stress-energy tensor appearing
in $\langle \hat{T}_{VV}\rangle_{\text{ren}}\simeq CV^{-2}$, is independent of the
coupling to curvature in the case of polar Kerr.

\end{abstract}

\maketitle

\section{Introduction}

In the absence of a viable theory of quantum gravity, in which both
matter fields and the gravitational field are quantized, the
semiclassical formulation of general relativity seeks to provide
insights into the backreaction of quantum fields onto
a classical spacetime. In this framework, the classical
(vacuum) Einstein equation is replaced by its semiclassical counterpart
\begin{equation}
G_{\mu\nu}=8\pi\langle \hat{T}_{\mu\nu}\rangle^{\Psi}_\text{ren},\label{eq:semiclassical-EE}
\end{equation}
where $G_{\mu\nu}$ is the Einstein tensor and $\langle \hat{T}_{\mu\nu}\rangle^{\Psi}_{\text{ren}}$ is the \emph{renormalized} expectation value of the
stress-energy tensor (SET) in a chosen state $\vert\Psi\rangle$
(the renormalized stress-energy tensor, RSET). 
In the standard quantum field theory in curved spacetimes approach,
Eq. (\ref{eq:semiclassical-EE}) is typically implemented iteratively: one initially fixes
a classical background metric, computes the
RSET on that geometry, and then incorporates its backreation on the metric \emph{a posteriori}; in principle, this process should then be repeated iteratively. 

However, even the first-order computation
of $\langle \hat{T}_{\mu\nu}\rangle^{\Psi}_\text{ren}$ (on a fixed zeroth-order background geometry)
is conceptually and technically subtle: the RSET is formally divergent (as is
any other in-coincidence observable nonlinear in the field operator)
and must be renormalized --  a particularly challenging task on four-dimensional black-hole (BH) backgrounds. A widely accepted local and covariant renormalization prescription is \emph{Hadamard point-splitting} (e.g.,~
\cite{wald1994quantum,wald1978trace}), in which
 renormalized quadratic field observables are obtained by splitting the field square into two separated points $x$ and $x'$, subtracting the \emph{Hadamard parametrix} $h(x,x')$, acting with covariant derivatives (if relevant) at the two points, and finally taking the coincidence limit $x'\to x$. The Hadamard
parametrix $h(x,x')$ captures the universal short-distance singularity structure
of any physical (Hadamard) state and is
determined covariantly by local geometric data along the geodesic
connecting $x$ and $x'$.
Accordingly, a basic ingredient for implementing point splitting is the Hadamard (symmetrized) two-point function (HTPF) defined by the expectation value of the anti-commutator of a scalar field operator $\hat{\Phi}$ in some quantum state $\vert\Psi\rangle$:
\begin{equation}
G^{\Psi}(x,x')\equiv\langle\Psi|\big\{\hat{\Phi}(x),\hat{\Phi}(x')\big\}\vert\Psi\rangle\equiv\langle\hat{\Phi}(x)\hat{\Phi}(x')+\hat{\Phi}(x')\hat{\Phi}(x)\rangle^{\Psi}.\label{eq:2ptf}
\end{equation}

The most elementary nonlinear observable is $\langle\hat{\Phi}^2\rangle$ (referred to as the \emph{field square} or \emph{vacuum polarization}), obtained from the renormalized coincident-point limit of (half) the HTPF. Specifically, the renormalized field square is given by
\begin{equation}\label{eq:2pt-def}
\langle\hat{\Phi}^2(x) \rangle^\Psi_{\text{ren}}\equiv\lim_{x'\to x}\left(\frac{1}{2}G^{\Psi}(x,x')-h(x,x')\right).
\end{equation}
A related observable, which is of interest for the current work, is
\begin{equation}
\langle\nabla_{y}\hat{\Phi}\nabla_{y}\hat{\Phi} (x)\rangle^\Psi_{\text{ren}}\equiv\lim_{x'\to x}\nabla_{y}\nabla_{y}'\left(\frac{1}{2}G^{\Psi}(x,x')-h(x,x')\right),\label{eq:hadamard-point-split}
\end{equation}
where $y \in \{u, v\}$ and $\nabla$ ($\nabla'$) is the covariant derivative acting on $x$ ($x'$) . \footnote{Further finite renormalization freedom exists in the prescription defined by Eq. \eqref{eq:2pt-def}-\eqref{eq:hadamard-point-split}. The ambiguity consists of local curvature terms (depending only on the spacetime geometry in a neighborhood of $x$) \cite{wald1994quantum}. However, in the case of interest, all such ambiguities vanish.}
To avoid clutter, we henceforth drop the subscript "ren" and write $\langle \hat{X} \rangle^\Psi$ for the renormalized expectation value of an observable $\hat{X}$ in a state $\vert\Psi\rangle$. When the state is not relevant or clear from context, we also omit the superscript $\Psi$. 

A central motivation for studying Eq.~(\ref{eq:semiclassical-EE}) comes from BH horizons, where quantum effects can accumulate and become physically significant. A celebrated example is BH evaporation via the emission of Hawking radiation \cite{hawking1975particle}, which involves a negative quantum influx across the event horizon (EH) \cite{unruh1976notes}, which in turn induces a slow drift of the BH parameters along the EH. For the Cauchy horizon (CH) of a charged or rotating BH
\footnote{In the physically realistic scenario of gravitational collapse, it is only the \emph{ingoing} section of the IH (denoted $\mathcal{CH^R}$ in Fig. \ref{fig:Penrose-diagram-fKerr}) which maintains the role of a CH, and it is therefore this section that we focus on here.}, quantum effects can be even more consequential: various key components of the RSET have been found to diverge (when expressed in coordinates regular at the CH)\footnote{On a fixed background, quantum energy-momentum fluxes at the CH typically diverge as $V^{-2}$, where $V$ is a Kruskal-type coordinate that is regular and vanishes at the
CH. This blow-up is stronger than the corresponding divergence of classical perturbations, which are known to generate only a weak \cite{Tipler:1977zza,Ori:2000fi} singularity at the CH (see Refs.  \cite{ori1992structure,DafermosLuk2017,gurriaran2026nonlinearinstabilitykerrcauchy,2026arXiv260404877L} in the asymptotically flat case). For backreaction, however, one must account for the fact that in the Unruh state the BH is evaporating -- an effect encoded already at the EH. A consistent treatment must therefore evolve Eq.~(\ref{eq:semiclassical-EE}) from the EH toward the CH, and this may alter the naive fixed-background expectation of a strong singularity at this order.} \cite{Lanir:2018vgb,Zilberman:2019buh,Hollands:2020qpe,PhysRevLett.132.121501,2022PhRvL.129z1102Z,Zilberman:2024jns,2021PhRvL.127w1301K,Alberti:2025mpg,2021PhRvD.104b5009K,McMaken:2024fvq}, suggesting that quantum effects may dramatically modify the classical spacetime picture. 
(See also earlier analytical indications of a divergence in quantum observables at the CH  \citep{Hiscock:1977qe,birrell1978falling,hiscock1980quantum,Ottewill:2000qh}. However, these works did not include an explicit computation of the relevant semiclassical quantities, and the issue therefore remains unresolved). 
The IH is therefore a particularly rich arena for studying the backreaction of quantum sources in the right-hand side of Eq.~(\ref{eq:semiclassical-EE}) onto the geometry in the left-hand side; see Ref.~\cite{Arad:2025imu} for a numerical exploration of near-IH backreaction in the spherical evaporating case, and Ref.~\cite{Zilberman:prep} for an analytical treatment of the nonevaporating case, also in spherical symmetry.

Although the RSET is the central quantity appearing directly in Eq. (\ref{eq:semiclassical-EE}), the simpler quantity $\langle \hat{\Phi}^2\rangle$ also offers information on quantum effects inside BHs (for example, it is directly related to the  trace of the RSET, see Ref. \cite{Lanir:2018vgb}). Studies of $\langle \hat{\Phi}^2\rangle$ in BH interiors go back to Ref.~\cite{Candelas:1985ip}, which computed this quantity in the interior of a Schwarzschild BH in the Hartle-Hawking (HH) state $\vert\text{H}\rangle$ (describing a BH in equilibrium with a bath of thermal radiation \cite{hartle1976path}), renormalizing via the $t$-splitting point-splitting technique of DeWitt \cite{dewitt1975quantum} and Christensen \cite{Christensen:1976vb}. Decades later, Ref.~\cite{Lanir:2018rap} computed $\langle \hat{\Phi}^2\rangle$ in the Schwarzschild interior in both the HH state and the physically-relevant Unruh state $\vert \text{U}\rangle$ (corresponding to an evaporating BH  \cite{unruh1976notes}) using a variant of \emph{pragmatic mode-sum regularization} (PMR). First introduced in Ref.~\cite{Levi:2015eea}, PMR is a point-splitting renormalization scheme requiring only the presence of one Killing field in the background, and in Ref.~\cite{Lanir:2018rap} the point split is taken in $\theta$, relying on spherical symmetry. In a later study \cite{Lanir:2018vgb}, restricted to a real scalar field in Reissner-Nordström (RN), some of the present authors together with collaborators computed $\langle \hat{\Phi}^2\rangle$ throughout the region between the horizons, with special emphasis on the near-IH regime, both in the HH and in the Unruh state, again using the $\theta$-splitting variant of PMR. The analysis in Ref.~\cite{Lanir:2018vgb} exposed  irregular behavior near the IH, revealing, after a few quickly decaying oscillations, inverse-power tails as one approaches the IH ($r_*^{-3}$ for Unruh and $r_*^{-2}$ for HH, where $r_*$ is the tortoise coordinate which asymptotes to $\infty$ at the IH). This naturally raises the question of whether analogous behavior persists at the Kerr IH (where the HH state does not exist for bosonic fields \cite{kay1991theorems,Ottewill:2000qh}, but the Unruh state does \cite{Hafner:2026zcl}) -- an issue that, as we explain below, has been difficult to resolve with existing point-splitting computations. 
This possibility was briefly noted in Sec. IV.B and footnote 27 of Ref.~\cite{Zilberman:2024jns}, where it was reported that a state subtraction analysis suggests $\langle\hat\Phi^2\rangle^{\mathrm U}_\text{ren}$ approaches a finite polar-IH value through an $r_*^{-3}$ tail, as in RN. The detailed analysis leading to this conclusion, among others, is presented in the present work.
Hereafter, $(t,r,\theta,\varphi)$ are the standard Boyer-Linquist coordinates in Kerr spacetime.

While they have enabled significant progress to be made through concrete computations, Hadamard point-splitting methods are inherently numerically demanding  and, in practice, can become ill-defined when the splitting direction becomes null. In Kerr, this obstructs certain natural variants (e.g. $\varphi$-splitting at the pole $\theta=0$, or $t$-splitting on the ergosurface, which includes the polar horizons). These difficulties have motivated  alternative regularization strategies. A particularly useful one is to consider differences of expectation values between two (Hadamard) states; we shall refer to this approach as \emph{state subtraction}.

Recent years have seen substantial progress on quantum effects near the IH of charged \cite{Zilberman:2019buh,Hollands:2020qpe,Hollands:2020qpe,Lanir:2018vgb,2021PhRvL.127w1301K,Alberti:2025mpg,2021PhRvD.104b5009K} and rotating \cite{2022PhRvL.129z1102Z,PhysRevLett.132.121501,Zilberman:2024jns,McMaken:2024fvq} BHs. In RN, most IH analyses of $\langle\hat{\Phi}^2\rangle$ and the RSET (with the key components being the so-called \emph{fluxes} $\langle \hat{T}_{vv} \rangle$ and $\langle \hat{T}_{uu} \rangle$, where $v$ and $u$ are the usual Eddington-Finkelstein coordinates) employ the PMR point-splitting method (with the exception of \cite{Alberti:2025mpg} using state subtraction); in RN-de Sitter the available works rely on state subtraction for the RSET (and the renormalized field current $\langle \hat{j}_v\rangle$). In Kerr-de Sitter, to our knowledge, the only existing work \cite{PhysRevLett.132.121501} employs state subtraction to regularize $\langle \hat{T}_{vv} \rangle$ (and $\langle \hat{T}_{v\varphi_-} \rangle$). For Kerr BHs, which are the focus of this paper, an early work \cite{2022PhRvL.129z1102Z} used both state subtraction and the $t$-splitting variant of PMR to regularize the IH quantum fluxes, finding excellent agreement between the two methods, and a subsequent study \cite{McMaken:2024fvq} extended the state-subtraction analysis across a broader parameter range. More recently, an extensive $t$-splitting study \cite{Zilberman:2024jns} investigated the quantum fluxes and $\langle \hat{\Phi}^2 \rangle$ throughout the Kerr interior along the axis of rotation, with particular focus on the near-IH domain (focusing on two specific spin-to-mass ratios, $a/M=0.8,0.9$). However, the aforementioned technical difficulty of approaching the polar IH with $t$-splitting restricted the computation to $r-r_-\gtrsim
10^{-5}M$, where $r_-$ is the $r$ value of the IH. While this range was sufficient to extract the limiting values of the fluxes at the IH (and verify their agreement with previous  \cite{2022PhRvL.129z1102Z}  state-subtraction results), it was insufficient to resolve the true asymptotic behavior of $\langle \hat{\Phi}^2 \rangle$.

The first achievement of the present paper is to close this gap. Using the aforementioned method of state subtraction\footnote{\label{ft:Hadamard_comp_state}
Our state-subtraction regularization employs a comparison state $\vert\underline{\text{U}}\rangle$ tailored to the IH vicinity, which was introduced originally in Ref.~\cite{2022PhRvL.129z1102Z}. Although $\vert\underline{\text{U}}\rangle$ has not been formally proven to be Hadamard, existing evidence suggests that it is, and we adopt this as a working assumption.}
, we compute the Unruh-state field square $\langle \hat{\Phi}^2\rangle $ (up to a residual smooth function of the coordinate $r$) 

at the very close vicinity of the Kerr IH along the axis of rotation. This allows us to expose the near-IH asymptotics of $\langle \hat{\Phi}^2\rangle $, previously inaccessible in Kerr: we find that its final approach to the (polar) IH limiting value is governed by inverse-power tails scaling as $r_*^{-2\ell-3}$ (per multipolar $\ell$ mode). We also derive the leading inverse-power tail analytically. This behavior is reminiscent of the classical late-time decay of scalar fields in BH exteriors, originally found by Price \cite{Price:1971fb} (in Schwarzschild's exterior) and later extended to BH interiors near the IH (see Ref. \cite{Ori:1996si} for RN and Ref. \cite{Ori:1998gp} for polar Kerr). We therefore refer to the corresponding inverse-power behavior in $r_*$ as a \emph{Price-like tail}.
As we shall see, the leading tail is only exposed at extraordinarily small  separations from the IH ($r - r_{-} \sim 10^{-45} M$ for $a/M = 0.8$). This, together with the inapplicability of the $\theta$-splitting PMR variant in Kerr, helps explain why earlier point-splitting computations near the polar Kerr IH could not resolve the true asymptotic behavior of $\langle \hat{\Phi}^2 \rangle$, and it highlights the practical advantage of state subtraction in this setting.
This method allows us to calculate, for the first time,
the exact (numerical) value of
$\langle\hat{\Phi}^{2}\rangle$ in the Unruh state on the polar Kerr IH (for $a/M = 0.8$).

Moreover, in the near-IH domain preceding the tails we observe a phase of quasinormal-mode (QNM) ringdown in the tortoise coordinate $r_*$. The complex ringing frequencies extracted numerically agree with twice the well-known QNM frequencies for scalar perturbations in Kerr. Taken together, the appearance of both QNM ringing and Price-like tails indicates that two characteristic features of classical BH perturbation theory also manifest themselves in the $r_*$-dependence of the quantum observable $\langle \hat{\Phi}^2\rangle $ inside the BH as one approaches the IH.

A second achievement of the current work is the generalization of previous IH RSET studies, which mainly focused on minimally coupled scalar fields (i.e., with coupling constant $\xi=0$), to arbitrary nonminimal coupling. This broader class includes, in particular, conformal coupling ($\xi=1/6$ for our case), which makes contact with other conformal fields such as the physically important electromagnetic field.
For minimally-coupled scalar fields, the classical SET is simpler (the last term in Eq. \eqref{eq: classical SET KG} below is absent). Allowing for a general coupling constant $\xi$ introduces additional terms (notably those involving second derivatives of $\langle\hat{\Phi}^2\rangle$), so that the near-IH behavior of $\langle \hat{\Phi}^2\rangle $ becomes the key missing ingredient for extending IH flux computations beyond $\xi=0$. Using our novel near-IH results for $\langle \hat{\Phi}^2\rangle $, we extend the Unruh-state IH-fluxes analysis to general $\xi$ and find, in particular, that the leading divergence of the polar IH fluxes (in coordinates regular at the IH) is in fact independent of $\xi$. In addition, we characterize the subleading $\xi$-dependent piece.

The paper is structured as follows. In Sec.~\ref{sec:Setup} we introduce the Kerr geometry and review canonical quantization of a real scalar field
in the Unruh state. Using state subtraction, we derive an expression for $\langle\hat{\Phi}^{2}(x)\rangle$ valid throughout the spacetime
(details of the comparison-state HTPF are given in Appendix~\ref{sec:appendixA}).
In Sec.~\ref{sec:Polar-vacuum-polarization} we evaluate the mode sum numerically at the pole and near the IH, revealing a Price-like tail
and a preceding QNM ringdown. These features are discussed in Subsecs. \ref{subsec:Price-tail}
and \ref{subsec:QNM-ringdown}, respectively. We additionally derive the leading inverse-power analytically; the analysis and requisite ingredients are provided in Apps.~\ref{sec:appendixC}-\ref{sec:appendixD}.
In Subsec.~\ref{subsec:Comparison-to-t-splitting} we compare (and combine) our state-subtraction results for $\langle\hat{\Phi}^2\rangle$ with those obtained via $t$-splitting
in Ref.~\cite{Zilberman:2024jns}. In Sec.~\ref{sec:Stress-energy-fluxes} we discuss implications for the IH stress-energy fluxes
$\langle \hat{T}_{vv}\rangle^\text{U}$ and $\langle \hat{T}_{uu}\rangle^\text{U}$ for a scalar field with general coupling $\xi$.
We conclude in Sec.~\ref{sec:Discussion} with a discussion and summary.
Throughout, we use units $c=G=1$ and signature $(-+++)$.
All figures and numerical results correspond to the spin parameter $a=0.8M$ unless explicitly stated otherwise; qualitative features are expected to persist for generic spin values.

\section{Setup}\label{sec:Setup}

\begin{figure}
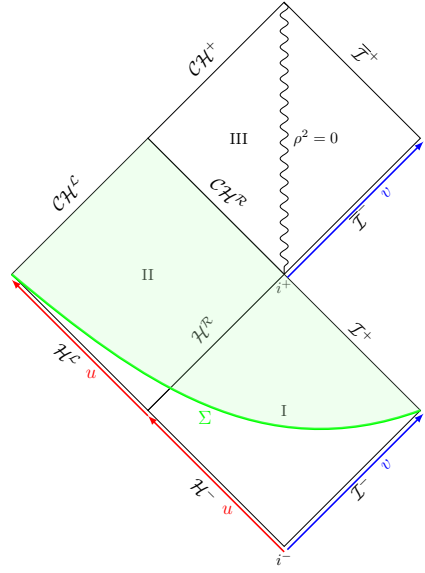

\begin{centering}
\includestandalone[scale=0.6]{Kerr-Penrose}
\par\end{centering}
\caption{Partial Penrose diagram for the subextremal Kerr BH. Region I denotes the
exterior region (where $r>r_+$) and II the interior of the BH  (where $r_-<r<r_+$). Region III is another asymptotically flat region (for $\theta\neq\pi/2)$ extending to negative $r$ values, and is relevant for the construction of the comparison state (in Appendix \ref{sec:appendixA}).
The wiggled line stands for the ring singularity at $\rho^{2}=0$ (corresponding to $r=0$, $\theta=\pi/2$).
The ranges of the Eddington-Finkelstein coordinates $u$ and $v$ are indicated
each by an arrow that points from $-\infty$ to $+\infty$. The green
line denoted $\Sigma$ indicates a Cauchy hypersurface in the region I\ensuremath{\cup}II,
and the (future) domain of dependence of $\Sigma$ corresponds to
the light-shaded green region.}\label{fig:Penrose-diagram-fKerr}
\end{figure}

We consider a real scalar field propagating in a Kerr BH spacetime,
with metric given in Boyer-Lindquist coordinates by
\begin{equation}
ds^{2}=-\biggl(1-\frac{2Mr}{\rho^{2}}\biggr)dt^{2}+\frac{\rho^{2}}{\Delta}dr^{2}+\rho^{2}d\theta^{2}+\biggl(r^{2}+a^{2}+\frac{2Mra^{2}}{\rho^{2}}\sin^{2}\theta\biggr)\sin^{2}\theta d\varphi^{2}-\frac{4Mra}{\rho^{2}}\sin^{2}\theta d\varphi dt,\label{eq:Kerr-metric}
\end{equation}
where $\rho^{2}\equiv r^{2}+a^{2}\cos^{2}\theta$ and $\Delta\equiv r^{2}-2Mr+a^{2}$.
$M$ and $J\equiv aM$ correspond to the mass and angular momentum
of the BH, respectively. For subextremal BHs, which we shall exclusively
consider here, the angular momentum per unit mass satisfies $\vert a\vert<M$.
The Penrose diagram for the analytic extension of Kerr spacetime is depicted in Fig. \ref{fig:Penrose-diagram-fKerr}.
The blocks I, II, III are defined as
\begin{equation}
\text{I}=\mathbb{R}_{t}\times(r_{+},\infty)\times S_{\theta,\varphi}^{2},\hspace{0.5cm}\text{II}=\mathbb{R}_{t}\times(r_{-},r_{+})\times S_{\theta,\varphi}^{2},\hspace{0.5cm}\text{III}=\mathbb{R}_{t}\times(-\infty,r_{-})\times S_{\theta,\varphi}^{2}.
\end{equation}
The function $\Delta$ has two roots $r_{\pm}=M\pm\sqrt{M^{2}-a^{2}}$,
corresponding to the inner ($r_{-})$ and outer ($r_{+})$ horizon.  In Kerr, the IH coincides with the Cauchy horizon ($\mathcal{CH}$).
While these two zeros of $\Delta$
are coordinate singularities, the zero of $\rho^ {}$
 is a true curvature (ring) singularity. This ring singularity
sits at $\left\{ r=0\right\} \cap\left\{ \theta=\pi/2\right\} $,
i.e., for any $\theta$ value other than $\pi/2$, the
singularity at $r=0$ is absent and region III in the Penrose diagram
in Fig. \ref{fig:Penrose-diagram-fKerr} can be naturally (and analytically)
extended to negative $r$ values. In contrast to the spherically symmetric
case, not only the coordinate $t$ (and $r_{*},$ see below) diverges at the horizons,
but also the azimuthal coordinate $\varphi$. To each horizon at $r=r_{\pm}$
one may associate a surface gravity $\kappa_{\pm}\equiv(r_{+}-r_{-})/2(r_{\pm}^{2}+a^{2})$,
an angular velocity $\Omega_{\pm}\equiv a/(2Mr_{\pm})$ and an
azimuthal coordinate $\varphi_{\pm}\equiv\varphi-\Omega_{\pm}t$ that
remains regular upon approaching the respective horizon. We also introduce
the tortoise coordinate, defined through the relation $dr/dr_{*}=\Delta/(r^{2}+a^{2})$
(up to an integration constant, which we fix like in Ref. \cite{2022PhRvD.106l5011Z},
see Eq. (2.5) therein -- we write the explicit form in Eq. \eqref{eq:tortoiseO}), and the standard Eddington-Finkelstein coordinates

 \begin{align}\label{eq:coords}
 & u=t-r_{*},\hspace{1em}v=t+r_{*} & \hspace{1em}\text{(region I)} \\
 & u=-t+r_{*},\hspace{1em}v=t+r_{*} & \hspace{1em}\text{(region II)}\label{eq:uv-coordinates}\\
 & u=-t+r_{*},\hspace{1em}v=-(t+r_{*}) & \hspace{1em}\text{(region III),}
\end{align}
which are always future-pointing in our convention. We note that although these coordinates are not generally null throughout the Kerr spacetime, they are null along the pole (and on the horizons for any polar angle). In order to extend
the metric analytically across the EH ($\mathcal{H}^\mathcal{R}$ in Fig. \ref{fig:Penrose-diagram-fKerr}), one introduces the Kruskal coordinate
$U\equiv\mp\kappa_{+}^{-1}\exp(\mp\kappa_{+}u)$ (upper sign for
$r>r_{+}$, lower sign for $r<r_{+}$). Similarly, for analyticity across the ingoing inner horizon ($\mathcal{CH}^\mathcal{R}$ in Fig. \ref{fig:Penrose-diagram-fKerr}), one introduces the Kruskal coordinate $V\equiv\mp\kappa_{-}^{-1}\exp({\mp\kappa_{-}v})$
(upper sign for $r>r_{-}$, lower sign for $r<r_{-}$).

On this background metric we consider a free massless scalar field $\Phi$ with a general coupling to the curvature $\xi$.  
The field is subject to the classical massless Klein-Gordon (KG) equation
\begin{equation}
\nabla_{a}\nabla^{a}\Phi=0,\label{eq:Klein-Gordon-eq}
\end{equation}
where the covariant derivative is taken with respect to the metric
(\ref{eq:Kerr-metric}). \footnote{Note that the coupling constant $\xi$, which would have entered the equation through the product $\xi R$, is absent from Eq. (\ref{eq:Klein-Gordon-eq}) due to the vanishing Ricci scalar of the background.} In view of the symmetries of the background, it is customary to use the Ansatz
\begin{equation}
\Phi_{\omega\ell m}^{\Lambda}(x)=\mathcal{N}_{\omega\ell m}^{\Lambda}\frac{\psi_{\omega\ell m}^{\Lambda}(r)}{\sqrt{r^{2}+a^{2}}}e^{-i\omega t}e^{im\varphi}S_{\ell m}^{\omega}(\theta),\hspace{1.5em}\Lambda\in\{\text{I},\text{II},\text{III}\},\label{eq:boulware-mode-ansatz}
\end{equation}
(indexed by the frequency $\omega\in\mathbb{R}$, the multipolar number
$\ell\in\mathbb{Z}_{\geq 0}$ and the azimuthal number $m\in\mathbb{N}_{\leq\vert\ell\vert}$)
in which case the wave equation \eqref{eq:Klein-Gordon-eq} separates into coupled radial and
angular equations \cite{teukolsky1972rotating}. The \emph{radial solution} $\psi_{\omega\ell m}^{\Lambda}(r)$ solves the radial equation
\begin{subequations}
    \begin{align}
 & \partial_{r_{*}}^{2}\psi_{\omega\ell m}^{\Lambda}(r)=-V_{\omega\ell m}(r)\psi_{\omega\ell m}^{\Lambda}(r)\hspace{4em}\Lambda\in\{\text{I},\text{II},\text{III}\}\label{eq:radial1}\\
 & V_{\omega\ell m}\equiv\frac{K_{\omega m}^{2}(r)-\lambda_{\ell m}(a\omega)\Delta}{(r^{2}+a^{2})^{2}}-G^{2}(r)-\frac{dG(r)}{dr_{*}}\label{eq:radial2}\\
 & K_{\omega\ell m}(r)\equiv(r^{2}+a^{2})\omega-am,\hspace{1em}\lambda_{\ell m}(a\omega)\equiv E_{\ell m}(a\omega)-2am\omega+a^{2}\omega^{2},\hspace{1em}G(r)\equiv\frac{r\Delta}{(r^{2}+a^{2})^{2}},\label{eq:radial3}
\end{align}\label{kg-radial-joined}
\end{subequations} and see Eq. (2.16) in \cite{2022PhRvD.106l5011Z} for the explicit form of the angular equation (which incorporates the spheroidal eigenvalue $E_{\ell m}(a\omega)$ appearing in Eq.~(\ref{eq:radial3})).
The latter has (real) solutions $S_{\ell m}^{\omega}(\theta)$, the so called spheroidal wave functions \cite{Berti:2005gp}, which
are implemented in Mathematica \cite{Mathematica14p1}. We shall
use the notation
\begin{equation}
Z_{\ell m}^{\omega}(\theta,\varphi)\equiv(2\pi)^{-\frac{1}{2}}e^{im\varphi}S_{\ell m}^{\omega}(\theta).
\end{equation}
In \eqref{eq:boulware-mode-ansatz}, the prefactor $\mathcal{N}_{\omega\ell m}^{\Lambda}$
is a suitable normalization constant to be chosen so that the modes are
appropriately normalized \cite{2023PhRvD.108l5017I}.
The large-$r_{*}$ asymptotics of the effective potential (\ref{eq:radial2})
as a function of the radial coordinate are 
\begin{equation}
V_{\omega\ell m}=\begin{cases}
\omega_{-}^{2}+\mathcal{O}(e^{-2\kappa_{-}r_{*}}) & r\to r_{-}\,\,\, (r_*\to\infty)\\
\omega_{+}^{2}+\mathcal{O}(e^{2\kappa_{+}r_{*}}) & r\to r_{+}\,\,\, (r_*\to-\infty)\\
\omega^{2}+\mathcal{O}(r_{*}^{-2}) & r\to\infty\,\,\, (r_*\to\infty)
\end{cases},\label{eq:potential-radial-eqn}
\end{equation}
where $\omega_{\pm}\equiv\omega-m\Omega_{\pm}.$ In
particular, the radial solution becomes free in all three asymptotic
$r_{*}$ regimes.

\textit{In} and \textit{up} radial solutions to \eqref{eq:radial1} can be defined in both the exterior (region I; $r>r_+$) and interior (region II; $r_-<r<r_+$) regions, which later serve to construct the \textit{Eddington-Finkelstein} family of mode-solutions to the wave equation \eqref{eq:Klein-Gordon-eq}. The construction is well-documented in the literature, see e.g. Sec. III in Ref. \cite{2022PhRvD.106l5011Z}. The exterior \textit{in} and \textit{up} radial solutions and corresponding scattering coefficients ($\tau_{\omega\ell m}^{\text{in}}$, $\tau_{\omega\ell m}^{\text{up}}$, $\rho_{\omega\ell m}^{\text{in}}$ and $\rho_{\omega\ell m}^{\text{up}})$ are defined according to the following asymptotic behaviors in region I,
\begin{align}
 & \psi_{\omega\ell m}^{\text{in-I}}(r)\simeq\begin{cases}
\tau_{\omega\ell m}^{\text{in}}e^{-i\omega_{+}r_{*}} & r_{*}\to-\infty\\
e^{-i\omega r_{*}}+\rho_{\omega\ell m}^{\text{in}}e^{i\omega r_{*}} & r_{*}\to\infty
\end{cases}\label{eq:inI-scattering}\\
 & \psi_{\omega\ell m}^{\text{up-I}}(r)\simeq\begin{cases}
e^{i\omega_{+}r_{*}}+\rho_{\omega\ell m}^\text{up} e^{-i\omega_+ r_*}& r_{*}\to-\infty\\
\tau_{\omega\ell m}^\text{up} e^{i\omega r_*} & r_{*}\to\infty
\end{cases},\label{eq:upI-scattering}
\end{align}
and similarly for the interior \textit{in} and \textit{up} radial solutions and corresponding scattering coefficients ($A_{\omega\ell m}$ and $B_{\omega\ell m}$) in region II, \footnote{In the language of Ref. \cite{2022PhRvD.106l5011Z}, $\psi_{\omega\ell m}^{\text{in-II}}$ corresponds to $\psi_{\omega\ell m}^\text{int}$ (as defined in Eq. (3.16) therein).}
\begin{align}
 & \psi_{\omega\ell m}^{\text{in-II}}(r)\simeq\begin{cases}
e^{-i\omega_{+}r_{*}} & r_{*}\to-\infty\\
A_{\omega\ell m}e^{+i\omega_{-}r_{*}}+B_{\omega\ell m}e^{-i\omega_{-}r_{*}} & r_{*}\to\infty
\end{cases}.\label{eq:inII-scattering}
\end{align}
The \emph{up}-II variant is obtained by complex conjugation. 

Throughout this paper, we use the term \emph{free approximation/variant} to denote that we are approximating the radial solution by a free solution (where applicable). We will do so particularly in the vicinity of the IH, where we will implement the second row of Eq. \eqref{eq:inII-scattering} for the form of $\psi_{\omega\ell m}^{\text{in-II}}$ (and similarly its complex conjugate for $\psi_{\omega\ell m}^{\text{up-II}}$). When we do not rely on such an approximation, i.e. when we employ the numerically obtained \emph{full} radial solution to Eq. \eqref{eq:radial1}, we will use the terminology of \emph{full variant}.

One can derive the following Wronskian relations relating the absolute values of the various scattering coefficients
\begin{subequations}\label{eq:wronskian-relations}
\begin{align}
    &\vert \rho_{\omega\ell m}^\text{in}\vert^2+\frac{\omega_+}{\omega}\vert\tau_{\omega\ell m}^\text{in}\vert^2=1 \\
    &\vert \rho_{\omega\ell m}^\text{up}\vert^2+\frac{\omega}{\omega_+}\vert\tau_{\omega\ell m}^\text{up}\vert^2=1\\
    &\vert B_{\omega\ell m}\vert^2-\vert A_{\omega\ell m}\vert^2=\frac{\omega_+}{\omega_-}.
\end{align}
\end{subequations} 
In particular, it follows that for $\omega \omega_+<0$ the absolute value of the reflection coefficient $\rho_{\omega\ell m}^\text{in/up}$ may exceed $1$. That is: there exist $m\neq 0$ modes (within the so-called \emph{superradiant band}) that are not only reflected by the potential outside the BH, but also amplified, therefore extracting energy (and angular momentum) from the BH. This is the renowned phenomenon of \emph{superradiance} \cite{starobinskii1973amplification,Zeldovich:1971ffh}. In this work, however, we focus on the axis of rotation, where only $m=0$ modes contribute and the potentially-amplifying factors (multiplying the exterior transmission coefficient in Eq. \eqref{eq:wronskian-relations}) reduce to $1$. In other words, at the pole of Kerr there are no superradiance effects, which simplifies the analysis and the underlying physics in that setting.

The Unruh vacuum (which is the state central to the present work), however, is defined in terms of a family of modes that does not decompose in the numerically-favorable form of Eq. \eqref{eq:boulware-mode-ansatz}: the \emph{in} and \emph{up} Unruh modes, which we shall denote by $H_{\omega\ell m}^{\text{in}}$ and $H_{\hat{\omega}\hat{\ell}\hat{m} }^{\text{up}}$. (A detailed description of these modes is given in Sec. III C in Ref. \cite{2022PhRvD.106l5011Z}. Here we merely provide a short review.)
The Unruh modes are solutions to the classical wave equation Eq. \eqref{eq:Klein-Gordon-eq}, defined by their initial data on the null Cauchy hypersurface $\mathcal{H}^{\mathcal{L}}\cup\mathcal{H}^{-}\cup\mathcal{I}^{-}$:
\begin{subequations}
\begin{align}
&H_{\omega\ell m}^{\text{in}}\simeq\frac{Z_{\ell m}^{\omega}(\theta,\varphi)}{\sqrt{4\pi\omega(r^{2}+a^{2})}}\begin{cases}
e^{-i\omega v} & \mathcal{I^{-}}\\
0 & \mathcal{H}^{\mathcal{L}}\cup\mathcal{H}^{-}
\end{cases},\label{eq:in-Unruh-modes}\\
&H_{\hat{\omega}\hat{\ell}\hat{m}}^{\text{up}}\simeq\frac{Y_{\hat{\ell}\hat{m}}(\theta,\varphi_+)}{\sqrt{4\pi\hat{\omega}(r^{2}+a^{2})}}\begin{cases}
0 & \mathcal{I^{-}}\\
e^{-i\hat{\omega}U} & \mathcal{H}^{\mathcal{L}}\cup\mathcal{H}^{-}
\end{cases},\label{eq:up-Unruh-modes}
\end{align}\label{eq:initial-data-Unruh-modes}
\end{subequations}
only for $\omega>0$ and $\hat{\omega}>0$, respectively. Note that following the notation of Ref. \cite{2022PhRvD.106l5011Z} we labeled the \emph{up} Unruh modes with frequency $\hat{\omega}$, in order to distinguish it from the Killing frequency $\omega$. Indeed, while the \emph{in} Unruh modes are decomposable in the usual sense of Eq. \eqref{eq:boulware-mode-ansatz}, the Kruskal-based \emph{up}-modes
will not yield fully-separable field equations. Hence, it is sufficient (and simplifies computations) to consider their angular dependence to be given by the spherical harmonics, indexed by $\hat{\ell}$ and $\hat{m}$, which have the additional benefit of being frequency-independent, 
\begin{equation}
    Y_{\hat{\ell}\hat{m} }(\theta,\varphi_+)=\sqrt{\frac{(2\hat{\ell}+1)(\hat{\ell}-\hat{m})!}{4\pi(\hat{\ell}+\hat{m})!}}P_{\hat{\ell}}^{\hat{m}}(\cos\theta)e^{i\hat{m}\varphi_+},
\end{equation}
with $P_{\hat{\ell}}^{\hat{m}}(\cos\theta)$ the usual associated Legendre polynomials. Moreover, it is useful to rewrite the \emph{up}-Unruh mode solutions in terms of the \emph{up} Eddington-Finkelstein modes (which additionally separate as \eqref{eq:boulware-mode-ansatz});
this is done by decomposing the Unruh \emph{up} initial conditions; see Secs. V and VI in Ref. \cite{2022PhRvD.106l5011Z} for a detailed discussion. 

Finally, to define the Unruh state, the quantum field operator $\hat\Phi$ is expanded in terms of Unruh-mode solutions as
\begin{equation}
\hat{\Phi}(x)=\sum_{\ell,m}\int_{0}^{\infty}d\omega\Biggl(H_{\omega\ell m}^{\text{in}}\hat{a}_{\omega\ell m}^{\text{in}}+H_{\omega\ell m}^{\text{in}\hspace{0.5em}*}\hat{a}_{\omega\ell m}^{\text{in}\hspace{0.5em}\dagger}\Biggr)+\sum_{\hat{\ell}\hat{m}}\int_{0}^{\infty}d\hat{\omega}\Biggl(H_{\hat{\omega}\hat{\ell}\hat{m}}^{\text{up}}\hat{a}_{\hat{\omega}\hat{\ell}\hat{m}}^{\text{up}}+H_{\hat{\omega}\hat{\ell}\hat{m}}^{\text{up}\hspace{0.5em}*}\hat{a}_{\hat{\omega}\hat{\ell}\hat{m}}^{\text{up}\hspace{0.5em}\dagger}\Biggr) \label{eq:QF-operator}
\end{equation}
The operators $\hat{a}_{\omega\ell m}^{\text{in}}$, $\hat{a}_{\omega\ell m}^{\text{in}\hspace{0.5em}\dagger}$ and $\hat{a}_{\hat{\omega}\hat{\ell} \hat{m}}^{\text{up}}$, $\hat{a}_{\hat{\omega}\hat{\ell}\hat{m} }^{\text{up}\hspace{0.5em}\dagger}$are the
standard annihilation and creation operators for the \emph{in} and \emph{up} modes, respectively, satisfying the usual canonical commutation relations (as reviewed in, e.g., Subsec. IV A in \cite{2022PhRvD.106l5011Z}). 

The Unruh state $\vert \text{U}\rangle$
is defined as the vacuum state of the construction, namely the state
for which 
\begin{equation}
\hat{a}_{\omega\ell m}^{\text{in}}\vert \text{U}\rangle=0\hspace{0.7em}\text{for all }\hspace{0.4em}\omega>0,\hspace{0.1em}\ell,m\,;
\end{equation}
\begin{equation}
\hat{a}_{\hat{\omega}\hat{\ell}\hat{ m}}^{\text{up}}\vert \text{U}\rangle=0\hspace{0.7em}\text{for all }\hspace{0.4em}\hat{\omega}>0,\hspace{0.1em}\hat{\ell},\hat{m}\,.
\end{equation}

\section{Polar renormalized $\langle\hat{\Phi}^{2}\rangle$}\label{sec:Polar-vacuum-polarization}

We are interested in computing the renormalized field square $\langle\hat{\Phi}^{2}\rangle$ in the Unruh
state in the interior region of a Kerr BH, focusing on the IH vicinity. The bare (i.e., prior to normalization) field square
is defined as (half) the coinciding-point limit of the HTPF \eqref{eq:2ptf}. The HTPF is a local quantity, quadratic in the field
operator, so its coincidence limit is well-known to be ill-defined.
The difference between the HTPFs in two Hadamard states, however,
is smooth even in the coincidence limit \cite{Hollands:2001nf}. This motivates the use of the state-subtraction method mentioned in the introduction: we renormalize by subtracting the HTPF of a "comparison" state $\vert\text{C}\rangle$, and only then take the coincidence limit.
Namely, as follows from \eqref{eq:2pt-def}, the (state-independent) Hadamard parametrix drops out and we are left with
\begin{equation}
\langle\hat{\Phi}^2(x)\rangle^{\text{U-C}}\equiv\lim_{x'\to x}\left(G^{\text{U}}(x,x')-G^{\text{C}}(x,x')\right),\label{eq:state-subtraction-definition}
\end{equation}
where $G^{\text{U}}(x,x')$ and $G^{\text{C}}(x,x')$ are the HTPFs in the Unruh state and in the comparison state, respectively. For our purpose here -- namely, capturing the asymptotic behavior of $\langle\hat{\Phi}^{2}\rangle^\text{U}$ in the IH vicinity -- a suitable choice for $\vert{\text{C}}\rangle$  is the so-called $\vert\underline{\text{U}}\rangle$ state, introduced by some of us in \cite{2022PhRvL.129z1102Z}. In practice, as standard, we implement the state-subtraction prescription mode by mode (as we did in our previous
work on quantum fluxes at the IH of a Kerr BH \cite{2022PhRvL.129z1102Z}).  
This requires mode-sum expressions for the HTPFs of the Unruh state and of the $\vert\underline{\text{U}}\rangle$ state inside Kerr, written in terms of numerically-amenable Eddington-Finkelstein modes (i.e., modes that are decomposable according to Eq. (\ref{eq:boulware-mode-ansatz})). The corresponding construction for the Unruh state was performed in Ref.~\cite{2022PhRvD.106l5011Z}, and the analogous construction for the  $\vert\underline{\text{U}}\rangle$  state is provided in  Appendix \ref{sec:appendixA} .

In the near-IH regime (where the radial solution becomes free, see Eq. \eqref{eq:inII-scattering}), we find (by subtracting \eqref{eq:GbarU} from Eq. (6.37) in Ref. \cite{2022PhRvD.106l5011Z}, and taking the coinciding point limit of the difference)
\begin{align}
\langle\hat{\Phi}^{2}(x)\rangle^{\text{U}-\underline{\text{U}}} & \simeq \frac{\hbar}{r_{-}^{2}+a^{2}}\sum_{\ell,m}\int_{0}^{\infty}d\omega\frac{\big[ S^\omega_{\ell m}(\theta)\big]^{2}}{8\pi^{2}}\Bigg\{\omega_{+}^{-1}E_{\omega\ell m}^{\text{up}(0)}+\omega^{-1}E_{\omega\ell m}^{\text{in}(0)}+2\Re\left\{ \Big(\omega_{+}^{-1}E_{\omega\ell m}^{\text{up}(1)}+\omega^{-1}E_{\omega\ell m}^{\text{in}(1)}\Big)e^{2i\omega_{-}r_{*}}\right\} \nonumber \\
 & \hspace{1em}-\omega^{-1}\vert\underline{\tau}_{\ \ \omega\ell m}^{\text{out}}\vert^{2}-\omega_{-}^{-1}\coth\left(\frac{\pi\omega_{-}}{\kappa_{-}}\right)(1+\vert\underline{\rho}_{\ \ \omega\ell m}^{\text{down}}\vert^{2})-2\text{\ensuremath{\omega_{-}^{-1}}csch\ensuremath{\left(\frac{\pi\omega_{-}}{\kappa_{-}}\right)}}\Re\left\{ \underline{\rho}_{\ \ \omega\ell m}^{\text{down}}e^{2i\omega_{-}r_{*}}\right\} \Biggr\}  \quad r\to r_-,\label{eq:mode-sum-phi2}
\end{align}
The underlined coefficients $\underline{\tau}_{\ \ \omega\ell m}^{\text{down}}$ and $\underline{\rho}_{\ \ \omega\ell m}^{\text{down}}$ correspond to the scattering problem of
the comparison state ($\underline{\text{U}}$) modes in region III and are defined explicitly
in the Appendix  (see Eq. \eqref{eq:f-down-psi}). We introduce the abbreviations
\begin{equation}\label{eq:integrands}
\begin{aligned}
 & E_{\omega\ell m}^{\text{up}(0)}=\coth\left(\frac{\pi\omega_{+}}{\kappa_{+}}\right)\left(\vert B_{\omega\ell m}\vert^{2}+\vert A_{\omega\ell m}\vert^{2}\right)\left(1+\vert\rho_{\omega\ell m}^{\text{up}}\vert^{2}\right)+4\text{csch}\left(\frac{\pi\omega_{+}}{\kappa_{+}}\right)\Re\left\{ \rho_{\omega\ell m}^{\text{up}}A_{\omega\ell m}B_{\omega\ell m}\right\} ,\\
 & E_{\omega\ell m}^{\text{up}(1)}=\coth\left(\frac{\pi\omega_{+}}{\kappa_{+}}\right)A_{\omega\ell m}B_{\omega\ell m}^{*}\left(1+\vert\rho_{\omega\ell m}^{\text{up}}\vert^{2}\right)+\text{csch}\left(\frac{\pi\omega_{+}}{\kappa_{+}}\right)\left(\rho_{\omega\ell m}^{\text{up}*}(B_{\omega\ell m}^{*})^{2}+\rho_{\omega\ell m}^{\text{up}}(A_{\omega\ell m})^{2}\right),\\
 & \hspace{2em}E_{\omega\ell m}^{\text{in}(0)}=\vert\tau_{\omega\ell m}^{\text{in}}\vert^{2}\left(\vert A_{\omega\ell m}\vert^{2}+\vert B_{\omega\ell m}\vert^{2}\right)\hspace{2em},\hspace{2em}E_{\omega\ell m}^{\text{in}(1)}=\vert\tau_{\omega\ell m}^{\text{in}}\vert^{2} A_{\omega\ell m}B_{\omega\ell m}^{*} .\\
\end{aligned}
\end{equation}
In this paper we exclusively focus on evaluation
of the mode-sum along the axis of rotation, i.e. $\theta=0$. In this
case, the spheroidal harmonics obey $S_{\ell m}^{\omega}(\theta=0)\propto\delta_{m0}$,
meaning that only the $m=0$ modes contribute (no superradiant effects) and, in
particular, $\omega_{+}=\omega=\omega_{-}$, which simplifies the
evaluation of the mode sum above.
Henceforth, all equations should be understood as holding only at the pole. In certain central equations, we indicate this explicitly by adding $(\theta=0)$ for emphasis.
The integrand in
Eq. \eqref{eq:mode-sum-phi2} naturally splits into an $r_{*}$-independent 
and an $r_{*}$-dependent part, and we compute both of them in this section.
The $r_*$-independent part of $\langle\hat{\Phi}^{2}\rangle^{\text{U}-\underline{\text{U}}}$ in the IH vicinity can be evaluated directly numerically (see Appendix~\ref{sec:NumCoeffs} for details) and yields a constant, which we denote by $c_-$, i.e. 
\begin{equation}
    c_-\equiv \frac{\hbar}{r_{-}^{2}+a^{2}}\sum_{\ell}\int_{0}^{\infty}d\omega\frac{\big[ S^\omega_{\ell 0}(\theta=0)\big]^{2}}{8\pi^{2}\omega}\Big\{E_{\omega\ell 0}^{\text{up}(0)}+E_{\omega\ell 0}^{\text{in}(0)} -\vert\underline{\tau}_{\ \ \omega\ell 0}^{\text{out}}\vert^{2}-\coth\left(\frac{\pi\omega_{-}}{\kappa_{-}}\right)(1+\vert\underline{\rho}_{\ \ \omega\ell 0}^{\text{down}}\vert^{2})\Big\},
 \end{equation}
where $E_{\omega\ell 0}^{\text{up}(0)}$ and $E_{\omega\ell 0}^{\text{in}(0)}$ are given in Eq. \eqref{eq:integrands}. In our case (of $a/M=0.8$),
\begin{equation}\label{cbar}
c_-=-0.003857694\,\hbar\,M^{-2}.
\end{equation}
We then write the remaining $r_*$-dependent contribution (at the polar IH vicinity) as
\begin{equation}
\langle\hat{\Phi}^{2}(x)\rangle^{\text{U}-\underline{\text{U}}}
\simeq c_{-}+\frac{\hbar}{r_{-}^{2}+a^{2}}\sum_{\ell=0}^{\infty}\int_{0}^{\infty}d\omega\,
\frac{\big[ S^\omega_{\ell 0}(\theta=0)\big]^{2}}{8\pi^{2}}\,
\Re\!\left(E_{\omega\ell0}\,e^{2i\omega r_{*}}\right)  \quad r\to r_-,
\label{eq:phi2-schematic}
\end{equation}
where the $r_*$-independent prefactor $E_{\omega\ell 0}$ is obtained from (the $\theta=0$ version of) Eqs.~\eqref{eq:mode-sum-phi2}-\eqref{eq:integrands} and depends on the $m=0$ scattering coefficients in regions I, II and III. I.e.,
\begin{equation}
   E_{\omega\ell 0}\equiv \frac{2}{\omega} \Big\{ E_{\omega\ell 0}^{\text{up}(1)}+E_{\omega\ell 0}^{\text{in}(1)}-\text{csch}\Big( \frac{\pi\omega}{\kappa_-}\Big)\underline{\rho}^\text{down}_{\hspace{2mm}\omega\ell 0}\Big\},\label{eq:Ewl0}
\end{equation}
where $E_{\omega\ell 0}^{\text{up}(1)}$ and $E_{\omega\ell 0}^{\text{in}(1)}$ are again given in Eq. \eqref{eq:integrands}.
Accordingly, we define the $r_*$-dependent part of $\langle\hat{\Phi}^{2}(x)\rangle^{\text{U}-\underline{\text{U}}}$ in the IH vicinity by
\begin{equation}\label{eq:delta-def}
\delta\langle\hat{\Phi}^{2}(x)\rangle^{\text{U}-\underline{\text{U}}}
\equiv
\langle\hat{\Phi}^{2}(x)\rangle^{\text{U}-\underline{\text{U}}}-c_- \, .
\end{equation}
Note that, due to the nontrivial oscillatory dependence of the integrand in \eqref{eq:phi2-schematic} on $r_*$, it is not a priori clear that the mode sum for  $\delta\langle \hat{\Phi}^{2}\rangle^{\text{U}-\underline{\text{U}}}$ (and thus for $\langle \hat{\Phi}^{2}\rangle^\text{U}_{\rm ren}$) converges to a finite limit as the IH is approached ($r_*\to\infty$). In what follows, we address this issue: we show that the limit is indeed finite, and we expose the intricate behavior of the field square as the IH is approached.
It is important to note that, while this (irregular) behavior is exposed in the difference $\vert\text{U}\rangle -\vert \underline{\text{U}}\rangle$, it originates entirely from the Unruh state (since the $\vert \underline{\text{U}}\rangle$ state is specifically constructed to be regular (Hadamard) across $\mathcal{CH}^\mathcal{R}$; see discussion to follow in Section \ref{subsec:Comparison-to-t-splitting}). Consequently, the renormalized field square in this state is smooth in $r$ at $r=r_-$. The Unruh state, in contrast, has been shown to be regular (Hadamard) up to, but not including, the IH \cite{Hafner:2026zcl}. Accordingly, the renormalized field square in the Unruh state is expected to be smooth for any $r<r_-$, but not necessarily at $r=r_-$ itself.

\subsection{Price-like tail}\label{subsec:Price-tail} 

Sufficiently close to the IH,
 $\langle\hat{\Phi}^{2}\rangle^{\text{U}-\underline{\text{U}}}$ is given by Eq.~\eqref{eq:phi2-schematic}, where the $r_{*}$-dependence only enters through the oscillatory
term $\propto e^{2i\omega r_{*}}$. Before turning to numerics for the integral term, we note that the large-$r_*$ behavior of the $\omega$ integral is controlled by the small-$\omega$ expansion
of the prefactor multiplying $\exp{(2i\omega r_*)}$ [proportional to $E_{\omega \ell 0}$; see Eq.~\eqref{eq:phi2-schematic}]. Indeed, after adding a small positive imaginary part to $r_*$ to regulate the integral (and taking it to zero at the end), one has \cite{Mathematica14p1} 
\begin{subequations}
\begin{align}
    \int_{0}^{\infty}d\omega\ \omega^{n}e^{2i\omega r_{*}}&=\frac{n!}{(-2ir_{*})^{n+1}},&\text{for integer }n\geq 0,\label{eq:small-w}\\
\int_{0}^{\infty}d\omega\ \omega^{n} \ln (\omega M)e^{2i\omega r_{*}}&=\frac{n!}{(-2ir_{*})^{n+1}}\Big(\frac{i\pi}{2}-\gamma+H_n -\ln(2r_*/M)\Big),&\text{for integer }n\geq 0,\label{eq:small-w-log}
\end{align}\label{eq:omega-power-integral}
\end{subequations}
where $\gamma$ is the Euler-Mascheroni constant and $H_n$ is the $n$-th harmonic number. Given Eqs.~\eqref{eq:phi2-schematic}  and \eqref{eq:small-w}, the asymptotic tails generated by non-logarithmic terms in the small-$\omega$ expansion of $E_{\omega\ell 0}$ are determined by either the real or the imaginary part of the relevant expansion coefficient, with only one of them contributing at a given order. By contrast, Eq.~\eqref{eq:small-w-log} implies that logarithmic terms generate tails receiving contributions from both the real and imaginary parts of the same coefficient.\footnote{In Eq. \eqref{eq:mode-sum-phi2} (and everywhere throughout this paper), the comparison state contribution is specific to the $\vert\underline{\text{U}}\rangle$ state. However, it is also possible to renormalize $\langle\hat{\Phi}^2\rangle$ using other classes of comparison states –– an analysis of a different family of comparison states which serves this purpose is currently in preparation \cite{Alberti:prep}. For each of these comparison states, the small-$\omega$ behavior of the analog of $\underline{\rho}_{\ \ \omega\ell m}^\text{down}$ (the related reflection coefficient in region III)
depends on the comparison state, therefore so does the integrand function. However, all the irregular features at the IH, including the inverse-power tails and the QNM ringing, are independent of the comparison state (as long as it is Hadamard at the IH).\label{ft:comparison-state-2}} 
We carried out this analysis analytically for the $\ell=0$ mode (which will dominate at sufficiently large $r_*$, see below) and found (i) that this mode has leading inverse-power tail $\propto r_*^{-3}$ (with a prefactor obtained here analytically for the first time), and (ii) that there are no $r_* \ln(r_*/M)$ terms at this order. Namely, 
\begin{equation}\label{eq:c0-def}
    \delta \langle\hat{\Phi}^{2}_{\ell=0}\rangle^{\text{U}-\underline{\text{U}}}=c_0r_*^{-3}+o(r_*^{-4}),
\end{equation}
where $\delta \langle\hat{\Phi}^{2}_{\ell=0}\rangle^{\text{U}-\underline{\text{U}}}$ is the $\ell=0$ contribution to $\delta \langle\hat{\Phi}^{2}\rangle^{\text{U}-\underline{\text{U}}}$ (defined in Eq. \eqref{eq:delta-def}), and $c_0$ is analytically found to be

\begin{equation}\label{eq:c0-analytical}
    c_0=-\hbar M^4\frac{11\kappa_-}{3\pi^2(r_-^2+a^2)}.
\end{equation}
Details of the computation can be found in Appendix \ref{sec:appendixC} and in \cite{Alberti:2026kerrNB}; Appendix \ref{sec:appendixD} describes the derivation of the necessary small-$\w$ expansions of the various scattering coefficients. The analytical result in  Eq. \eqref{eq:c0-analytical} is plotted against the numerics in Fig.~\ref{fig:l0tail-c0} in Appendix~\ref{sec:appendixC} for two values of $a/M$. As shown in \cite{2022PhRvD.106l5011Z} (see in particular Eqs.~(B49)--(B50) therein), in the IH limit $r_*\to\infty$, the mode-sum expressions for the bare fluxes $\langle \hat{T}_{uu}\rangle_{\text{bare}}$ and $\langle \hat{T}_{vv}\rangle_{\text{bare}}$ (in the minimal coupling case) become independent of $r$: in particular, the oscillatory contributions $\propto e^{\pm 2 i \omega_- r_*}$ (originating from terms quadratic in the radial  functions and their derivatives) drop out.
As evident from Eq. (\ref{eq:mode-sum-phi2}), this simplification does \emph{not} occur for $\langle\hat{\Phi}^{2}\rangle$ (nor, for example, for $\langle \hat{T}_{uv}\rangle$ or $\langle \hat{T}_{\theta\theta}\rangle$): the mode sum contains a rapidly-oscillatory dependence on $r_*$. Accordingly, in order to evaluate the $r_{*}$-dependent piece
of the mode-sum \eqref{eq:phi2-schematic}, we perform the computation for a sequence of finite, increasingly large values of $r_*$, and only then infer the asymptotic behavior as $r_{*}\to\infty$. The mode-sum computation involves both an  $\ell$ sum and an $\omega$ integral, with an integrand that decays exponentially in both $\ell$ and $\omega$. 

\begin{figure}
\begin{centering}
\includegraphics[scale=0.7]{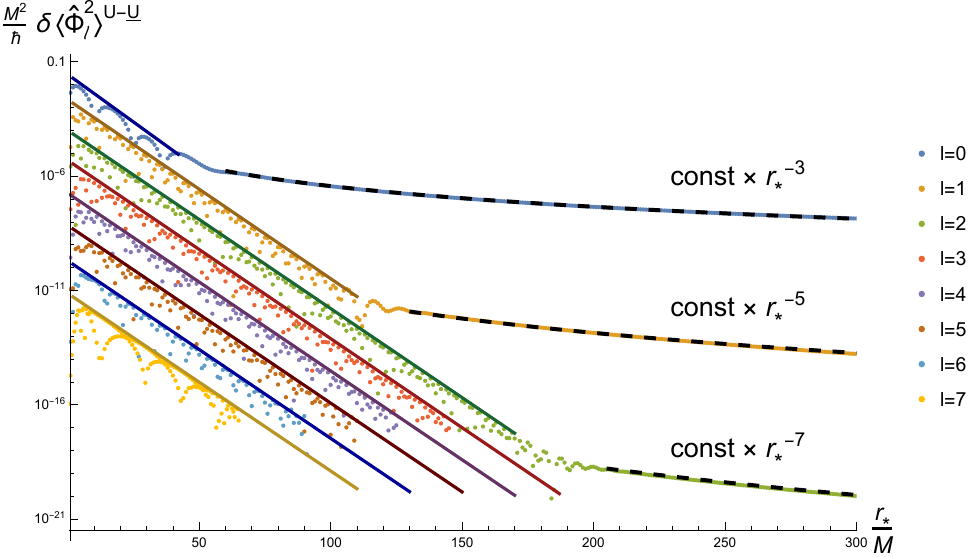}
\par\end{centering}
\begin{centering}
\caption{Near-IH individual $\ell$ contributions to $\delta \langle\hat{\Phi}^{2}\rangle^{\text{U}-\underline{\text{U}}}$ (defined in Eq. \eqref{eq:delta-def}). Each curve shows a transient 
 ringdown followed by an inverse-power tail. The points correspond to numerically computed values, and the tails are accompanied by their respective powers, displayed in dashed black lines. As detailed in Subsec. \ref{subsec:QNM-ringdown} below, the ringdown is fitted with the fundamental $(n=0)$, $m=0$ QNMs according to Eq. \eqref{eq:qnm-fit}.  (See also footnote \ref{fn:artificial}.) Colorful solid lines correspond to the fitted exponentially decaying envelope. }\label{fig:QNM-ringdown-and-tail}
\par\end{centering}
\end{figure}

\begin{figure}
\begin{centering}
\includegraphics[scale=0.7]{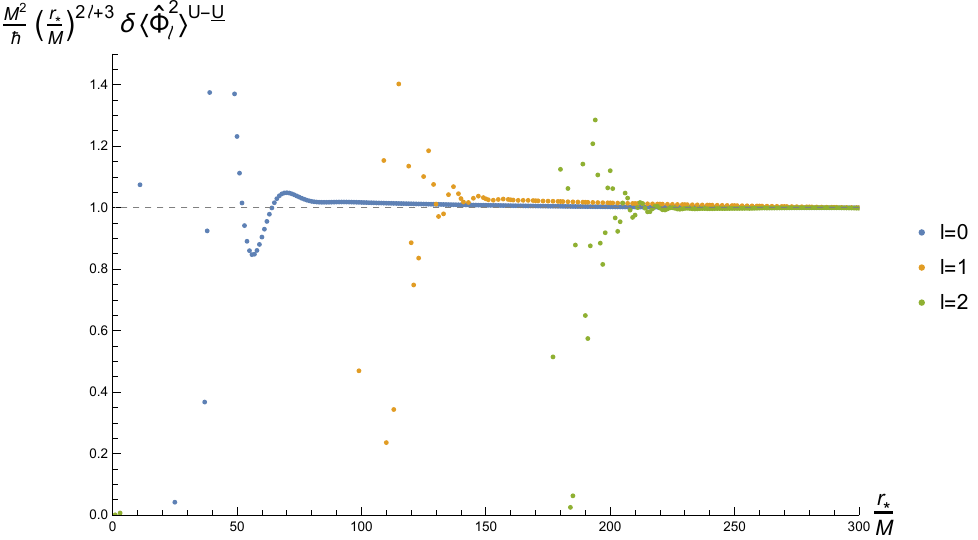}
\par\end{centering}
\caption{Individual $\ell$ contributions to $\delta \langle\hat{\Phi}^{2}\rangle^{\text{U}-\underline{\text{U}}}$ divided by their respective 
inverse power tails $(r_*/M)^{-n_\ell}$ with
$n_{\ell}=2\ell+3$, illustrating the approach to a constant at large $r_*$ (which is normalized here to 1).}
\label{fig:Inverse-power-tails-normalized}

\end{figure}

We find
it more insightful to display the results corresponding to first integrating over $\omega$
for fixed $\ell$, and to display the individual $\ell$ contributions before performing the $\ell$ sum, see Fig.~\ref{fig:QNM-ringdown-and-tail}. The
results portrayed in this figure show, for every $\ell$, exponentially decaying oscillations
followed by an inverse-power tail of the form $r_{*}^{-2\ell-3}$,
clearly reminiscent of late-time Price tails \cite{Price:1971fb} (which arise in the context of classical perturbations, upon
the substitution $t\rightarrow r_{*}$). As evident from the figure, the magnitude of these tails decreases with $\ell$, and the transient regime where the ringdown dominates before giving away to the final asymptotic tail increases with $\ell$ (hence, only the $\ell=0,1,2$ tails are exposed here).
The tails, rescaled by their respective inverse powers, are presented in Fig. \ref{fig:Inverse-power-tails-normalized}.  This figure allows to corroborate the tails (via the approach to a plateau at large $r_*$, normalized to 1) as well as to appreciate the quality of the asymptotic tail behavior. Deviations from the plateau at large $r_{*}$ are expected to arise due
to numerical artifacts, which can be understood as follows. The integrand oscillates with a characteristic wavelength which shrinks like $1/r_*$.
Then, to reliably resolve these oscillations numerically (e.g., by sampling
$\mathcal{O}(10)$ points per oscillation), the scattering coefficients
must be computed on an $\omega$-grid with a spacing of $d\omega=\mathcal{O}(1/(10r_{*}))$. This becomes increasingly challenging at very large $r_{*}$ values. The numerical strategy as to how to overcome this is presented in Appendix \ref{sec:appendixB}.
This limitation is essentially $\ell$-independent, but it is more pronounced for larger $\ell$ because the plotted quantities are multiplied by larger powers of
$r_{*}$, which amplifies any residual error. Moreover, exposing the $\ell=2$ and higher tails
is intrinsically more challenging because it requires higher precision
in the scattering coefficients (see vertical scale in Fig. \ref{fig:QNM-ringdown-and-tail}).

\subsection{QNM ringdown}\label{subsec:QNM-ringdown}

In the pre-tail regime, we observe exponentially decaying oscillations (as visible in Fig. \ref{fig:QNM-ringdown-and-tail}), whose amplitudes are
exponentially suppressed with increasing $\ell$. 

\begin{figure}
\hfill{}\subfloat[\label{fig:l0-ring}$\ell=0$ ]{
\centering{}\includegraphics[width=0.45\columnwidth]{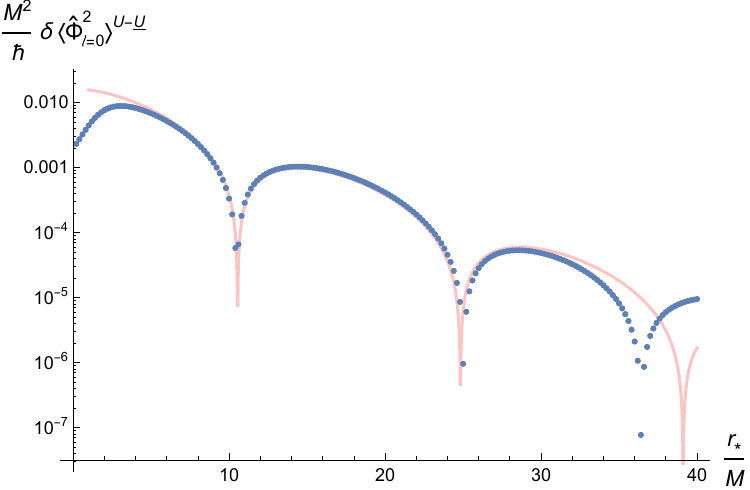}}\hfill{}\subfloat[\label{fig:l1-ring}$\ell=1$]{
\centering{}\includegraphics[width=0.45\columnwidth]{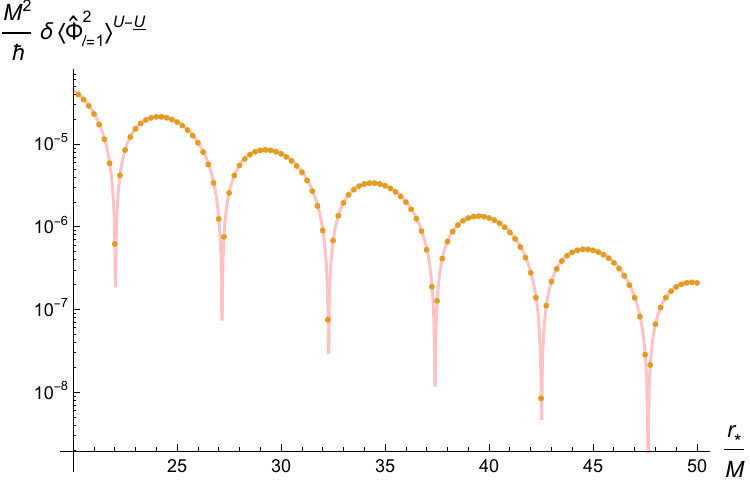}}\hfill{}

\subfloat[\label{fig:Estrella-1}$\ell=2$]{
\centering{}\includegraphics[width=0.45\columnwidth]{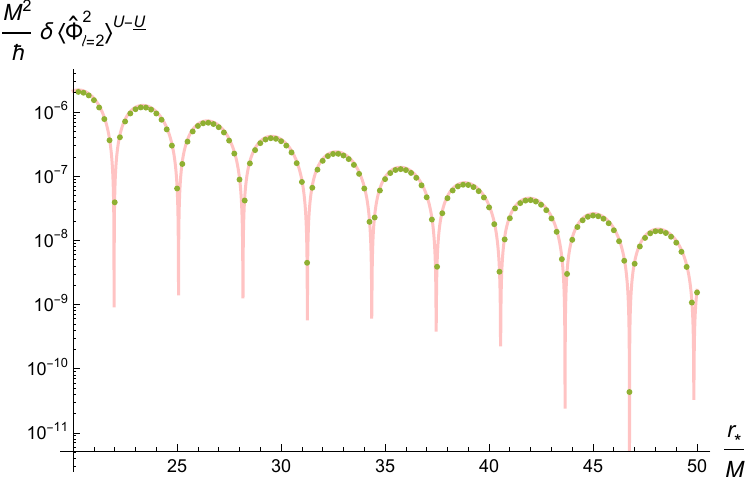}}\hfill{}\subfloat[\label{fig:l3-ring}$\ell=3$]{
\centering{}\includegraphics[width=0.45\columnwidth]{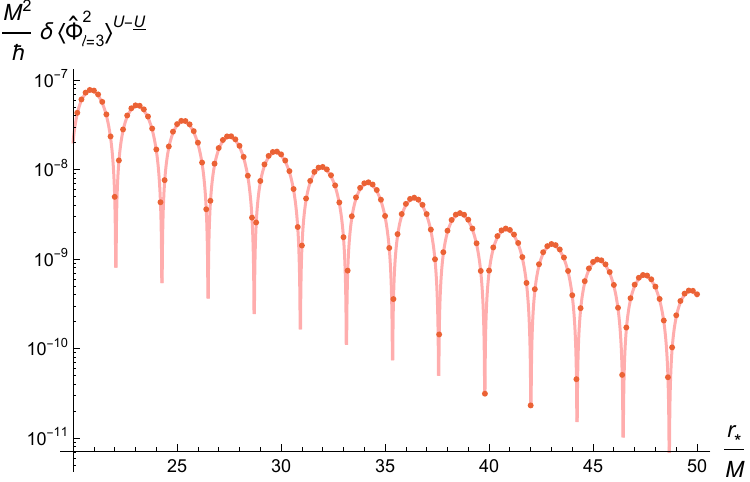}}\hfill{}

\caption{Zoomed-in view of the ringdown phase in the individual $\ell$ contributions $\delta\langle\hat{\Phi}_\ell^2\rangle^{\text{U}-\underline{\text{U}}}$ for $\ell=0,1,2,3$. In each plot the points are numerically computed and the solid curve  displays a fit according to Eq. \eqref{eq:qnm-fit}, namely  $A_\ell\exp{(-2i\omega_{n=0,\ell,m=0}^{\text{QNM}}r_*)}$, with numerically-determined complex amplitudes $A_\ell$. In particular, as is well-known for QNM frequencies, as $\ell$ increases in the eikonal limit (see, e.g., \cite{Dolan:2010wr}) the oscillation frequency grows linearly, while the decay rate approaches a known constant.
} \label{fig:ringdown}
\end{figure}

Remarkably, for all the $\ell$
values we computed ($\ell\leq7$), the (complex) frequencies of these oscillations
were numerically found to match\footnote{Note that in Fig. \ref{fig:QNM-ringdown-and-tail}, the artificial ringing observed for $\ell=7$ (and less notably for
$\ell=2)$ is a sampling artifact associated with the discrete spacing $\Delta r_{*}/M=1/2$.
The real part of the ``physical'' QNM frequency is, for $\ell=7$, $\Re\{\omega^{\text{QNM}}_{070}\}\approx1.151/M$, yielding a period of about $4M$ in $r_*$.} \label{fn:artificial}\emph{twice }the known fundamental ($n=0$) polar ($m=0$) QNM frequencies for
scalar perturbations in Kerr (which can be found e.g., in \cite{emmanuele-berti-QNM}). 
This observation is illustrated for $\ell\leq3$ in Fig. \ref{fig:ringdown} . I.e., the observed behavior of $\delta\langle\hat{\Phi}_\ell^2\rangle^{\text{U}-\underline{\text{U}}}$ (at $\theta=0$) in the ringdown $r_*$-regime (i.e., sufficiently near the IH but still before the leading asymptotic tail) is well described by 
\begin{equation}\label{eq:qnm-fit}
\approx \Re\big\{A_\ell\exp{(-2i\omega_{n=0,\ell,m=0}^{\text{QNM}}r_*)}\big\},
\end{equation}
where $\omega_{n=0,\ell,m=0}^{\text{QNM}}$ denotes the fundamental $m=0$ scalar QNM frequency and $A_{\ell}$ represents the numerically-fitted complex amplitude.
We refer to Appendix \ref{sec:appendixB} for details on the accuracy of this result.

\subsection{Comparison to $t$-splitting results}\label{subsec:Comparison-to-t-splitting}

In the subsections above we have computed $\langle\hat{\Phi}^{2}\rangle^{\text{U}-\underline{\text{U}}}$, the difference between  $\langle\hat{\Phi}^{2}\rangle$ in
the Unruh state and in the comparison state $\vert \underline{\text{U}}\rangle$, and we have found that this quantity consists, up to regular terms, of a superposition of QNM ringings and Price-like tails, both in $r_*$. Two key questions arise: (1) how does the choice of comparison state influence the results, and (2) are the observed ringing and tail features inherent to the (physically relevant) Unruh state? These two questions are intimately related; their resolution relies on the regularity of the comparison
state. Indeed, the (only) motivation for constructing a comparison state for our purpose is that it will be Hadamard across $\mathcal{CH}^{\mathcal{R}}$, see Penrose diagram in Fig. \ref{fig:Penrose-diagram-fKerr}. This Hadamard property implies that, near the IH, expectation values are smooth functions of any coordinate regular across the IH (e.g., $r$). Considering the near-IH asymptotics of 
$r_{*}\simeq(-2\kappa_{-})^{-1}\ln \vert (r-r_-)/M\vert$,
we find that neither the ringdown in Eq. \eqref{eq:qnm-fit} nor the inverse-power tails are smooth functions of $r$. Therefore, since by construction the comparison state cannot introduce non-smooth behavior (in $r$), we conclude that the ringing and tails are features of the Unruh state \footnote{This claim is further supported by the fact that we repeated the numerical analysis
with several different comparison states  used as regulators, and in each case obtained identical results up to regular terms (i.e., same values of $A_\ell$ and $c_\ell$, as defined in \eqref{eq:IH-phi2-rstar}). A detailed account of these additional computations is beyond the scope of the current paper.}. Then, to summarize our findings, the (regularized)  individual-$\ell$ contribution to the Unruh-state field square near the IH admits the asymptotic form 
\begin{equation}
\langle\hat{\Phi}_{\ell}^{2}\rangle^{\text{U}}\simeq \text{const.}+c_{_\ell}r_{*}^{-2\ell-3}+ \Re\big\{A_{\ell}e^{-2i\omega_{0\ell0}^{\text{QNM}}r_{*}}\big\}+(...) \hspace{4em}(r\to r_{-}),\label{eq:IH-phi2-rstar}
\end{equation}
where $c_\ell$ is a real constant (with $c_0$ given analytically in Eq. \eqref{eq:c0-analytical}), $A_\ell$ was introduced in Eq. \eqref{eq:qnm-fit}, and $(...)$ is expected to include overtones as well as higher-order inverse-power tails in $r_*$ (such tails may also be accompanied by logarithmic factors of $\ln(r_*/M)$; see, e.g., Ref.~\cite{PhysRevD.94.124053}, where analogous  $\ln(t/M)$ factors arise in the late-time behavior outside the EH). 

For fixed $\ell$, the ringing dominates the irregular
behavior only over a finite range of $r_*$, and not asymptotically (similar to what happens to the classical field outside the BH as a function of $t$ instead of $r_*$), 
see Fig. \ref{fig:QNM-ringdown-and-tail}. The ringing is therefore an intermediate-domain phenomenon, which, particularly for low $\ell$, does not include many cycles. As such, its numerical investigation does not  allow us to exactly determine the leading complex frequency. However, fitting the oscillations within the domain where they constitute the dominant behavior, we find good agreement with the known fundamental quasinormal frequencies for scalar perturbations in Kerr; see Appendix \ref{sec:appendixB} for numerical details. The sum over $\ell$ of the individual 
$\langle\hat{\Phi}^{2}_\ell\rangle^{\text{U}}$ contributions is dominated by the $\ell=0$ mode (because of the $r_*^{-2\ell-3}$ asymptotic behavior), as shown in Fig. \ref{fig:QNM-ringdown-and-tail} and Subfig. \ref{fig:l0 dominance}.
Concretely, defining $\langle\hat{\Phi}^{2}\rangle_{-}^{\text{U}}$ as the polar limiting value of $\langle\hat{\Phi}^{2}\rangle^{\text{U}}$ at the IH, we find
\begin{equation}
\langle\hat{\Phi}^{2}\rangle^{\text{U}}=\langle\hat{\Phi}^{2}\rangle_{-}^{\text{U}}+c_{0}r_{*}^{-3}+o(r_{*}^{-4})\hspace{4em}(r\to r_-),\label{eq:phi2-sum-ell}
\end{equation}
where $c_0$ is given in \eqref{eq:c0-analytical}. Notably, this final tail is exposed only extremely close to the IH, at $r_*\approx 60M$ (corresponding to $r-r_- \sim10^{-40}M$ for $a=0.8M$), as seen in Fig. \ref{fig:QNM-ringdown-and-tail}.

Determination of the 
const. term in Eq. \eqref{eq:IH-phi2-rstar} cannot be done within our state-subtraction method, since it would require knowledge of the near-IH regular behavior of the comparison state $\vert \underline{\text{U}}\rangle$ itself. Fortunately, some
of us have recently studied $\langle\hat{\Phi}^{2}\rangle$ in the Unruh
state in the interior region of Kerr along the axis of rotation using more direct renormalization
techniques. Specifically, in Ref. \cite{Zilberman:2024jns} $\langle\hat{\Phi}^{2}\rangle^{\text{U}}$
was computed throughout (except for on and asymptotically near the IH) the polar Kerr BH interior using a point-splitting
regularization scheme -- namely, the $t$-splitting variant of PMR \cite{Levi:2015eea,Levi:2016esr,Levi:2016quh}, extended to the Kerr interior in \cite{Zilberman:2024jns}. 
 However, such a splitting direction becomes null on approaching the polar IH, 
 making the numerical implementation of the method challenging in that regime.
In addition to the general difficulty associated with the $t$-splitting direction becoming null near the polar IH, extracting the limiting value of $\langle\hat{\Phi}^2\rangle$ is further complicated by the nontrivial, nonmonotonic behavior investigated here.  For these reasons, we find that the near-IH behavior of observables like
$\langle\hat{\Phi}^{2}\rangle$ is most efficiently explored by supplementing the point-splitting approach with other methods, as we do here. In the following, we describe how to combine the results obtained for $\langle\hat{\Phi}^2\rangle^\text{U}$  in Ref. \cite{Zilberman:2024jns} (through $t$-splitting) and the novel results for $\langle\hat{\Phi}^2\rangle^{\text{U}-\underline{\text{U}}}$ obtained here (through state-subtraction)
to isolate the regular contribution of the comparison state, and ultimately extract the limiting polar IH value $\langle\hat{\Phi}^{2}\rangle^\text{U}_-$  (as defined in Eq. \eqref{eq:phi2-sum-ell}).

\begin{figure}
\begin{centering}
\hfill{}\subfloat[\label{fig:comparison-tsplitting} 
State-subtraction results for near-IH $\langle\hat{\Phi}^{2}\rangle^{\text{U}-\underline{\text{U}}}$ (orange points)
 along with $t$-splitting results from Ref.
\cite{Zilberman:2024jns} for near-IH $\langle\hat{\Phi}^{2}\rangle^\text{U}$ (blue crosses) in their overlap domain.
The state-subtraction results in the displayed $r$-regime were computed using
the \emph{full} radial solution (rather than the \emph{free} approximation; see paragraph below Eq. \eqref{eq:inII-scattering}).]{
\centering{}\includegraphics[width=0.46\columnwidth]{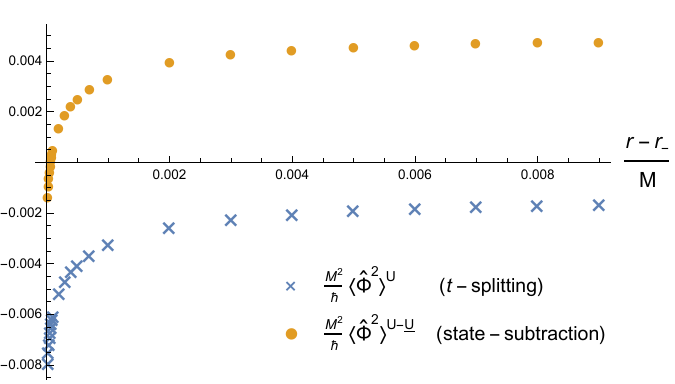}}\hfill{}\subfloat[Results for near-IH $\langle\hat{\Phi}^{2}\rangle^{\underline{\text{U}}}$
obtained according to Eq. \eqref{comparison-state-contribution}. Blue points correspond to numerically implementing the \emph{full} variant -- plotted in panel (a)-- in the state-subtraction contribution to Eq. \eqref{comparison-state-contribution}, whereas orange points correspond to implementing the \emph{free} variant. The solid lines show linear fits in $(r-r_{-})/M$, consistent with the expected regular behavior of the comparison state near the IH. The difference between the \emph{free} and \emph{full} variants manifests as an additional linear deviation in $(r-r_{-})/M$ that vanishes in the IH limit.  \label{fig:phi2-comparison-state}
]{\begin{centering}
\includegraphics[scale=0.8]{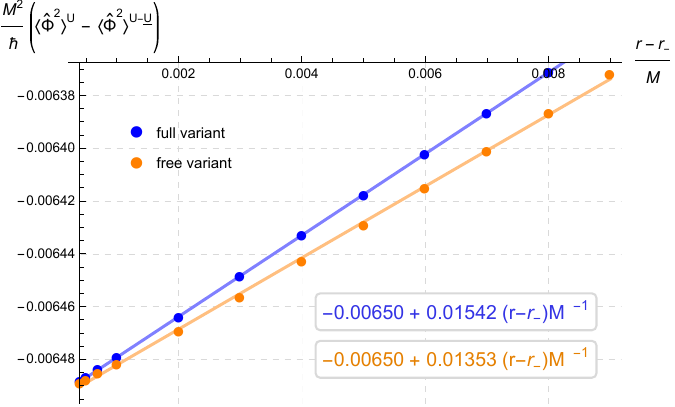}
\par\end{centering}
}
\par\end{centering}
\caption{Comparison between $t$-splitting results for $\langle\hat{\Phi}^{2}\rangle^\text{U}$ from Ref. \cite{Zilberman:2024jns} and
our state-subtraction 
results for $\langle\hat{\Phi}^{2}\rangle^{\text{U}-\underline{\text{U}}}$ (obtained using Eq.~\eqref{eq:phi2-schematic} with \eqref{cbar}), as functions of $(r-r_-)/M$ in their overlap $r$-domain. The left panel shows both quantities, while the right panel shows their difference (which corresponds via Eq. \eqref{comparison-state-contribution} to $\langle\hat{\Phi}^{2}\rangle^{\underline{\text{U}}}_\text{ren}$).}
\end{figure}

\begin{figure}
\begin{centering}
\hfill{}\subfloat[\label{fig:l0 dominance} 
$ \langle\hat\Phi^2\rangle^\text{U}$, defined as the difference between $ \langle\hat\Phi^2\rangle^\text{U}$ and its asymptotic value at the IH, denoted $ \langle\hat\Phi^2\rangle_{-}^\text{U}$ and given in Eq. \eqref{eq:phi2(-)}, i.e., $ \langle\hat\Phi^2\rangle^\text{U}\equiv\langle\hat{\Phi}^{2}\rangle^{\text{U}}-\langle\hat{\Phi}^{2}\rangle_{-}^{\text{U}}$. The $\ell=0$, $\ell=1$ and "total" (summed up to $\ell=7$) contributions are shown separately in blue, green, and orange, respectively. The points and crosses are numerically computed, whereas the solid lines are interpolated. The orange crosses are seen to nearly coincide with the blue points, illustrating the dominance of $\ell=0$ in the $\ell$-sum.]{
\centering{}\includegraphics[width=0.45\columnwidth]{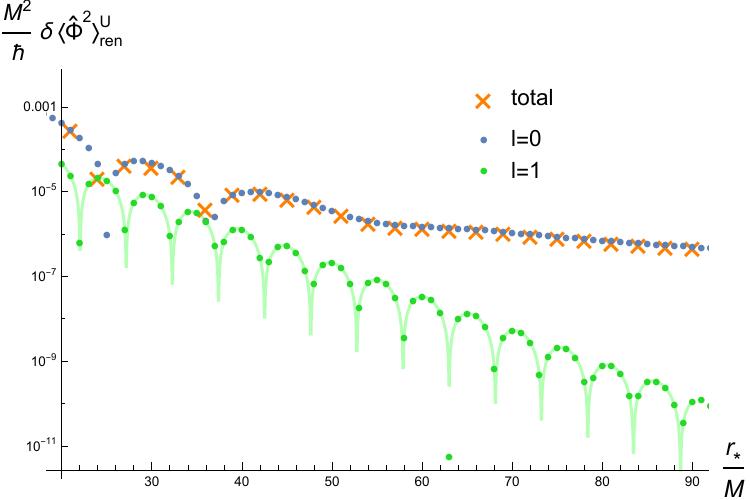}}\hfill{}\subfloat[\label{fig:phi2-maxima} $\langle\hat{\Phi}^{2}\rangle^{\text{U}}$ in the innermost near-IH
regime. The maximum and minimum, marked "2" and "3", correspond to those labeled in Fig. \ref{fig:phi2-throughout-interior}.
The orange crosses are numerically computed and the solid line is interpolated.
]{\begin{centering}\includegraphics[width=0.46\columnwidth]{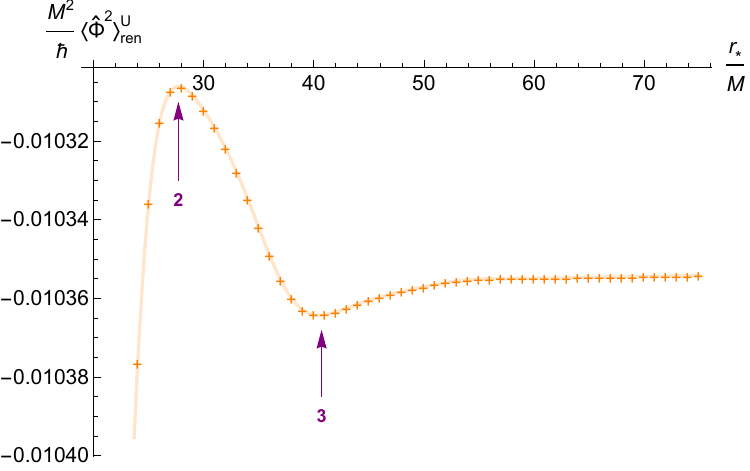}
\par\end{centering}

}
\par\end{centering}
\caption{Further features of $\langle\hat{\Phi}^{2}\rangle^{\text{U}}$ at the innermost near-IH regime (here, $r_*\approx 20M$ corresponds to $r-r_-\sim 10^{-13}M$ and $r_*\approx 90M$ corresponds to $r-r_-\sim10^{-60}M$).
The left panel illustrates the dominance of the $\ell=0$ mode in the innermost near-IH regime; the ringdown of this mode is, in turn, responsible for the extrema seen in the right panel.}

\end{figure}

In the closest approach to the IH for which point-splitting results are available (see Fig.~17 in Ref. \cite{Zilberman:2024jns} for  $a/M=0.8$), our state-subtraction results (for $\langle\hat{\Phi}^{2}\rangle^{\text{U}-\underline{\text{U}}}$) 
are expected
to agree with the point-splitting results
(for $\langle\hat{\Phi}^{2}\rangle^\text{U}$)
up to regular contributions (coming
from the comparison state). 
In order to recover such (regular) contribution from the comparison state, we subtract the numerical state-subtraction results from the $t$-splitting results (both plotted in Fig. \ref{fig:comparison-tsplitting}), i.e. (retaining the "ren" subscript for clarity)
\begin{equation}\label{comparison-state-contribution}
\langle\hat{\Phi}^{2}\rangle_{\text{ren}}^{\underline{\text{U}}}=\underbrace{\langle\hat{\Phi}^{2}\rangle_{\text{ren}}^{\text{U}}}_{\text{t-splitting}}-\underbrace{\left(\langle\hat{\Phi}^{2}\rangle_{\text{ren}}^{\text{U}}-\langle\hat{\Phi}^{2}\rangle_{\text{ren}}^{\underline{\text{U}}}\right)}_{\text{state-subtraction}}=\langle\hat{\Phi}^{2}\rangle_{\text{ren}}^{\text{U}}-\langle\hat{\Phi}^{2}\rangle^{\text{U}-\underline{\text{U}}},
\end{equation}
where $\langle\hat{\Phi}^{2}\rangle^{\text{U}-\underline{\text{U}}}$ was \emph{directly} obtained via Eq. \eqref{eq:phi2-schematic}. The above difference (between the $t$-splitting and state-subtraction results in the available overlap domain)  is plotted in Fig. \ref{fig:phi2-comparison-state}.  As expected, $\langle\hat{\Phi}^{2}\rangle_{\text{ren}}^{\underline{\text{U}}}$ in the IH vicinity is numerically found to be regular, and in the $r$-regime considered it is dominated by the linear term in a series expansion in $(r-r_-)/M$; see Fig. \ref{fig:phi2-comparison-state}. I.e., $\langle\hat{\Phi}^{2}\rangle_{\text{ren}}^{\underline{\text{U}}}$ in the IH vicinity is well-described by 
\begin{equation}\label{eq:phi2-uunderbar}
\langle\hat{\Phi}^{2}\rangle_{\text{ren}}^{\underline{\text{U}}}(r)\simeq\langle\hat{\Phi}^{2}\rangle_{-}^{\underline{\text{U}}}+a_1\frac{r-r_-}{M}+\mathcal{O}\bigg(\left(\frac{r-r_-}{M}\right)^2\bigg)\\,\hspace{2cm}r\to r_-,
\end{equation}
for numerically fitted parameters $\langle\hat{\Phi}^{2}\rangle_{-}^{\underline{\text{U}}}$ (corresponding to the asymptotic polar IH value of $\langle\hat{\Phi}^{2}\rangle_{\text{ren}}^{\underline{\text{U}}}$) and $a_1$, both given in Fig. \ref{fig:phi2-comparison-state}. Lastly, $\langle\hat{\Phi}^2\rangle^\text{U}(r)$ in the IH vicinity may be extracted by adding the novel state-subtraction results for $\langle\hat{\Phi}^{2}\rangle^{\text{U}-\underline{\text{U}}}$ to Eq. \eqref{eq:phi2-uunderbar}. This procedure is better than simply fitting the t-splitting data for $\langle\hat{\Phi}^{2}\rangle_{\text{ren}}^{\text{U}}$ directly because it is easier to fit data for the smooth-in-$r$ $\langle\hat{\Phi}^{2}\rangle_{\text{ren}}^{\underline{\text{U}}}$ rather than data for the non-smooth-in-$r$ $\langle\hat{\Phi}^{2}\rangle_{\text{ren}}^{\text{U}}$.

\begin{figure}
\begin{centering}
\includegraphics[scale=0.7]{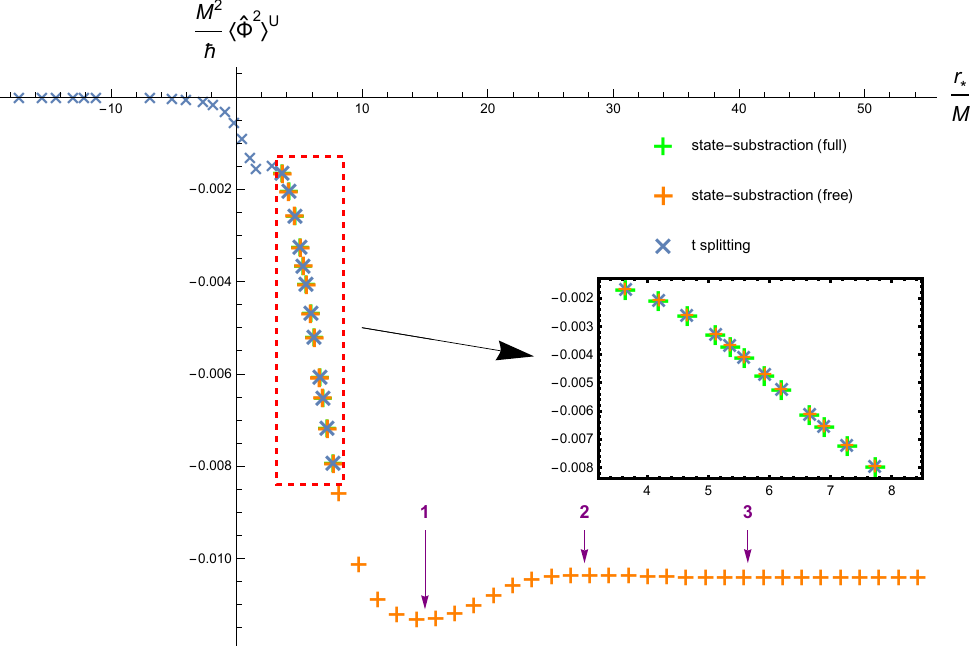}
\par\end{centering}
\caption{$\langle\hat{\Phi}^{2}\rangle^{\text{U}}$ throughout the (polar) interior
of the Kerr BH. The blue crosses denote the $t$-splitting results from Ref.
\cite{Zilberman:2024jns} (see Figs. 15 and 17 therein), which extend to the left displaying
the EH asymptotics. The orange pluses correspond to our novel results (obtained through the state-subtraction
method, consequently using Eq. \eqref{comparison-state-contribution}  to recover  $\langle\hat{\Phi}^{2}\rangle_{\text{ren}}^{\underline{\text{U}}}$ and adding \eqref{eq:phi2-uunderbar} to $\langle\hat{\Phi}^{2}\rangle^{\text{U}-\underline{\text{U}}}$ at the final stage) and extend towards the right displaying the IH asymptotics.
The three arrows correspond to the maxima/minima of the oscillations,
predominantly due to the $\ell=0$ mode, see Fig. \ref{fig:phi2-maxima}. The inset figure further demonstrates the
agreement between the $t$-splitting results and
the state-subtraction results (with its two different variants displayed, see paragraph after Eq. \eqref{eq:inII-scattering}).}\label{fig:phi2-throughout-interior}
\end{figure}

Altogether, by combining the $t$-splitting results of  Ref. \cite{Zilberman:2024jns} for $\langle\hat{\Phi}^2\rangle^\text{U}$ with our state-subtraction results for $\langle\hat{\Phi}^2\rangle^{\text{U}-\underline{\text{U}}}$, we were able to: 
(i) determine
the constant $\langle\hat{\Phi}^{2}\rangle_{-}^{\text{U}}$ appearing in Eq. \eqref{eq:phi2-sum-ell} (this completes Table I in Ref.~\cite{Zilberman:2024jns} for the case $a=0.8M$);
(ii)
extend the $\langle\hat{\Phi}^{2}\rangle^{\text{U}}$ results of Ref. \cite{Zilberman:2024jns}
deep into the near-IH regime (see Fig. \ref{fig:phi2-throughout-interior}
and its caption; blue crosses correspond to $t$-splitting results from \cite{Zilberman:2024jns},
while the orange and green pluses denote our new results); and
(iii) demonstrate excellent agreement between the $t$-splitting
and state-subtraction methods in the overlap region (see the inset in Fig. \ref{fig:phi2-throughout-interior}), thereby providing validation for both approaches. 

Explicitly, we find the limiting polar IH value of $\langle\hat{\Phi}^{2}\rangle^{\text{U}}$, as defined in Eq. \eqref{eq:phi2-sum-ell},
to be (for $a/M=0.8$)
\begin{equation}
\langle\hat{\Phi}^{2}\rangle_{-}^{\text{U}}=-0.01035\ \hbar M^{-2}.\label{eq:phi2(-)}
\end{equation}
(This can be compared with the analogous result in RN for $Q/M=0.8$ given in Ref. \cite{Lanir:2018vgb}, $\langle\hat{\Phi}^{2}\rangle_{-}^{\text{U}}= -0.07258\ \hbar M^{-2}$.) 

Finally, it is insightful to rewrite our main result in Eddington-Finkelstein
coordinates \eqref{eq:uv-coordinates}. Along a $u=u_{0}$ ray
approaching $\mathcal{CH}^\mathcal{R}$, $\langle\hat\Phi^2\rangle^\text{U}$  behaves asymptotically like 
\begin{equation}
\langle\hat{\Phi}^{2}(u_{0},v)\rangle^{\text{U}}=\langle\hat{\Phi}^{2}\rangle_{-}^{\text{U}}+\alpha v^{-3}+o(v^{-4}) \hspace{1.5cm}(\theta=0),\label{eq:phi2-uv}
\end{equation}
for some real constant $\alpha$. We expect the subleading corrections in Eq. \eqref{eq:phi2-uv} to include a QNM ringdown, as specified per $\ell$-mode in \eqref{eq:IH-phi2-rstar}.
Interestingly, the leading-order tail (i.e., $v^{-3}$) in Eq.~\eqref{eq:phi2-uv} is also known to  govern the asymptotic behavior of \emph{classical} scalar-field modes at the early portion of the CH in Kerr, see Eqs. (93), (11) in Ref. \cite{Ori:1998gp} (and note $l_0=0$ and $n_0=3$ in our case). Here, the relevant modes are the $m=0$ axially symmetric modes, which are the only modes contributing at the pole.


\section{Stress-energy fluxes }\label{sec:Stress-energy-fluxes}

The (classical) SET of the massless KG field with curvature coupling $\xi$ in a vacuum spacetime can be written as
\begin{align}
T_{ab} & =\nabla_{a}\Phi\nabla_{b}\Phi-\frac{1}{2}g_{ab}\Bigl(\nabla^{c}\Phi\nabla_{c}\Phi-2\xi\nabla_{c}\nabla^{c}\Phi^{2}\Bigr)-\xi\nabla_{a}\nabla_{b}\Phi^{2}\label{eq: classical SET KG}\\
 & =\nabla_{a}\Phi\nabla_{b}\Phi-\left(\frac{1}{4}-\xi\right)g_{ab}\nabla_{c}\nabla^{c}\Phi^{2}-\xi\nabla_{a}\nabla_{b}\Phi^{2}.\nonumber 
\end{align}
In the second equality, we used the fact that $\nabla^{a}\Phi\nabla_{a}\Phi=\frac{1}{2}\nabla_{a}\nabla^{a}\Phi^{2}$
for solutions to Eq.  \eqref{eq:Klein-Gordon-eq}.
In this paper we focus
on the axis of rotation $(\theta=0)$ in Kerr, where (in double-null coordinates) both $g_{uu}$
and $g_{vv}$ vanish identically. Recalling the notation $y \in \{u, v\}$, we may therefore write
\begin{equation}
T_{yy}=\nabla_{y}\Phi\nabla_{y}\Phi-\xi\nabla_{y}\nabla_{y}\Phi^{2}\hspace{3em}(\text{Kerr},\ \theta=0).
\end{equation}
Like $\Phi^2$, the classical stress-energy
tensor is local and quadratic in the field, hence it requires renormalization
in the quantum theory. Then (with the subscript "ren" suppressed to conform with our notation throughout the paper),
\begin{equation}
\langle \hat{T}_{yy}\rangle^{\text{U}}=\langle\nabla_{y}\hat{\Phi}\nabla_{y}\hat{\Phi}\rangle^{\text{U}}-\xi\nabla_{y}\nabla_{y}\langle\hat{\Phi}^{2}\rangle^{\text{U}}\hspace{3em}(\theta=0).\label{eq:Tyy_general_xi}
\end{equation}
The first term on the right-hand side, denoted below by $\langle \hat{T}_{yy}\rangle^{\text{U,}\xi=0}$,
corresponds to the minimally coupled ($\xi=0$) fluxes and has
been recently analyzed  and computed at the IH \cite{2022PhRvL.129z1102Z} and throughout the interior of Kerr BHs  \cite{Zilberman:2024jns} by some of us. One achievement of the current paper is to extend the aforementioned minimal-coupling results for the polar IH fluxes to arbitrary $\xi$. For this purpose, the only additional input missing is the near-IH asymptotic behavior of $\langle\hat{\Phi}^{2}\rangle$ (entering through the $\propto\xi$ term in Eq. \eqref{eq:Tyy_general_xi}). In Subsecs. \ref{subsec:Price-tail} and \ref{subsec:QNM-ringdown} we found that, in the vicinity of the IH, this behavior consists of
a superposition of QNM ringings followed by an $r_{*}^{-3}$ tail;
see Eq. \eqref{eq:phi2-sum-ell}. 

In the following, we derive an expression
for the polar fluxes valid for generic $r$ values,
taking the IH limit only later.  Taking Eq. \eqref{eq:Tyy_general_xi} as our starting point,
\begin{align}
\langle \hat{T}_{yy}\rangle^{\text{U}} & =\langle \hat{T}_{yy}\rangle^{\text{U,}\xi=0}-\xi\nabla_{y}\nabla_{y}\langle\hat{\Phi}^{2}\rangle^{\text{U}} \\
 & =\langle \hat{T}_{yy}\rangle^{\text{U,}\xi=0}-\xi\left(\partial_{y}^{2}-\Gamma_{yy}^{\mu}\partial_{\mu}\right)\langle\hat{\Phi}^{2}\rangle^{\text{U}} \\
 & =\langle \hat{T}_{yy}\rangle^{\text{U,}\xi=0}-\xi\left(\partial_{y}^{2}-\Gamma_{yy}^{y}\partial_{y}-\Gamma_{yy}^{\overline{y}}\partial_{\overline{y}}\right)\langle\hat{\Phi}^{2}\rangle^{\text{U}},
\end{align}
where the Christoffel symbols ${\Gamma^\mu}_{\rho\nu}$  are evaluated at $\theta 
=0$, and where  $\overline{y}$  denotes the "complementary" null coordinate of $y$; that is, $\overline{y} = v$ if $y = u$, and $\overline{y} = u$ if $y = v$.
We used the fact that $\langle\hat{\Phi}^{2}\rangle^{\text{U}}$ cannot depend
on $\varphi$ (due to axial symmetry), and we assumed its $\theta$-dependence to be regular. In particular, axial symmetry implies that near the pole,
the dependence on $\theta$ can only appear through even powers of $\theta$, so that $\partial_\theta\langle\hat{\Phi}^2\rangle^\text{U}\vert_{\theta=0}=0$. Moreover, using $\Gamma_{yy}^{\overline{y}}=0$
and $\Gamma_{yy}^{y}=\frac{1}{2}g^{uv}\partial_{r_{*}}g_{uv}$ at the pole, one obtains
\begin{equation}
\langle \hat{T}_{yy}\rangle^{\text{U}}=\langle \hat{T}_{yy}\rangle^{\text{U,}\xi=0}-\xi\left(\frac{1}{4}\partial_{r_{*}}^{2}-\frac{1}{2}\frac{M(r^{2}-a^{2})}{(r^{2}+a^{2})^{2}}\partial_{r_{*}}\right)\langle\hat{\Phi}^{2}\rangle^{\text{U}}.
\end{equation}
Evaluating the expression above in the IH vicinity, using Eq.  \eqref{eq:phi2-sum-ell}, we find
\begin{equation}\label{eq:RSET}
\left.
\begin{aligned}
\langle \hat{T}_{yy}\rangle^{\text{U}}
&=\langle \hat{T}_{yy}\rangle^{\text{U,}\xi=0}
-\xi\left(\frac{1}{4}\partial_{r_{*}}^{2}+\frac{\kappa_{-}}{2}\partial_{r_{*}}\right)\langle\hat{\Phi}^{2}\rangle^{\text{U}} \\
&=\langle \hat{T}_{yy}\rangle^{\text{U,}\xi=0}
+\frac{3\xi\kappa_{-}}{2}c_{0}r_{*}^{-4}
+o\!\left(r_{*}^{-5}\right)
\end{aligned}
\right.
\qquad (\theta=0),
\end{equation}
with the constant $c_{0}$ given in Eq. \eqref{eq:c0-analytical}.
We conclude that the polar IH fluxes $\langle T_{uu}\rangle^{\text{U}}$
and $\langle T_{vv}\rangle^{\text{U}}$ (obtained by taking the limit $r_*\to\infty$ in the above) are \emph{independent} of the coupling
to the curvature $\xi$. In particular, the polar IH flux results of \cite{2022PhRvL.129z1102Z}, computed for a minimally-coupled scalar field, in fact apply for arbitrary $\xi$. The $\xi$-dependent approach to the IH value is governed by an
$r_{*}^{-4}$ inverse-power tail. Translating to the Kruskal coordinate $V$ (regular and vanishing at $\mathcal{CH^R}$),
we find that the leading divergence $\langle T_{VV}\rangle^{\text{U}}\sim CV^{-2}$
is likewise independent of the coupling to curvature (including the coefficient $C$). Our analysis also points to a potential divergence of the near-IH polar trace (however, a detailed analysis of the trace requires going off-pole and is postponed to a later paper  \cite{Alberti:off-pole}).

\section{Discussion}\label{sec:Discussion}

This paper was motivated by the computation of the quantum energy fluxes $\langle \hat{T}_{uu}\rangle^\text{U}$ and $\langle \hat{T}_{vv}\rangle^\text{U}$  (which are expected to drive
near-IH backreaction \cite{Zilberman:2019buh}) for a real massless scalar field in the Unruh state at the IH of a Kerr BH along the axis of rotation, extending the minimally coupled ($\xi=0$) results of \cite{2022PhRvL.129z1102Z} to arbitrary curvature coupling $\xi$. A key ingredient in this computation is the near-IH asymptotic behavior of $\langle\hat{\Phi}^{2}\rangle^\text{U}$, which we investigate in detail using the state-subtraction method. 

Our numerical analysis reveals that $\langle\hat{\Phi}^{2}\rangle^\text{U}$ approaches a finite IH limit in a distinctive two-stage manner: for each $\ell$-mode, the approach is marked by a ringdown followed by an inverse-power tail, both in the tortoise coordinate $r_*$ (see Eq. \eqref{eq:IH-phi2-rstar} and Figs. \ref{fig:QNM-ringdown-and-tail}, \ref{fig:Inverse-power-tails-normalized}
and \ref{fig:ringdown}).
Two analogies with the well-known classical  behavior of fields outside the BH can be drawn. First, the
complex ringing frequencies are numerically found to match twice\footnote{The factor of $2$ appears when the ringing is expressed in terms of the coordinate $r_*$; when expressed in terms of the coordinate $v$ (along a $u=\text{const.}$ ray) the ringing frequency corresponds to the classical, fundamental QNM frequency $\omega^\text{QNM}$, i.e., one schematically has the behavior  $\sim\exp(-i\omega^\text{QNM} v)$. 
In fact, one may indeed expect such a QNM ringdown for ($m=0$, axially symmetric) classical field modes at the IH by the following argument. Quasinormal modes are defined by the boundary condition that they be purely ingoing at the EH \cite{Berti:2009kk}, i.e. $\sim\exp(-i\omega^\text{QNM} v)$ there. An incoming wave from past null infinity therefore rings down on the EH with this profile; then, propagating this data through the interior (according to Eq. \eqref{kg-radial-joined}, following the asymptotic behaviors given in Eq.~\eqref{eq:inII-scattering}), one expects the same $\sim\exp(-i\omega^\text{QNM} v)$ ringing behavior to persist as the IH is approached.} the well-known
classical fundamental QNM frequencies for classical scalar perturbations in Kerr (available, e.g., in \cite{emmanuele-berti-QNM}); see Eq.\eqref{eq:qnm-fit}. Second, the inverse-power tails exhibit Price-like behavior $r_{*}^{-2\ell-3}$,
resembling the late-time decay of classical fields in the BH exteriors. 
Our results for $\langle\hat{\Phi}^{2}\rangle^\text{U}$ also resemble previous results in RN, obtained in \cite{Lanir:2018vgb}, where $\langle\hat{\Phi}^2\rangle^\text{U}$ approaches its asymptotic value via an $r_*^{-3}$ tail preceded by oscillations in $r_*$ (however, the frequencies of these oscillations, as well as the individual $\ell$ contributions to the tails, were not investigated in that work). Remarkably, in both the present polar Kerr analysis as well as the RN one, the inverse-power-law behavior is exposed only extremely close to the IH: in Kerr (and for $a/M=0.8$) the leading $\ell=0$ tail becomes visible only at $r_*\approx70 M$, corresponding to  $r-r_-\sim 10^{-45}M$, whereas in RN  (and for $Q/M=0.8$) the summed (over $\ell$) inverse-power tail becomes visible only at $r_*\approx400M$, corresponding to $r-r_-\sim 10^{-175}M$ (see Fig. 3 of  \cite{Lanir:2018vgb}). In addition, we derive here the leading inverse-power tail analytically -- both the exponent and its prefactor (given in Eq. \eqref{eq:c0-analytical}).

The above investigation of the near-IH polar behavior of $\langle\hat{\Phi}^{2}\rangle^\text{U}$, itself a $\xi$-independent quantity, was crucial (through Eq. \eqref{eq:Tyy_general_xi}) for uncovering the behavior of the fluxes $\langle \hat{T}_{uu}\rangle^\text{U}$
and $\langle \hat{T}_{vv}\rangle^\text{U}$ for general coupling $\xi\neq 0$:  both their asymptotic IH values, as well as their behavior on approaching these values. Our results generalize those of the previous study \cite{2022PhRvL.129z1102Z} in that they (i) allow for generic coupling to curvature $\xi$ (at the pole), and
(ii) extend application of the state-subtraction method used in both works from the IH to its close vicinity. Our analysis reveals that the limiting IH values of both flux components remain unaltered, and hence so does the corresponding divergence when expressed in a regular (Kruskal) coordinate at $\mathcal{CH^R}$. That is, the prefactor $C$ appearing in $\langle \hat{T}_{VV}\rangle^\text{U}\approx CV^{-2}$ is independent of the coupling to curvature $\xi$. 
Moreover, our results indicate that for non-minimal coupling, the leading $\xi$-dependent divergence on approaching the IH is softened by a logarithmic factor, i.e. $\sim c_0V^{-2}\ln \vert\kappa_-V\vert^{-4}$, with the prefactor $c_0$ given analytically in Eq. \eqref{eq:c0-analytical}. 
While our analysis (including the above statements) is on a fixed Kerr background, the results for $\langle \hat{T}_{uu}\rangle^\text{U}$
and $\langle \hat{T}_{vv}\rangle^\text{U}$ provide key input for a future backreaction analysis through the semiclassical Einstein equation \eqref{eq:semiclassical-EE}.
Our results were obtained by applying the state-subtraction method in the close IH vicinity\footnote{While state-subtraction was previously used directly at the IH \cite{2022PhRvL.129z1102Z}, as well as compared against $t$-splitting results there, to the best of our knowledge this is the first work where the state-subtraction method was extended to the \emph{vicinity} of the IH. In particular, this extension required new numerical input that was not required exactly at the IH, that is, the reflection coefficient related to the $\vert\underline{\text{U}}\rangle$ state,  $\underline{\rho}_{\ \ \omega \ell 0}^\text{down}$  (as defined in Eq. \eqref{eq:f-down-psi}).}, employing a comparison state (denoted $\vert \underline{\text{U}}\rangle$)  introduced in \cite{2022PhRvL.129z1102Z}. Since the comparison state is regular across $\mathcal{CH}^\mathcal{R}$  (as supported by our numerics, see Fig. \ref{fig:phi2-comparison-state}; see also footnote \ref{ft:Hadamard_comp_state}), the irregular features in the state difference $\vert\text{U}\rangle -\vert \underline{\text{U}}\rangle$ on approaching the IH must originate from the Unruh state.

Moreover, since $\langle \hat{T}_{vv}\rangle^{\underline{\text{U}}}=0$ at  $\mathcal{CH}^\mathcal{R}$ from regularity of $\vert \underline{\text{U}}\rangle$, state-subtraction also suffices to uncover the asymptotic value of  $\langle \hat{T}_{vv}\rangle^{{\text{U}}}$ at  $\mathcal{CH}^\mathcal{R}$ (and hence also the coefficient of the divergence of $\langle \hat{T}_{VV}\rangle^{{\text{U}}}$). 
However, since regularity does not fix $\langle\hat{\Phi}^2 \rangle^{\underline{\text{U}}}$ at $\mathcal{CH}^\mathcal{R}$, it is (only) when combining the state-subtraction results with the $t$-splitting results that we can uncover the asymptotic value of $\langle\hat{\Phi}^{2}\rangle^\text{U}$ (see Eq. \eqref{eq:phi2(-)}). Moreover, combining both methods yields $\langle\hat{\Phi}^{2}\rangle^\text{U}$ throughout the entire interior region, as illustrated in Fig. \ref{fig:phi2-throughout-interior}.

A closely related issue is that the Kerr IH is particularly challenging to probe with standard point-splitting methods: as one approaches the polar IH, natural splitting directions become null and the numerical implementation becomes ill-defined, preventing access to the extremely close IH vicinity. This practical obstruction is precisely what makes state subtraction effective here, allowing us to reach the deep near-IH regime where the QNM ringdown and Price-like tails are exposed.

It would be natural and desirable to extend this analysis away from the pole, where $m\neq0$ modes (and associated superradiant effects) enter the computation. As preliminary analysis already suggests, this may significantly modify the behavior of $\langle\hat{\Phi}^{2}\rangle^\text{U}$ and, correspondingly, the general-$\xi$ fluxes. An additional appealing prospect for future research would be the analytical derivation of both the ringing and the inverse-power-law asymptotic behavior (obtained here numerically), which would offer a more complete understanding of the near-IH dynamics of spinning BHs.

Furthermore, as mentioned in the Introduction, extending our results to arbitrary curvature coupling allows us to draw immediate conclusions for a conformally-coupled scalar field ($\xi=1/6$). This is of particular interest in view of the physically important electromagnetic field, which is conformally invariant in four spacetime dimensions. Extending the present analysis from the scalar to the electromagnetic case is therefore a natural and promising direction for future work.

\begin{acknowledgments} 
We are grateful to Amos Ori for substantial contributions to the development of this work.
M.A. was supported by the Israel Science Foundation under Grant No. 1676/25, with additional support provided by the Max Planck Society through the Max Planck-Israel Program. M.A. thanks Stefan Hollands and Jochen Zahn for their hospitality during her stay at the MPI-MiS and for useful discussions.
N.Z. acknowledges the support of Fulbright Israel. N.Z. also thanks the Technion, and Amos Ori in particular, for their hospitality during her visits.

\end{acknowledgments}

\appendix 

\section{The $\vert\underline{\text{U}}\rangle$ two-point function}
\label{sec:appendixA}

In this Appendix we derive the mode-sum expression for the HTPF associated with the comparison state $\vert\underline{\text{U}}\rangle$ (originally introduced in \cite{2022PhRvL.129z1102Z}). We remark, however, that the Hadamard property of the $\vert\underline{\text{U}}\rangle$ state remains an (empirically motivated) assumption; see foonote \ref{ft:Hadamard_comp_state}. 
The derivation of the $\vert\underline{\text{U}}\rangle$-state HTPF mode sum closely follows the analogous derivation of the Unruh-state HTPF mode sum, carried out in Ref. \cite{2022PhRvD.106l5011Z}. We begin the analysis in region III (see Fig. \ref{fig:Penrose-diagram-fKerr}), which is an asymptotically flat region that plays the role of the $\vert\underline{\text{U}}\rangle$-state analog of region I (see Ref. \cite{2022PhRvL.129z1102Z} for more details). We then continue to region II, the BH interior, where (by explicitly performing a point-split in $\theta$ and $t$, and combining with the known Unruh-state counterpart) we derive Eqs.~ \eqref{eq:mode-sum-phi2}-\eqref{eq:integrands} for the mode-sum difference used in the main text.

\subsection{Region III}
The scattering problem in region III (extending to $r\to-\infty$) is mathematically well-posed, as the effective potential \eqref{eq:radial2} remains well behaved in the limit $r\to0$ (indeed, it approaches a constant and can be extended smoothly to negative $r$ values). The presence of a singularity and closed timelike curves  (and, consequently, the absence of well-defined Cauchy evolution) in region III does not interfere with the evolution of the individual Eddington-like modes, which still separate as \eqref{eq:boulware-mode-ansatz}, are mode-wise specified by their final data on $\mathcal{CH}^{+}\cup\mathcal{\overline{\mathcal{I}}}^{+}$ (see Fig. \ref{fig:Penrose-diagram-fKerr}) and satisfy a well-posed scattering problem. Then, the derivation of the $\vert\underline{\text{U}}\rangle$ mode-sum expression for the HTPF proceeds similarly to that of the Unruh state, which is given in detail in \cite{2022PhRvD.106l5011Z}.  

First, we expand the field operator in terms of \emph{out} and \emph{down} modes $g_{\omega\ell m}^{\text{out}}$ and $g_{\hat{\omega}\hat{\ell}\hat{ m}}^{\text{down}}$: these are solutions to Eq. \eqref{eq:Klein-Gordon-eq}, defined by their final data

\begin{align}
   & g_{\omega\ell m}^{\text{out}}\simeq\frac{Z_{\ell (-m)}^{-\omega}(\theta,\varphi)}{\sqrt{4\pi \omega(r^{2}+a^{2})}}\begin{cases}
0, & \mathcal{CH}^{\mathcal{L}}\cup\mathcal{CH}^{+}\\
e^{-i\omega u},& \overline{\mathcal{I}}^{+}
\end{cases}\label{eq:C-modes-out}\\
&g_{\hat{\omega}\hat{\ell} \hat{m}}^{\text{down}}\simeq\frac{Y_{\hat{\ell} \hat{m}}(\theta,\varphi_-)}{\sqrt{4\pi\hat{\omega}(r^{2}+a^{2})}}\begin{cases}
e^{-i\hat{\omega}V}, & \mathcal{CH}^{\mathcal{L}}\cup\mathcal{CH}^{+}\\
0, & \overline{\mathcal{I}}^{+}
\end{cases}\label{eq:C-modes-down}
\end{align}
for $\omega, \hat{\omega}>0$ (this is analogous to the \emph{in} and \emph{up} Unruh modes defined implicitly through Eq. \eqref{eq:initial-data-Unruh-modes}, with the only difference that we redefined the labels for the angular functions so that the \emph{out} modes will still separate as \eqref{eq:boulware-mode-ansatz})\footnote{As mentioned, region III in the Penrose diagram in Fig. \ref{fig:Penrose-diagram-fKerr} is not globally hyperbolic, so (strictly speaking) the standard definition of the KG inner product does not apply for the modes given by the final data (\ref{eq:C-modes-out},\ref{eq:C-modes-down}). However, with the standard extension of the KG inner product to the union of null hypersurfaces $\overline{\mathcal{I}}^+\cup\mathcal{CH}^+\cup\mathcal{CH}^\mathcal{L}$ (see, e.g., \cite{Frolov:Novikov}), these \emph{down} and \emph{out} modes are chosen to be normalized with a positive norm for positive $\omega$ and $\hat\omega$, correspondingly. Consequently, these modes form a complete KG-orthonormal set of solutions to the wave equation when restricted to the null boundary.
}.

In terms of the above modes, we write the field operator (in regions II and III) as 
\begin{equation}
\hat{\Phi}(x)=\sum_{\ell,m}\int_{0}^{\infty}d\omega\left(g_{\omega\ell m}^{\text{out}}(x)\hat{c}_{\omega\ell m}^{\text{out}}+g_{\omega\ell m}^{\text{out}\ *}(x)\hat{c}_{\omega\ell m}^{\text{out}\hspace{0.5em}\dagger}\right)+
\sum_{\hat{\ell},\hat{m}}\int_{0}^{\infty}d\hat{\omega}\left(g_{\hat{\omega}\hat{\ell}\hat{m}}^{\text{down}}(x)\hat{c}_{\hat{\omega}\hat{\ell}\hat{m}}^{\text{down}}+g_{\hat{\omega}\hat{\ell}\hat{m}}^{\text{down}\ *}(x)\hat{c}_{\hat{\omega}\hat{\ell}\hat{m}}^{\text{down}\hspace{0.5em}\dagger}\right).
\end{equation}
Here,  $\hat{c}_{\omega\ell m}^{\text{out}\hspace{0.5em}\dagger}$, $\hat{c}_{\hat{\omega}\hat{\ell}\hat{m}}^{\text{down}\hspace{0.5em}\dagger}$ and $\hat{c}_{\omega\ell m}^{\text{out}}$, $\hat{c}_{\hat{\omega}\hat{\ell}\hat{m}}^{\text{down}}$ are the standard creation and annihilation operators satisfying the canonical commutation relations (see, e.g., Sec. IV.A in Ref. \cite{2022PhRvD.106l5011Z}). The $\vert\underline{\text{U}}\rangle$ state is defined as the vacuum state of this construction (i.e., the state annihilated by all  $\hat{c}_{\hat{\omega}\hat{\ell}\hat{m}}^{\text{down}}$ and $\hat{c}_{\omega\ell m}^{\text{out}}$ operators). Using the canonical commutation relations, the HTPF can be rewritten (for spacetime points $x$ and $x'$) as 
\begin{equation}\label{eq:2pt-function-comparison}
    G^{\underline{\text{U}}}(x,x')\equiv\langle\left\{ \Phi(x),\Phi(x')\right\} \rangle^{\underline{\text{U}}} \equiv G^{\underline{\text{U}}\text{,out}}(x,x')+G^{\underline{\text{U}}\text{,down}}(x,x'),
\end{equation}
with
\begin{subequations}
    \begin{align}
     G^{\underline{\text{U}}\text{,out}}(x,x')&=\hbar\sum_{\ell,m}\int_{0}^{\infty}d\omega\left\{ g_{\omega\ell m}^{\text{out}}(x),g_{\omega\ell m}^{\text{out}\ *}(x')\right\}\label{gdown1},\\
     G^{\underline{\text{U}}\text{,down}}(x,x')&=\hbar\sum_{\hat{\ell},\hat{m}}\int_{0}^{\infty}d\hat{\omega}\left\{ g_{\hat{\omega}\hat{\ell}\hat{m}}^{\text{down}}(x),g_{\hat{\omega}\hat{\ell}\hat{m}}^{\text{down}\ *}(x')\right\}.
     \end{align}\label{GIII-out-down}
\end{subequations}
\begin{figure}
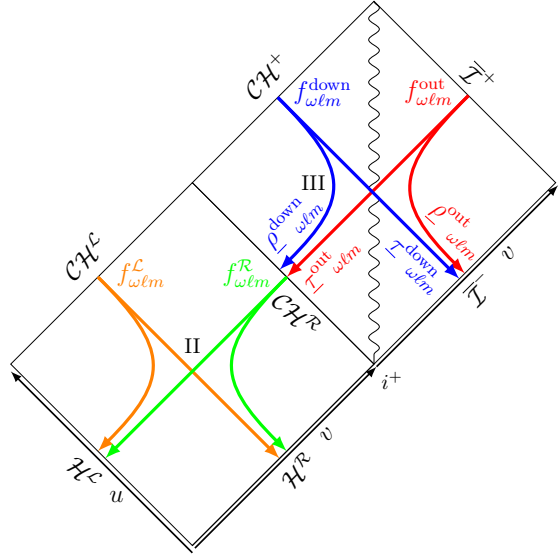

\begin{centering}
\includestandalone[scale=0.8]{U-underbar}
\par\end{centering}
\caption{Scattering problem of the \emph{out} Eddington-type modes $f_{\omega\ell m}^{\text{out}}$ (in red, defined through Eq. \eqref{eq:out-f}-\eqref{eq:out-psi}) and \emph{down} Eddington-type modes $f_{\omega\ell m}^{\text{down}}$ (in blue, defined through Eq. \eqref{eq:f+down-modes}-\eqref{eq:f-down-psi}) in region III. Likewise, in region II we depict the  $f_{\omega\ell m}^{\mathcal{R}}$ modes (in green, defined through Eq. \eqref{eq:f-LR}-\eqref{eq:f-L}) and the $f_{\omega\ell m}^{\mathcal{L}}$ modes (in orange, defined through Eq. \eqref{eq:f-LR}-\eqref{fR-modes}).}\label{fig:C-modes}
\end{figure}
As usual, it is advantageous to work with Eddington-type modes, i.e., modes that separate as \eqref{eq:boulware-mode-ansatz}. We may define Eddington-\emph{out} type modes in region III by
\begin{equation}\label{eq:out-f}
    f_{\omega\ell m}^{\text{out}}\equiv\frac{1}{\sqrt{4\pi\vert \omega\vert(r^{2}+a^{2})}}e^{-i\omega t}Z_{\ell m}^{\omega}(\theta,\varphi)\psi_{\omega\ell m}^{\text{out}}(r)
\end{equation}
with the radial solution obeying
\begin{equation}\label{eq:out-psi}
\psi_{\omega\ell m}^{\text{out}}(r)\simeq\begin{cases}
e^{i\omega r_{*}} + \underline{\rho}_{\ \ \omega\ell m}^{\text{out}}e^{-i\omega r_{*}} & r_{*}\to -\infty\\
\underline{\tau}_{\ \ \omega\ell m}^{\text{out}}e^{i\omega_- r_{*}} & r_{*}\to \infty\ ,
\end{cases}\\
\end{equation}
where the coordinate $r_*$ asymptotes to $+\infty$ at the IH and to $-\infty$ at negative radial infinity ($r\to-\infty$). The scattering problem of these modes is schematically depicted in Fig. \ref{fig:C-modes}. The scattering coefficients satisfy the Wronskian relation
\begin{equation}\label{eq:wronskian-out}
    \vert \underline{\rho}_{\ \ \omega\ell m}^{\text{out}}\vert^2+\frac{\omega_-}{\omega}\vert \underline{\tau}_{\ \ \omega\ell m}^{\text{out}}\vert^2=1.
\end{equation}
For future reference, we explicitly write the asymptotic behavior for these modes at $\mathcal{CH^R}$ 
\begin{equation}\label{eq:f-out-CHR}
    f_{\omega\ell m}^{\text{out}}\bigg\vert_{\mathcal{CH^R}}=\frac{1}{\sqrt{4\pi\vert \omega\vert(r^{2}+a^{2})}} Z_{\ell m}^{\omega}(\theta,\varphi_-)\underline{\tau}_{\ \ \omega\ell m}^{\text{out}}e^{i\omega_- u}
\end{equation}
and at $\overline{\mathcal{I}}^+$
\begin{equation}\label{eq:f-out-I+}
    f_{\omega\ell m}^{\text{out}}\bigg\vert_{\overline{\mathcal{I}}^+}=\frac{(-1)^m}{\sqrt{4\pi\vert \omega\vert(r^{2}+a^{2})}}\Big( Z_{\ell (-m)}^{-\omega}(\theta,\varphi)e^{-i\omega u}\Big)^*,
\end{equation}
where in Eq. \eqref{eq:f-out-I+} we used the property that $Z_{\ell m}^{\omega\ *}(\theta,\varphi)=(-1)^mZ_{\ell (-m)}^{-\omega}(\theta,\varphi)$ to facilitate comparison with Eq. ~\eqref{eq:C-modes-down}. Then, by comparing Eqs. ~\eqref{eq:C-modes-down} and \eqref{eq:f-out-I+}, we find 
\begin{equation}\label{eq:outc-outEddIII}
    g_{\omega\ell m}^{\text{out}}\bigg\vert_{\overline{\mathcal{I}}^+}=(-1)^m f_{\omega\ell m}^{\text{out}\ *}\bigg\vert_{\overline{\mathcal{I}}^+}\,.
\end{equation}
In addition, both $ g_{\omega\ell m}^{\text{out}}$ and $ f_{\omega\ell m}^{\text{out}}$ vanish on $\mathcal{CH}^+$ and satisfy the wave equation ~\eqref{eq:Klein-Gordon-eq}, and hence the above relation is satisfied throughout region III. Thus, $g^\text{out}_{\omega\ell m}$ are of Eddington type (i.e., decompose according to ~\eqref{eq:boulware-mode-ansatz}).

Unlike the  $g^\text{out}_{\omega\ell m}$ modes, the $g_{\hat{\omega}\hat{\ell} \hat{m}}^{\text{down}}$  modes do not decompose as Eq. ~\eqref{eq:boulware-mode-ansatz}. To proceed, we express each such mode \eqref{eq:C-modes-down} as a sum of two other mode solutions, which we denote $g_{\hat{\omega}\hat{\ell} \hat{m}}^{\text{down,}\mathcal{L}}$ and $g_{\hat{\omega}\hat{\ell} \hat{m}}^{\text{down,+}}$, initially supported on $\mathcal{CH}^{\mathcal{L}}$ and  $\mathcal{CH}^{+}$, respectively, and both vanishing at $\overline{\mathcal{I}}^{+}$. I.e., these solutions obey the asymptotic relations
\begin{equation}
\begin{aligned}
    g_{\hat{\omega}\hat{\ell} \hat{m}}^{\text{down},\mathcal{L}}	&\simeq\frac{1}{\sqrt{4\pi \hat{\omega}(r^{2}+a^{2})}}Y_{\hat{\ell} \hat{m}}(\theta,\varphi_{-})\begin{cases}
e^{-i\hat{\omega}V} & \mathcal{CH}^{\mathcal{L}}\\
0 & \mathcal{CH}^{+}\\
0 & \overline{\mathcal{I}}^{+}
\end{cases},\\
g_{\hat{\omega}\hat{\ell} \hat{m}}^{\text{down,+}}	&\simeq\frac{1}{\sqrt{4\pi \hat{\omega}(r^{2}+a^{2})}}Y_{\hat{\ell} \hat{m}}(\theta,\varphi_{-})\begin{cases}
0 & \mathcal{CH}^{\mathcal{L}}\\
e^{-i\hat{\omega}V} & \mathcal{CH}^{+}\\
0 & \overline{\mathcal{I}}^{+}
\end{cases} 
\end{aligned}
\end{equation} 
and the following holds
\begin{equation}\label{eq:g_sum}
 g_{\hat{\omega}\hat{\ell} \hat{m}}^{\text{down}}= g_{\hat{\omega}\hat{\ell} \hat{m}}^{\text{down,}\mathcal{L}}+ g_{\hat{\omega}\hat{\ell} \hat{m}}^{\text{down,+}}.
\end{equation}
In the following, we reexpress the Kruskal-based \emph{down} modes $g_{\hat{\omega}\hat{\ell} \hat{m}}^{\text{down}}$ in terms of Eddington-type \emph{down} modes $f_{\omega\ell m}^{\text{down}}$ in region III, which separate as Eq. \eqref{eq:boulware-mode-ansatz}, i.e.
\begin{equation}\label{eq:f+down-modes}
    f_{\omega\ell m}^{\text{down}}=\frac{1}{\sqrt{4\pi\vert\omega_{-}\vert(r^{2}+a^{2})}}Z_{\ell m}^{\omega}(\theta,\varphi)e^{-i\omega t}\psi_{\omega\ell m}^{\text{down}}(r)\,,
\end{equation}
with the radial solution $\psi_{\omega\ell m}^{\text{down}}(r)$ obeying the asymptotic behavior 
\begin{equation}\label{eq:f-down-psi}
    \psi_{\omega\ell m}^{\text{down}}(r)\simeq\begin{cases}
\underline{\tau}_{\ \ \omega\ell m}^{\text{down}}e^{-i\omega r_{*}}, & r_{*}\to-\infty\\\
e^{-i\omega_{-}r_{*}} + \underline{\rho}_{\ \ \omega\ell m}^{\text{down}}e^{i\omega_{-}r_{*}}, & r_{*}\to\infty.
\end{cases}\\
\end{equation}
The scattering problem for the Eddington modes in region III is schematically depicted in Fig. \ref{fig:C-modes}. We explicitly give the asymptotic behavior for the \emph{down} modes (which can be derived by using Eqs. \eqref{eq:f+down-modes}-\eqref{eq:f-down-psi}, along with \eqref{eq:coords}) 
 \begin{equation}\label{eq:fdownCHR}
    f_{\omega\ell m}^{\text{down}}\simeq\frac{1}{\sqrt{4\pi\vert\omega_{-}\vert(r_{-}^{2}+a^{2})}}Z_{\ell m}^{\omega}(\theta,\varphi_-)\begin{cases}
    e^{i\omega_-v}\hspace{2cm}\mathcal{CH}^+,\\
    \underline{\rho}_{\ \ \omega\ell m}^{\text{down }}e^{i\omega_{-}u}\hspace{.85cm} \mathcal{CH}^\mathcal{R}.
        
    \end{cases}
\end{equation}

The scattering coefficients $\underline{\rho}_{\omega\ell m}^{\text{down }}$ and $\underline{\tau}_{\ \ \omega\ell m}^{\text{down}}$ are defined with respect to the above asymptotic behavior and satisfy the Wronskian relation 
\begin{equation}
    \vert
    \underline{\rho}_{\ \ \omega\ell m}^{\text{down }}\vert^{2}+\frac{\omega}{\omega_{-}}\vert\underline{\tau}_{\ \ \omega\ell m}^{\text{down }}\vert^{2}=1.
\end{equation}
Furthermore, the \emph{down} and \emph{out} scattering coefficients are related by
\begin{equation}\label{eq:wronskian-comparison}
    \begin{aligned}
        \underline{\tau}_{\ \ \omega\ell m}^{\text{down }}&=\frac{\omega_-}{\omega}\underline{\tau}_{\ \ \omega\ell m}^{\text{out }},\\
        \frac{\underline{\rho}_{\ \ \omega\ell m}^{\text{out }*}}{\underline{\rho}_{\ \ \omega\ell m}^{\text{down }}}&=-\frac{\underline{\tau}_{\ \ \omega\ell m}^{\text{out }*}}{\underline{\tau}_{\ \ \omega\ell m}^{\text{out }}}\,.
    \end{aligned}
\end{equation}
In accordance with the analogous derivation for the Unruh state in Ref. \cite{2022PhRvD.106l5011Z}, we will use the notation $F_{\omega_{-}\ell m}^{(-)}\equiv F_{\omega(\omega_{-},m)\ell m}$ for indexed quantities $F_{\omega\ell m}$, where $\omega(\omega_-,m)\equiv \omega_-+m\Omega_-$. 
We will also use $J$ to collectively denote $\ell m$. We start by rewriting the Kruskal-based initial data in terms of Eddington-based data; we aim for expansion coefficients that relate the angular and $(t,r_{*})$ initial functions of the two mode families, i.e., coefficients denoted $\alpha_{\hat{\omega}\omega_{-}}^{+}$ and  $C_{\hat{J}J}^{\omega_-}$ that obey
\begin{equation}
    \begin{aligned}
        e^{-i\hat{\omega}V}	&=\frac{1}{2\pi}\int_{\mathbb{R}}d\omega_{-}\alpha_{\hat{\omega}\omega_{-}}^{+}e^{i\omega_{-}v},\\
        Y_{\hat{J}}(\theta,\varphi_{-})	&=\sum_{J}C_{\hat{J}J}^{\omega_{-}}Z_{J}^{\omega(\omega_{-},m)}(\theta,\varphi_{-}).
    \end{aligned}
\end{equation}
Applying an inverse Fourier transform one may write $\alpha_{\hat{\omega}\omega_{-}}^{+}=\int_{\mathbb{R}}dv\ e^{-i\omega_{-}v}e^{-i\hat{\omega}V(v)}$. Inserting the expression for $V(v)$ which is valid in region III (given in the text below Eq. \eqref{eq:uv-coordinates})
one finds\footnote{This integral can be carried out as in Eq. (3.3) in Ref. \cite{Lanir:2017oia}. Our $\alpha_{\hat{\omega}\omega_{-}}^{+}$ may be obtained from $\alpha_{\omega\tilde{\omega}}^\text{past}$, given in Eq. (3.3) therein and defined above it, upon substituting $\kappa_+\to\kappa_-$, $\omega\to-\hat{\omega}$ and $\tilde{\omega}\to\omega_-$.})
\begin{equation}\label{eq:bog+}
    \alpha_{\hat{\omega}\omega_-}^+=\frac{1}{\kappa_-}\Big(\frac{\hat{\omega}}{\kappa_-} \Big)^{i\frac{\omega_-}{\kappa_-}}e^{\frac{-\pi\omega_-}{2\kappa_-}}\Gamma\Big( \frac{-i\omega_-}{\kappa_-}\Big).
    \end{equation}
The coefficients $C_{\hat{J}J}^{\omega_-}$ relating the orthonormal functions on the two-sphere are defined by Eq. (5.2) in \cite{2022PhRvD.106l5011Z} (with the replacement $\omega_+\to \omega_-$, $\varphi_+\to \varphi_-$) and satisfy the completeness relation given in Eq. (5.17) therein. It follows that the $g_{\hat{\omega}\hat{\ell} \hat{m}}^{\text{down,+}}$ modes and the (separable) $f^{\text{down}(-)}_{\omega_-J}$ modes are related on $\mathcal{CH}^+$ by 
\begin{equation}\label{eq:out-decomposition-+}
    \sqrt{\hat{\omega}}g_{\hat{\omega}\hat{J}}^{\text{down,+}}\bigg\vert_{\mathcal{CH}^+}=\frac{1}{2\pi}\sum_J\int_\mathbb{R}d\omega_- \sqrt{\vert \omega_-\vert }C_{\hat{J}J}^{\omega_-}\alpha_{\hat{\omega}\omega_-}^+ f^{\text{down}(-)}_{\omega_-J}\bigg\vert_{\mathcal{CH}^+}.
\end{equation}
Both the \emph{down},+ Kruskal modes   $g_{\hat{\omega}\hat{\ell}\hat{m}}^{\text{down},+}$  and the Eddington-type \emph{down} modes $f_{\omega\ell m}^{\text{down}}$ vanish on $\overline{\mathcal{I}}^+$ and satisfy the same wave equation, hence the decomposition in Eq.  \eqref{eq:out-decomposition-+} is satisfied not only at $\mathcal{CH}^+$ but everywhere throughout region III. Inserting this as well as Eq. \eqref{eq:outc-outEddIII}  into the expression for the HTPF in region III (given in Eq. \eqref{GIII-out-down}),
and using
\begin{equation}
   \int_0^\infty \frac{d\hat\omega}{\hat\omega}\alpha_{\hat\omega\omega_-}^+(\alpha_{\hat\omega\omega_-'}^+)^*=\frac{4\pi^2}{\omega_-}\frac{1}{e^{2\pi\omega_-/\kappa_-}-1}\delta(\omega_--\omega_-')
\end{equation}
as well as the completeness relation in (5.17) therein, one arrives at the result 
\begin{equation}\label{eq:g-folding}
    G^{\underline{\text{U}}}
    (x,x')=\hbar\sum_{J}\int_{\mathbb{R}} d\omega_-
\text{sign}(\omega_-)\frac{1}{e^{\frac{2\pi\omega_-}{\kappa_-}}-1}\big\{f^{\text{down}(-)}_{\omega_-J}(x),f^{\text{down}(-)\ *}_{\omega_-J}(x')\big\}+\hbar\sum_{J}\int_0^\infty d\omega \big\{f^{\text{out}}_{\omega J}(x),f^{\text{out}\ *}_{\omega J}(x')\big\},
\end{equation}
which is valid for $x,x'$ in region III. The first integral can be "folded" (using the symmetries of the modes, i.e. $f^{\text{down}(-)}_{(-\omega_-)\ell(-m)}=(-1)^mf^{\text{down}(-)\ *}_{\omega_-\ell m}$) around $\omega_-=0$ to yield\footnote{It should be noted that the folding procedure mixes the contributions of pairs of $\pm \vert m\vert$ modes.}
\begin{equation}\label{2pt-function-III-folded}
    G^{\underline{\text{U}}}(x,x')=\hbar\sum_{J}\int_{0}^\infty d\omega_-
\coth\Big(\frac{\pi\omega_-}{\kappa_-} \Big)\big\{f^{\text{down}(-)}_{\omega_-J}(x),f^{\text{down}(-)\ *}_{\omega_-J}(x')\big\}+\hbar\sum_{J}\int_0^\infty d\omega \big\{f^{\text{out}}_{\omega J}(x),f^{\text{out}\ *}_{\omega J}(x')\big\}.
\end{equation}
Finally, returning to the standard $\omega$-indexed notation, we may write the above as
\begin{equation}\label{2pt-function-III-folded}
    G^{\underline{\text{U}}}(x,x')=\hbar\sum_{\ell,m}\int_{0}^\infty d\omega_-
\coth\Big(\frac{\pi\omega_-}{\kappa_-} \Big)\big\{f^{\text{down}}_{\omega\ell m}(x),f^{\text{down}\ *}_{\omega\ell m}(x')\big\}+\hbar\sum_{\ell ,m}\int_0^\infty d\omega \big\{f^{\text{out}}_{\omega \ell m}(x),f^{\text{out}\ *}_{\omega  \ell m}(x')\big\}.
\end{equation}
It is worth comparing this expression with the known expression for the HTPF in the Unruh state in the BH exterior (region I), see e.g. Eq. (5.29) in Ref. \cite{2022PhRvD.106l5011Z}. As expected, the expressions are identical under the replacements $f^\text{in}_{\omega\ell m}\mapsto f^{\text{out}}_{\omega \ell m}$, $f^\text{up}_{\omega\ell m}\mapsto f^{\text{down}}_{\omega \ell m}$,  $\omega_+\mapsto \omega_-$ and $\kappa_+\mapsto \kappa_-$.

\subsection{Region II}
We now consider the HTPF \eqref{eq:2pt-function-comparison} in the case that $x,x'$ are in region II, aiming at an expression in terms of the Eddington-type modes which are computationally advantageous. For this purpose, we will first define the Eddington-type modes in region II: \emph{left} modes $f_{\omega\ell m}^{\mathcal{L}}$ emerging from $\mathcal{CH}^{\mathcal{L}}$, and \emph{right} modes  $f_{\omega\ell m}^{\mathcal{R}}$ emerging from $\mathcal{CH}^{\mathcal{R}}$, as depicted in Fig.~\ref{fig:C-modes}. Both families of modes separate as Eq. \eqref{eq:boulware-mode-ansatz} (in what follows, we denote their respective radial solutions by $\psi_{\omega\ell m}^{\mathcal{L}}$ and $\psi_{\omega\ell m}^{\mathcal{R}}$), 
\begin{equation}\label{eq:f-LR}
    f_{\omega\ell m}^{\mathcal{R}/\mathcal{L}}\equiv\frac{1}{\sqrt{4\pi\vert\omega_{-}\vert(r^{2}+a^{2})}}e^{-i\omega t}Z_{\ell m}^{\omega}(\theta,\varphi)\psi_{\omega\ell m}^{\mathcal{R}/\mathcal{L}}(r).
\end{equation}
The $\mathcal{L}$-modes are defined by their final data
\begin{equation}
    f_{\omega\ell m}^{\mathcal{L}}	\simeq\frac{1}{\sqrt{4\pi\vert\omega_{-}\vert(r^{2}+a^{2})}}Z_{\ell m}^{\omega}(\theta,\varphi_{-})\begin{cases}
e^{-i\omega_{-}v} & \mathcal{CH}^{\mathcal{L}}\\
0 & \mathcal{CH}^{\mathcal{R}}
\end{cases},\label{eq:f-L}
\end{equation}
(in particular, $\psi_{\omega\ell m}^{\mathcal{L}}\simeq e^{-i\omega_{-}r_{*}}$ at $r\to r_{-})$. The $\mathcal{R}$-modes are defined through the final data
\begin{equation}\label{fR-modes}
    f_{\omega\ell m}^{\mathcal{R}}\simeq\frac{1}{\sqrt{4\pi\vert\omega_{-}\vert(r^{2}+a^{2})}}Z_{\ell m}^{\omega}(\theta,\varphi_{-})\begin{cases}
0 & \mathcal{CH}^{\mathcal{L}}\\
e^{i\omega_{-}u} & \mathcal{CH}^{\mathcal{R}}
\end{cases}
\end{equation}
(in particular, $\psi_{\omega\ell m}^{\mathcal{R}}\simeq e^{+i\omega_{-}r_{*}}$ at $r\to r_{-})$. The prefactor of the modes is chosen such that the KG inner product of $f_{\omega\ell m}^{\Lambda}$ and $f_{\omega'\ell' m'}^{\Lambda'}$, where $\Lambda,\Lambda'$ take the values $\mathcal{L},\mathcal{R}$, is $\text{sign}(\omega_-)\delta_{\Lambda\Lambda'}\delta_{\ell\ell'}\delta_{mm'}\delta({\omega-\omega'})$. Since the radial equation is real, we can write $\psi_{\omega\ell m}^{\mathcal{R}}=\psi_{\omega\ell m}^{\mathcal{L}*}$ and express everything in terms of a single radial solution.
Using Eq.~\eqref{eq:g_sum}, we rewrite the \emph{down} contribution to the HTPF $  G^{\underline{\text{U}}\text{,down}}(x,x')$ in region II as
\begin{equation}\label{eq:4-int}
    \begin{aligned}
        G^{\underline{\text{U}}\text{,down}}(x,x')&=\hbar\sum_{\hat{l},\hat{m}}\int_{0}^{\infty}d\hat{\omega}\Big(\left\{ g_{\hat{\omega}\hat{\ell} \hat{m}}^{\text{down,}\mathcal{L}}(x),g_{\hat{\omega}\hat{\ell} \hat{m}}^{\text{down,}\mathcal{L}\ *}(x'){}^{}\right\} +\left\{ g_{\hat{\omega}\hat{\ell} \hat{m}}^{\text{down,}\mathcal{L}}(x),g_{\hat{\omega}\hat{\ell} \hat{m}}^{\text{down,}+\ *}(x')\right\} \\&\hspace{3em}+\left\{ g_{\hat{\omega}\hat{\ell} \hat{m}}^{\text{down,}+}(x),g_{\hat{\omega}\hat{\ell} \hat{m}}^{\text{down,}\mathcal{L}\ *}(x'){}^{}\right\} +\left\{ g_{\hat{\omega}\hat{\ell} \hat{m}}^{\text{down,}+}(x),g_{\hat{\omega}\hat{\ell} \hat{m}}^{\text{down,}+\ *}(x'){}^{}\right\} \Big).
    \end{aligned}
\end{equation}
Comparing Eqs. \eqref{eq:fdownCHR} and \eqref{fR-modes}, as well as Eqs. \eqref{eq:f-out-CHR} and \eqref{fR-modes}, we note that on $\mathcal{CH^{\mathcal{R}}}$ 
\begin{equation}\label{eq:relations-RightCH}
\begin{aligned}
     f_{\omega\ell m}^{\text{down}}\bigg\vert_{\mathcal{CH^{\mathcal{R}}}}&=\underline{\rho}_{\ \ \omega\ell m}^{\text{down }}f_{\omega\ell m}^{\mathcal{R}}\bigg\vert_{\mathcal{CH^{\mathcal{R}}}}\\
     f_{\omega\ell m}^{\text{out}}\bigg\vert_{\mathcal{CH^{\mathcal{R}}}}&=\sqrt{\frac{\vert\omega_-\vert}{\vert \omega\vert }}\underline{\tau}_{\ \ \omega\ell m}^{\text{out}}f_{\omega\ell m}^{\mathcal{R}}\bigg\vert_{\mathcal{CH^{\mathcal{R}}}}
\end{aligned}
\end{equation}
and by similar arguments as above, these relations actually hold everywhere in region II. Our next goal is to find the coefficients relating the \emph{down}-$\mathcal{L}$
Kruskal-type modes  $g_{\hat{\omega}\hat{\ell}\hat{m}}^{\text{down},\mathcal{L}}$  and the \emph{left} Eddington-type modes $f_{\omega\ell m}^{\mathcal{L}}$ in region II. Namely, we want to fix the coefficients $\alpha_{\hat{\omega}\omega_{-}}^{\mathcal{L}}$ and $C_{\hat{J}J}^{\omega_{-}}$ in the expansion
\begin{equation} \label{eq:outL-eddingtonII}
\sqrt{\hat{\omega}}g_{\hat{\omega}\hat{J}}^{\text{down,}\mathcal{L}}(x)=\frac{1}{2\pi}\sum_{J}\int_{\mathbb{R}}d\omega_{-}\sqrt{\vert\omega_{-}\vert}C_{\hat{J}J}^{\omega_{-}}\alpha_{\hat{\omega}\omega_{-}}^{\mathcal{L}}f_{\omega_{-}J}^{\mathcal{L}(-)}(x).
\end{equation}
Re-expressing the initial Kruskal data $e^{-i\hat{\omega}V}$ in terms of the Eddington-type data $e^{-i\omega_{-}v}$ and performing an inverse Fourier transform (like we did in region III), we find (the resulting expression is that in Eq. (6.10) in Ref.  \cite{2022PhRvD.106l5011Z} upon replacing $\kappa_+\to\kappa_-$, $\hat\omega\to-\hat\omega$ and $\omega_+\to\omega_-$)
\begin{equation}
\alpha_{\hat{\omega}\omega_{-}}^{\mathcal{L}}=\alpha_{\hat{\omega}\omega_{-}}^{+}e^{\pi\omega_-/\kappa_-},
\end{equation}
with $\alpha_{\hat{\omega}\omega_{-}}^{+}$ given in Eq.  \eqref{eq:bog+} and angular coefficients $C_{\hat{J}J}^{\omega_{-}}$ unchanged. Using Eq. \eqref{eq:relations-RightCH}, we additionally have 
\begin{equation}
\begin{aligned}
     \sqrt{\hat{\omega}}g_{\hat{\omega}\hat{\ell} \hat{m}}^{\text{down,+}}(x)&=\frac{1}{2\pi}\sum_J\int_\mathbb{R}d\omega_- \sqrt{\vert \omega_-\vert }C_{\hat{J}J}^{\omega_-}\alpha_{\hat{\omega}\omega_-}^+ \underline{\rho}_{\ \ \omega_-\ell m}^{\text{down}(-)}f_{\omega_-\ell m}^{\mathcal{R}(-)}(x)\\
     \sqrt{\omega}g_{\omega\ell m}^{\text{out}}(x)&=(-1)^m \sqrt{\vert \omega_-\vert}\underline{\tau}_{\ \ \omega_-\ell m}^{\text{out}(-)\ *}f_{\omega_-\ell m}^{\mathcal{R}(-)\ *}(x)
\end{aligned}
\end{equation}
in region II. These expressions, along with Eq. \eqref{eq:outL-eddingtonII} may be inserted in Eqs. \eqref{gdown1}, \eqref{eq:4-int}, and using the same strategy used in region III, we obtain the result 
\begin{equation}
    \begin{aligned}
        G^{\underline{\text{U}}}(x,x')&=\hbar\sum_J\int_\mathbb{R}d\omega_-\ \text{sign}(\omega_-)\frac{1}{1-e^{\frac{-2\pi\omega_-}{\kappa_-}}}\{f_{\omega_-J}^{\mathcal{L}(-)}(x) ,f_{\omega_-J}^{\mathcal{L}(-)\ *}(x') \}\\
        &+\hbar\sum_J\int_\mathbb{R}d\omega_-\ \text{sign}(\omega_-)\frac{1}{e^{\frac{\pi\omega_-}{\kappa_-}}-e^{\frac{-\pi\omega_-}{\kappa_-}}}2\Re\Big( \underline{\rho}_{\ \ \omega_-J}^{\text{down}(-)\ *}\{f_{\omega_-J}^{\mathcal{L}(-)}(x) ,f_{\omega_-J}^{\mathcal{R}(-)\ *}(x') \}\Big)\\
        &+\hbar\sum_J\int_\mathbb{R}d\omega_-\ \text{sign}(\omega_-)\frac{1}{e^{\frac{2\pi\omega_-}{\kappa_-}}-1}\vert\underline{\rho}_{\ \ \omega_-J}^{\text{down}(-)}\vert^2\{f_{\omega_-J}^{\mathcal{R}(-)}(x) ,f_{\omega_-J}^{\mathcal{R}(-)\ *}(x') \}\\
         &+\hbar \sum_J\int_0^\infty d\omega\frac{\vert \omega_-\vert }{\omega}\vert \underline{\tau}_{\ \ \omega_-J}^{\text{out}(-)}\vert^2 \{f_{\omega_-J}^{\mathcal{R}(-)}(x),f_{\omega_-J}^{\mathcal{R}(-)\ *}(x') \},
\end{aligned}
\end{equation}
where we can already identify the hyperbolic cosecant in the second line. Folding the first $3$ integrals and dropping the $(-)$ notation at the final stage yields the result
\begin{equation}
    \begin{aligned}
        G^{\underline{\text{U}}}(x,x')&=\hbar\sum_J\int_0^\infty d\omega_- \coth\Big(\frac{\pi\omega_-}{\kappa_-} \Big)\Big(
        \{f_{\omega J}^{\mathcal{L}}(x) ,f_{\omega J}^{\mathcal{L}\ *}(x') \}+\vert\underline{\rho}_{\ \ \omega J}^{\text{down}}\vert^2\{f_{\omega J}^{\mathcal{R}}(x) ,f_{\omega J}^{\mathcal{R}\ *}(x') \}\Big)\\
        &+\hbar\sum_J\int_0^\infty d\omega_-\ 2\text{csch} \Big(\frac{\pi\omega_-}{\kappa_-} \Big) \Re\Big( \underline{\rho}_{\ \ \omega J}^{\text{down}}\{f_{\omega J}^{\mathcal{L}\ *}(x) ,f_{\omega J}^{\mathcal{R} }(x') \}\Big)\\
     &+\hbar \sum_J\int_0^\infty d\omega \frac{\vert \omega_-\vert }{\omega}\vert \underline{\tau}_{\ \ \omega J}^{\text{out}}\vert^2 \{f_{\omega J}^{\mathcal{R}\ }(x),f_{\omega J}^{\mathcal{R}\ *}(x') \}. 
\end{aligned}
\end{equation}
In particular, it follows that the Unruh HTPF in region II (given explicitly in Eq. (6.37) in Ref. \cite{2022PhRvD.106l5011Z}) and the $\vert\underline{\text{U}}\rangle$-state HTPF in region II are related by the simple mapping $\omega_{+}\mapsto\omega_{-}$, $\kappa_{+}\mapsto\kappa_{-}$, $\rho_{\omega\ell m}^{\text{up}}\mapsto\underline{\rho}_{\ \ \omega\ell m}^{\text{down}}$, $\tau_{\omega\ell m}^{\text{in}}\mapsto\underline{\tau}_{\ \ \omega\ell m}^{\text{out}}$ and $f_{\omega\ell m}^{R/L}\mapsto f_{\omega\ell m}^{\mathcal{R}/\mathcal{L}}$ (with the $f_{\omega\ell m}^{R/L}$ functions defined in the reference) .

As long as the points are separated (and not connected by a null geodesic) the HTPFs in the two states $\vert\text{U}\rangle$ and $\vert\underline{\text{U}}\rangle $ are finite and individually well-defined. The coinciding point limit of their difference is also well-defined, as the difference of two HTPFs is smooth. In Ref. \cite{Zilberman:2024jns}, which developed $t$-splitting methods applicable inside a Kerr BH along the axis, an  intermediate split in $\theta$ was required in addition to the split in $t$. Accordingly, here we consider two distinct points $x$ and $x'$, separated by $\Delta t$ and $\Delta\theta$, i.e., $x=(t,r,\theta,\varphi)$ and $x'=(t+\Delta t,r,\theta+\Delta\theta,\varphi)$. In this case 
\begin{equation}\label{eq:GbarU}
    \begin{aligned}
        G^{\underline{\text{U}}}(x,x')=&\frac{\hbar}{8\pi^{2}(r^{2}+a^{2})}\sum_{\ell m}\int_{0}^{\infty}\frac{d\omega_{-}}{\omega_{-}}\Bigg[\coth\left(\frac{\pi\omega_{-}}{\kappa_{-}}\right)\left(\vert\psi_{\omega\ell m}^{\mathcal{R}}(r)\vert^{2}+\vert \underline{\rho}_{\ \ \omega \ell m}^{\text{down}} \vert^{2}\vert\psi_{\omega\ell m}^{\mathcal{R}}(r)\vert^{2}\right)\\
        &+2\text{csch}\left(\frac{\pi\omega_{-}}{\kappa_{-}}\right)\Re\left(\underline{\rho}_{\ \ \omega \ell m}^{\text{down}}\psi_{\omega\ell m}^{\mathcal{R}}(r)^{2}\right)\Bigg]S_{\ell m}^{\omega}(\theta)S_{\ell m}^{\omega}(\theta+\Delta\theta)\cos(\omega\Delta t)\\
        &+\frac{\hbar}{8\pi^{2}(r^{2}+a^{2})}\sum_{\ell m}\int_{0}^{\infty}\frac{d\omega}{\omega}\vert\underline{\tau}_{\ \ \omega\ell m}^{\text{out}}\vert^{2}\vert\psi_{\omega\ell m}^{\mathcal{R}}(r)\vert^{2}S_{\ell m}^{\omega}(\theta)S_{\ell m}^{\omega}(\theta+\Delta\theta)\cos(\omega\Delta t),
    \end{aligned}
\end{equation}
and a similar expression holds for $G^\text{U}(x,x')$, derived from Eq. (6.37) in Ref. \cite{2022PhRvD.106l5011Z}.
In deriving these forms for $ G^{\underline{\text{U}}}(x,x')$ and $ G^{\text{U}}(x,x')$, we used the following result\footnote{In the original reference \cite{2022PhRvD.106l5011Z}, only the $\omega_+$ case was discussed, but the arguments exposed there trivially extend to the $\omega_-$ case.
}: Given a function $E_{m}(\omega)$ with the property $E_{(-m)}(-\omega)=E_{m}(\omega)$, we may write $\sum_{m=-\ell}^{\ell}\int_{0}^{\infty}d\omega_{\pm}E_{m}(\omega)=\sum_{m=-\ell}^{\ell}\int_{0}^{\infty}d\omega E_{m}(\omega)$. 
Finally, inserting the asymptotic IH behavior $\psi_{\omega\ell m}^{\mathcal{R}}\simeq e^{i\omega_-r_*}$ 
(and correspondingly for the radial solutions entering the HTPF in the Unruh state) and taking the limit $x'\to x$ (i.e., $\Delta t\to 0$ and $\Delta \theta \to 0$) in the HTPF difference $G^{\text{U}-\underline{\text{U}}}(x,x')$ yields Eqs.~ \eqref{eq:mode-sum-phi2}-\eqref{eq:integrands} in the main text.

\section{Numerical methods}\label{sec:appendixB}

\subsection{Computation of $\rho_{\omega \ell 0}^\text{up}$, $A_{\omega \ell 0}$, $B_{\omega \ell 0}$, $\underline{\rho}_{\ \ \omega \ell 0}^\text{down}$ and $S_{ \ell 0}^\omega$}\label{sec:NumCoeffs}

The numerical ingredients necessary to evaluate the mode-sum \eqref{eq:mode-sum-phi2}-\eqref{eq:integrands} at the pole $(\theta=0)$ are the exterior scattering reflection coefficient $\rho_{\omega \ell 0}^\text{up}$, the interior scattering coefficients $A_{\omega \ell 0}$, $B_{\omega \ell 0}$, the spheroidal harmonics $S_{ \ell 0}^\omega$ and the reflection scattering coefficient associated with the comparison $\vert\underline{\text{U}}\rangle$ state, $\underline{\rho}_{\ \ \omega \ell 0}^\text{down}$ 
\footnote{Note that Eqs. \eqref{eq:mode-sum-phi2},\eqref{eq:integrands}  (evaluated at the pole) also involve the transmission coefficients  $\tau_{\omega \ell 0}^\text{in}$ and  $\underline{\tau}_{\ \ \omega\ell 0}^{\text{out}}$, but only through their absolute values, which are related to the corresponding reflection coefficients listed above through the Wronskian relations \eqref{eq:wronskian-relations}, \eqref{eq:wronskian-out} and \eqref{eq:wronskian-comparison}.}. We obtained all ingredients, with the exception of $\underline{\rho}_{\ \ \omega \ell 0}^\text{down}$, as detailed in Appendix C, Sec. 1 in Ref. \cite{Zilberman:2024jns}. The $\vert\underline{\text{U}}\rangle$ reflection coefficient $\underline{\rho}_{\ \ \omega \ell 0}^\text{down}$ was obtained by numerically solving the related radial equation in Mathematica \cite{Mathematica14p1}, within a precision of $10^{-17}$ for $\ell=0$ and $\ell=1$, and $10^{-20}$ for $\ell=2$. Higher precision is required for larger $\ell$ in order to resolve the inverse-power tails (see vertical scale in Fig. \ref{fig:QNM-ringdown-and-tail}). For $\ell>2$, our $\underline{\rho}_{\ \ \omega \ell 0}^\text{down}$ has a precision of 17 significant figures. We did not attempt to increase the precision for any $\ell>2$, and therefore did not expose the tails in these higher-$\ell$ cases. Regarding the ringing in Fig. \ref{fig:ringdown}, the numerically obtained ringing frequencies are found to match the $\ell=0$ to $\ell=3$ fundamental $m=0$ mode frequencies to a (relative) precision of $2\%$ for $\ell=0$, $0.1\%$ for $\ell=1$, and $0.02\%$ for $\ell=2$ and $\ell=3$. The ringing frequencies can only be determined exactly if the ringing itself extends to infinity (in $r_*$), which is not the case here, due to the presence of the tails. However, the agreement with the classical QNM frequencies extends to the accuracy to which we were able to numerically determine the ringing frequencies. 

For the figures throughout the paper, the accuracy of all points is such that the error would not be visible on the scale presented.

\subsection{Strongly-oscillating integrand}
The mode-sum expression for $ \langle\Phi^{2}\rangle^{\text{U}-\underline{\text{U}}}$ in the (polar) IH vicinity, given through Eqs. \eqref{eq:delta-def} and \eqref{eq:phi2-schematic}, has an integrand function which oscillates like $\exp(2i\omega r_*)$. At large $r_*$ (corresponding to the close vicinity of the IH) the integrand hence oscillates very rapidly (the wavelength of the oscillations in $\omega$ being $\sim r_*^{-1}$). In order to adequately resolve the integrand, we proceed as follows: we compute the envelope of the oscillations (see Eq. \eqref{eq:phi2-schematic}) at an \textit{initial} spacing $d\omega_i$, which is determined by the spacing of the scattering coefficients and which ranges from $d\omega_i=(400M)^{-1}$ to $(200M)^{-1}$ in our analysis. The envelope is well-behaved and decays exponentially at large $\omega$. We then interpolated this envelope and refined it by sampling the interpolating function at a \textit{refined} $d\omega_{\text{ref}}$ spacing of our choice (a typical value $d\omega_{\text{ref}}=(1000M)^{-1}$ was sufficient to produce the figures in this document). The refined integrand is then trivially multiplied by the oscillating (discrete) exponential function $\exp{(2i\omega_{\text{ref}}r_*)}$, allowing one to resolve the oscillations and accurately evaluate the integral. In doing so, we considered $\omega_{\text{max}}$ ranging from $10M^{-1}$ to $15M^{-1}$. The  (absolute) error associated with $\omega_\text{max}$ is $\mathcal{O
}(10^{-20})$, which is negligible with respect to other errors present in the analysis. The $\ell_\text{max}$ error, apparent in quantities involving the sum over $\ell$, like \eqref{eq:phi2(-)}, induces an error $\mathcal{O}(10^{-11})$, which is smaller than the finite $d\omega$ error, being $\mathcal{O}(10^{-9})$ in the same quantity. Hence, the state-subtraction results in Subsecs. \ref{subsec:Price-tail} and \ref{subsec:QNM-ringdown}, such as $c_-$ appearing in Eq. \eqref{cbar}, carry absolute error $\mathcal{O}(10^{-9})$ (corresponding to $0.001\%$ relative precision). When the state-subtraction results are combined with the near-IH $t$-splitting results in Ref.  \cite{Zilberman:2024jns}, the limited precision of the latter analysis (which is of order $\mathcal{O}(10^{-5})$) constitutes the limiting factor.
 Therefore, quantities in Subsec. \ref{subsec:Comparison-to-t-splitting} (e.g.,  Eq. \eqref{eq:phi2(-)}) carry larger absolute error, of order $\sim  \mathcal{O}(10^{-5})$ (corresponding to $0.5\%$ relative precision).

\section{Analytic derivation of the leading inverse-power tail in $\langle\hat\Phi^2\rangle^\text{U}$ on approaching the polar IH}\label{sec:appendixC}

In this Appendix, we follow the approach outlined below Eq.~\eqref{eq:omega-power-integral} and employ the small-frequency expansions of the various scattering coefficients entering the mode-sum in \eqref{eq:phi2-schematic}-\eqref{eq:Ewl0} to analytically derive the leading inverse-power tail for the $\ell=0$ mode contribution to  $\langle\hat\Phi^2\rangle^\text{U}$. Specifically, the analysis below establishes (i) the leading inverse-power for this mode (found to be $r_*^{-3}$), (ii) the associated prefactor $c_0$, defined in Eq.~\eqref{eq:c0-def} and given explicitly in Eq.~\eqref{eq:c0-analytical}, and (iii) the absence of $r_*\ln (r_*/M)$ terms at this order. As discussed in \ref{subsec:Comparison-to-t-splitting}, the tail in  $\langle\hat\Phi^2\rangle^{\text{U}-\underline{\text{U}}}$ entirely originates from  $\langle\hat\Phi^2\rangle^\text{U}$. Hence, like elsewhere in the manuscript, we can focus on $\langle\hat\Phi^2\rangle^{\text{U}-\underline{\text{U}}}$ in the present tail analysis.\footnote{The analytical analysis presented in this Appendix is specific for using the  $\vert \underline{\text{U}}\rangle$ state as regulator. However, the result for $c_0$ can be checked to hold for a different class of states for which the analog of the scattering coefficient  $\underline{\rho}_{\ \ \omega 00}^{\text{down}}$ has a small-$\omega$ expansion of the type $\sum_{n=0}^\infty\underline{\rho}_n(i\omega M)^n$, with real coefficients $\underline{\rho}_n$. It is, however, the $\vert \underline{\text{U}}\rangle$ state that we use in this work, and hence we shall not elaborate further on such other computations.  See also footnote \ref{ft:comparison-state-2}.}

In the following, we give the small-$\omega$ expansion of $\rho^\text{up}_{\omega00}$, defined in Eq.~\eqref{eq:upI-scattering} (we take $\ln(z)$ to denote the principal branch, so that $\ln(-i)=-i\pi/2$):
\begin{equation}\label{eq:rho-small-w}
\rho^\text{up}_{\omega00}=\rho_0+\rho_1(i\omega M)+\rho_2(i\omega M)^2+(i\omega M)^3\big(\rho_3+\tilde{\rho}_3\ln(-i\omega M) \big)+(i\omega M)^4\big(\rho_4+\tilde{\rho}_4\ln(-i\omega M)\big)+o\big((\omega M)^5\big),
\end{equation}
with real dimensionless coefficients $\rho_i$, $i\in\{0,4 \}$ and $\tilde{\rho}_i$, $i\in\{3,4 \}$. Numerical evidence indicates, however, that the relatively simple structure in Eq. \eqref{eq:rho-small-w} breaks down already at next order $o\big((\omega M)^5\big)$. 

One can show that the interior scattering coefficients $A_{\omega 0 0}$ and $B_{\omega 0 0}$, defined via Eq.~\eqref{eq:inII-scattering}, admit the series expansions 
\begin{equation}\label{eq:AB-small-w}
    A_{\omega 00}=\sum_{n=0}^\infty A_n(i\omega M)^n, \hspace{2cm}B_{\omega 00 }=\sum_{n=0}^\infty B_n(i\omega M)^n,
\end{equation}
for real $A_n$ and $B_n$, for all $n\geq0$. In fact, the form of this simple power-series expansion can be shown to hold for any $\ell\geq0,m=0$. In doing so, one needs to justify (i) the absence of log terms (unlike the scattering coefficients in the exterior, see Eq. \eqref{eq:rho-small-w}), and (ii) the reality of the prefactors in the expansion. Statement (i) follows since $A_{\omega \ell 0}$ and $B_{\omega\ell 0}$ are coefficients for the scattering between two regular points of the radial differential equation \eqref{kg-radial-joined} (corresponding to the horizons). Regarding (ii), the reality of the coefficients in the expansion follows from the symmetry $A_{-\omega \ell 0}^*=A_{\omega\ell 0}$, for $\omega\in\mathbb{R}$, (and likewise for $B_{\omega\ell 0}$), which is induced by the radial operator in Eq.~\eqref{kg-radial-joined} (with $m=0$), together with the initial data in Eq.~\eqref{eq:inII-scattering}.

\begin{figure}
\begin{centering}
\hfill{}\includegraphics[width=0.45\columnwidth]{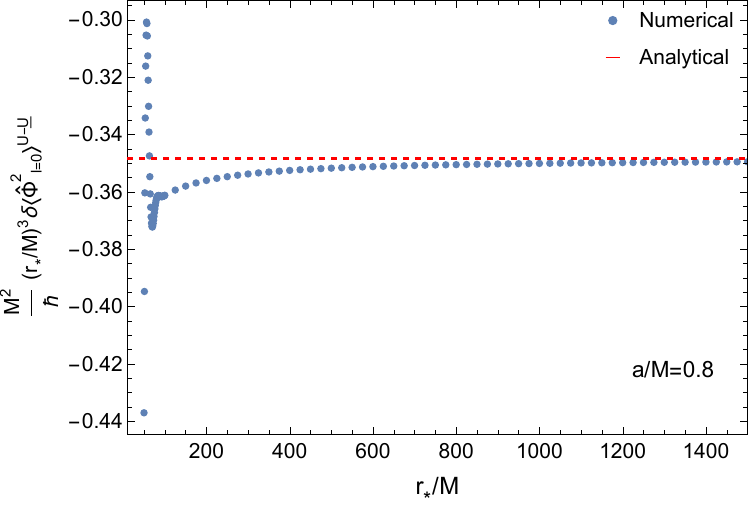}\hfill{}\includegraphics[width=0.46\columnwidth]{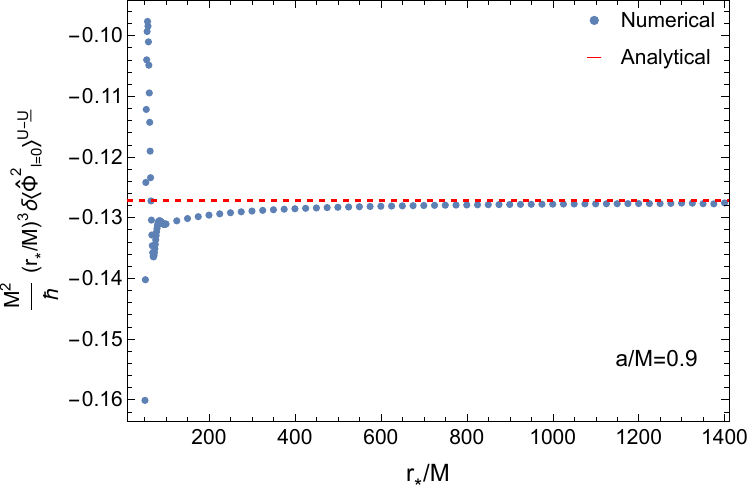}\hfill{}
\par\end{centering}
\caption{Numerical results (blue points) for $\delta\langle\Phi^2_{\ell=0}\rangle^{\text{U}-\underline{\text{U}}}$ multiplied by $(r_*/M)^3$ as a function of $r_*/M$, shown for $r_*/M>50$, for $a/M=0.8$ (left) and $a/M=0.9$ (right). The oscillations on the left of both panels correspond to the QNM ringdown, discussed in detail in Subsec. \ref{subsec:QNM-ringdown}. The dashed red line indicates the analytical asymptotic value $c_0$ (given explicitly in Eq. \eqref{eq:c0-analytical}). The numerical results converge to the analytical prediction on approaching the IH ($r_*/M\to\infty$).}\label{fig:l0tail-c0}
\end{figure}

In the following, we include the explicit coefficients which are necessary to compute the $\ell=0$ leading tail. For $\rho_{\omega00}^\text{up}$, only the coefficients
\begin{subequations}\label{eq:small-w-rho-expl}
    \begin{align}
        \rho_0&=-1,\\
        \rho_1&=-2\frac{r_+}{M},\\
        \rho_2&=2\big(\frac{a^2}{M^2}-4\frac{r_+}{M} \big),\\
        \tilde{\rho}_3&=-16\frac{r_+}{M},\\
        \tilde{\rho}_4&=-\frac{16}{3}\big(23\frac{r_+}{M}-6\frac{a^2}{M^2}\big)
    \end{align}
\end{subequations}
explicitly appear \footnote{We note that the coefficient $\tilde\rho_4$ appears in \eqref{eq:small-w-rho-expl} (and so is relevant for the calculation of $c_0$) whereas the lower order coefficient $\rho_3$ does not. This is due to the  reality of these coefficients combined with the  specific way in which $\rho_{\omega \ell m}^\text{up}$ appears in the integrand in Eq.~\eqref{eq:phi2-schematic} together with the integrals in \eqref{eq:omega-power-integral}.}. For the interior scattering coefficients, the relevant coefficients are
\begin{subequations}\label{eq:small-w-AB-expl}
    \begin{align}
        A_0&=\frac{1}{2}\Big( \sqrt{\frac{\kappa_+}{\kappa_-}}-\sqrt{\frac{\kappa_-}{\kappa_+}}\Big),&B_0=\frac{1}{2}\Big( \sqrt{\frac{\kappa_+}{\kappa_-}}+\sqrt{\frac{\kappa_-}{\kappa_+}}\Big),\\
        A_1&=-B_1=\frac{5}{3}A_0.
    \end{align}
\end{subequations}
Using the expansion coefficients in Eqs. \eqref{eq:small-w-rho-expl}-\eqref{eq:small-w-AB-expl}, the integrand in Eq. \eqref{eq:phi2-schematic} (and as specified in Eqs. \eqref{eq:integrands}-\eqref{eq:Ewl0}) can be expanded in a small-frequency series\footnote{The small-frequency expansion for $\rho_{\omega00}^\text{up}$, and the present analogies between the $\text{U}$ and $\underline{\text{U}}$ state (discussed in Appendix \ref{sec:appendixA}) together suggest an expansion for the scattering coefficient $\underline{\rho}_{\ \ \omega 0 0}^\text{down}$ with the same structure as that in Eq. \eqref{eq:rho-small-w}. I.e., real expansion coefficients when expanded in $i\omega M$, and $\ln(-i\omega M)$ terms at $3^\text{rd}$ and $4^\text{th}$ order.  Under this assumption (which we verified numerically), it can then be shown \cite{Alberti:2026kerrNB} that the $\ell=0$ tail prefactor is independent of the specific values of the expansion coefficients of  $\underline{\rho}_{\ \ \omega 0 0}^\text{down}$. Note, however, that the expansion to linear order $\underline{\rho}_{\ \ \omega 0 0}^\text{down}=-1+(2r_-/M)(i\omega M)+\mathcal{O}((\omega M)^2)$, plays an important role in establishing continuity (and in particular integrability) of the integrand factor in Eq.~\eqref{eq:Ewl0} at small frequencies.}. The mode-integral can then be computed term by term using the integral relations \eqref{eq:small-w-log}, and taking into account the parity of the prefactors, the result in \eqref{eq:c0-analytical} follows. The derivation can be found in Ref. \cite{Alberti:2026kerrNB}. The analytical result is plotted against the numerics in Fig. \ref{fig:l0tail-c0} for $a/M=0.8,0.9$. In both cases, the numerically obtained prefactor matches the analytical prediction \eqref{eq:c0-analytical} to $0.1\%$ relative precision (consistent with the error limitations of the numerics).

\section{Small-frequency expansions of scattering coefficients}
\label{sec:appendixD}

In this appendix we briefly describe how we analytically calculated the  small-frequency expansions of the interior scattering coefficients $A_{\omega \ell 0}$ and $B_{\omega \ell 0}$, defined via Eq.~\eqref{eq:inII-scattering},
and the exterior coefficient $\rho^\text{up}_{\omega00}$, defined in Eq.~\eqref{eq:upI-scattering}.
In order to obtain these expansions, we used the so-called MST method
for the radial Teukolsky equation for generic integer spin $s=0,\pm 1,\pm 2$ (see Eq.~(4.9) in~\cite{Teukolsky:1973ha}). The MST method is reviewed in Ref.~\cite{Sasaki:2003xr} for radial solutions and scattering in the exterior of Kerr and we readily extended to the interior (for the interior, see also Ref.~\cite{2024PhRvD.109f5023M}, with which we have slight disagreements for $s\neq 0$, which we note below).

The MST method essentially yields double-ended infinite-series  representations for the scattering coefficients and the radial solutions.
The MST infinite series are naturally adapted to providing small-frequency expansions up to arbitrary order.
While Ref.~\cite{Sasaki:2003xr} (and references therein as well as Ref.~\cite{2024PhRvD.109f5023M}) provide  MST, infinite-series  representations for the scattering coefficients (and radial solutions), their actual  small-frequency expansions are obtained by carefully expanding the summands for small $M\omega$ and truncating the infinite series at the appropriate order.
In the case of the scalar field,
Ref.~\cite{PhysRevD.94.124053} derived the explicit small-frequency expansions of the exterior scattering coefficients (with a  normalization of the radial solution different from the one in this paper) as well as of various quantities which are necessary for calculating the expansions of the interior coefficients. 
Thus, to obtain the explicit expansion for $\rho^\text{up}_{\omega00}$, we use the expansions of the exterior coefficients given in \cite{PhysRevD.94.124053} and adapt them to our normalization here; 
in its turn, to obtain the explicit expansions for $A_{\omega \ell 0}$ and $B_{\omega \ell 0}$, we use the infinite-series representations which we derive together with the  expansions of the building blocks given in \cite{PhysRevD.94.124053}.
Below we give the explicit MST series representations for $\rho^\text{up}_{\omega \ell m}$, $A_{\omega \ell m}$ and $B_{\omega \ell m}$
and provide their explicit small-frequency expansions for $m=0$: for $\ell=0$ in the case of $\rho^\text{up}_{\omega \ell m}$ and for generic $\ell$ in the case of $A_{\omega \ell m}$ and $B_{\omega \ell m}$.

Before doing so, it is useful to provide a dictionary between the notation used in Ref.~\cite{Sasaki:2003xr} and that used in this Appendix.
The MST expansions  in~\cite{Sasaki:2003xr} are described in terms of variables
 \begin{alignat}{3}
 &q\equiv \frac{a}{M},
 &&\epsilon\equiv2 M \omega, 
 &&\kappa\equiv\sqrt{1-q^2}, \\
 &x\equiv \frac{\omega (r_{+}-r)}{\epsilon\kappa}, \qquad
&&\tau\equiv \frac{\epsilon-m q}{\kappa},\qquad
&&\epsilon_{\pm}\equiv \frac{\epsilon\pm \tau}{2}.
 \end{alignat}
 These MST variables connect to variables we have defined in this paper, such as
 \begin{equation}
\omega_{\pm}\equiv \omega-m\Omega_{\pm}, 
\quad \Omega_{\pm}\equiv \frac{a}{2Mr_{\pm}},
\quad \kappa_{\pm}\equiv \frac{r_{+}-r_{-}}{4Mr_{\pm}},
 \end{equation}
via
 \begin{equation}
  \epsilon_{\pm}=\pm \frac{2Mr_{\pm}}{r_+-r_-}\omega_{\pm}=\pm \frac{\omega_{\pm}}{2\kappa_\pm},
 \quad
 2M\kappa= r_+-r_-=4Mr_\pm\kappa_{\pm}, 
 \quad
x= \frac{r_{+}-r}{r_{+}-r_{-}}.
  \end{equation}
We also note that Refs.~\cite{Sasaki:2003xr,2024PhRvD.109f5023M} make a different choice for the integration constant for the tortoise coordinate $r_*$ defined via $\dfrac{d\tort}{dr}=\dfrac{(r^2+a^2)}{\Delta}$. We shall denote by $\rsST$ the  tortoise coordinate used in Refs.~\cite{Sasaki:2003xr,2024PhRvD.109f5023M}.
Specifically, they choose
\begin{align}\label{eq:tortoiseST}
\rsST
&=r+\frac{2M}{r_+-r_-}\left\{r_+\ln\left|\frac{r-r_+}{r_+ + r_-}\right|-r_-\ln\left|\frac{r-r_-}{r_+ + r_-}\right|\right\}, \\
\end{align}
whereas in this paper we  use (see Eq.~(2.5) in Ref.~\cite{2022PhRvD.106l5011Z})
\begin{equation}\label{eq:tortoiseO}
\tort=
r+\frac{2M}{r_+-r_-}\left\{r_+\ln\left|\frac{r-r_+}{r_+ - r_-}\right|-r_-\ln\left|\frac{r-r_-}{r_+ - r_-}\right|\right\} .
\end{equation}
The two definitions differ by a constant
\begin{equation}\label{eq:tortoisediff}
\delta\tort\equiv \tort^\text{ST}-\tort=2M \log \kappa,
\end{equation}
which may lead to a different phase in the definition of  scattering coefficients.

We next consider two solutions to the radial Teukolsky equation, Eq.~(4.9) in~\cite{Teukolsky:1973ha}, for generic integer spin $s=0,\pm 1,\pm 2$ and at the end we specialize to $s=0$.
For $s=0$, the first solution will define the interior scattering coefficients $A_{\omega \ell m}$ and $B_{\omega \ell m}$, and the second solution the exterior coefficient $\rho^\text{up}_{\omega \ell m}$.

\subsection{Ingoing radial solution}

The first  solution to the general-spin radial Teukolsky equation that we consider is readily defined in the interior and the exterior of the black hole and is that given by (see Eq.~(116) in~\cite{Sasaki:2003xr} after placing an absolute value around $(-x)$, and Eq.~(A8) in~\cite{2024PhRvD.109f5023M})
\begin{align}
\Rin&=e^{i \epsilon \kappa x}|-x|^{-s-i\epsilon_+}(1-x)^{i\epsilon_-}p_{\text{in}}^\nu(x), \label{Eq:Rin 2F1} \\
p_{\text{in}}^\nu&\equiv\sum_{n=-\infty}^{\infty}\an{n} p_{n+\nu}(x), \nn\\
p_{n+\nu}(x)&\equiv{}_2F_1\left(n+\nu+1-i\tau,-n-\nu-i\tau;1-s-2i\epsilon_+;x\right).\nn 
\end{align}
The  coefficients $a^\nu_n$ satisfy a  three-term recurrence relation (see Eq.~(123) in Ref.~\cite{Sasaki:2003xr}):
\begin{equation}
\alpha_n^\nu \an{n+1}+\beta_n^\nu \an{n}+\gamma_n^\nu \an{n-1}=0,
\label{Eq:anrecursion}
\end{equation}
where
\begin{align}
\alpha_n^\nu&\equiv\frac{i\epsilon\kappa(n+\nu+1+s+i\epsilon)(n+\nu+1+s-i\epsilon)(n+\nu+1+i\tau)}{(n+\nu+1)(2 n+2 \nu+3)} ,\label{Eq:alpha}\\
\beta_n^\nu&\equiv-{}_{s}\lambda_{\ell m\omega}-s(s+1)+(n+\nu)(n+\nu+1)+\epsilon^2+\epsilon(\epsilon-m q)+\frac{\epsilon(\epsilon-m q)(s^2+\epsilon^2)}{(n+\nu)(n+\nu+1)},
\nonumber
\\
\gamma_n^\nu&\equiv-\frac{i\epsilon\kappa(n+\nu-s+i\epsilon)(n+\nu-s-i\epsilon)(n+\nu-i\tau)}{(n+\nu)(2 n+2 \nu-1)}, 
\nonumber
\end{align} 
with the normalization $a^\nu_{0}=1$, and 
where ${}_{s}\lambda_{\ell m\omega}$ is the eigenvalue of the  general-spin angular Teukolsky equation (Eq.~(4.10) in \cite{Teukolsky:1973ha}; it is ${}_{s=0}\lambda_{\ell m\omega}=\lambda_{\ell m}(a\omega)$).
In its turn, the so-called renormalized angular momentum parameter $\nu$ satisfies
an implicit equation in terms of infinite continued fractions  (see Eq.~(133) in Ref.~\cite{Sasaki:2003xr}):
\begin{equation}
R_nL_{n-1}=1,
\end{equation}
for an arbitrary choice of $n\in\mathbb{Z}$,
where
\begin{equation}
R_n=-\frac{\gamma_n^\nu}{\beta_n^\nu-\frac{\alpha_n^\nu\gamma_{n+1}^\nu}{\beta_{n+1}^\nu-\frac{\alpha_{n+1}^\nu\gamma_{n+2}^\nu}{\beta_{n+2}^{\nu}-\dots}}},\quad
L_n=-\frac{\alpha_n^\nu}{\beta_n^\nu-\frac{\alpha_{n-1}^\nu\gamma_{n}^\nu}{\beta_{n-1}^\nu-\frac{\alpha_{n-2}^\nu\gamma_{n-1}^\nu}{\beta_{n-2}^{\nu}-\dots}}}.
\end{equation}

The  radial Teukolsky solution \eqref{Eq:Rin 2F1} satisfies the following boundary conditions: 
\begin{align}\label{eq:Rin bc}
\Rin(r,\omega)&\sim \left\{\begin{array}{l l}
\RintraInt\Delta^{-s}e^{-i \omega_-\rsST}+\RinrefInt e^{+i \omega_-\rsST}, & r\rightarrow r_-, \\
\Btra\Delta^{-s}e^{-i \omega_+\rsST}, & r\rightarrow r_+, \\
r^{-2s-1} \Bref e^{i \omega \rsST} +r^{-1} \Binc e^{-i \omega \rsST}, & r \rightarrow \infty,
\end{array}
\right. \end{align}
where the exterior coefficients $\Btra$,  $\Binc$ and  $\Bref$  are given by, respectively, Eqs.~(167), (168) and (169) 
in~\cite{Sasaki:2003xr}.
By taking the asymptotics of Eq.~\eqref{Eq:Rin 2F1} as $r\to r_-$ and comparing with \eqref{eq:Rin bc}, we readily find the following representations for the interior coefficients:

\begin{align}\label{eq:Bintreftra}
\RintraInt&=(-1)^s  (2\kappa  M)^{2 s} e^{i
   (\kappa  \epsilon -\phi) } \Gamma
   (1-s-2i \epsilon_+) \Gamma
   (s+2i \epsilon_-) \sum
   _{n=-\infty}^{\infty}
   \frac{a_n}{\Gamma
   (-\nu-n-i \tau ) \Gamma
   (1+\nu+n-i \tau)},\nn
\\
\RinrefInt&=e^{i (\kappa  \epsilon + \phi) } \Gamma (1-s-2i \epsilon_+) \Gamma
   (-s-2i \epsilon_- ) \sum _{n=-\infty}^{\infty}
   \frac{a_n}{\Gamma (-\nu-n-s-i \epsilon ) \Gamma
   (1+\nu+n-s-i \epsilon)},
   \end{align}
   where
   \begin{equation}
\phi\equiv \epsilon_- \kappa\left(1+\frac{2\ln\kappa}{1-\kappa}\right).
   \end{equation}
   The expressions in \eqref{eq:Bintreftra} agree with
Eqs.~(A14) and (A15) in~\cite{2024PhRvD.109f5023M} except for the following: (i) Eqs.~(A14) and (A15) in~\cite{2024PhRvD.109f5023M} contain factors $H(\pm s)$ (where $H(s)$ is equal to 1 for $s\geq 0$ and to 0 otherwise); (ii) our $\RintraInt$ in \eqref{eq:Bintreftra} contains an extra factor $(-1)^s$ with respect to $B_{\text{int}}^{\text{trans}}$ (which corresponds to $\RintraInt$ here) in Eq.~(A15) in~\cite{2024PhRvD.109f5023M}.
We note, however, that these discrepancies do not affect the case $s=0$ of relevance in this paper.

It is convenient to define a new solution which is also purely ingoing into the event horizon but   with transmission coefficient equal to one:
 \begin{align}\label{eq:hatted slns}
 \Rinhat\equiv \frac{\Rin}{\Btra}.
 \end{align}
Consequently, from \eqref{eq:Rin bc}, it satisfies the following boundary conditions:
\begin{align} \label{eq:f,near hor}
\Rinhat(r,\omega)& \sim
\left\{\begin{array}{l l}
\RintraIntNorm\Delta^{-s}e^{-i \omega_-\rsST}+\RinrefIntNorm e^{+i \omega_-\rsST}, & r\rightarrow r_-, \\
\Delta^{-s}e^{-i \omega_+\rsST}, & r\rightarrow r_+, 
\end{array}
\right.
\end{align}
where 
$\hat{B}^{\text{tra/ref}}_-\equiv B^{\text{tra/ref}}_-/\Btra$.
By combining the  expressions for 
$B^{\text{tra/ref}}_-$  in \eqref{eq:Bintreftra} and for $\Btra$ in Eq.~(167) in~\cite{Sasaki:2003xr}, we obtain the following infinite series representations for the interior coefficients:
\begin{align}\label{eq:hat Bref/tra}
\RintraIntNorm=&
(-1)^s
\kappa^{-2im\kappa/q}
\Gamma\left(1-s-2i\epsilon_+\right)\Gamma\left(s+2i\epsilon_-\right)\frac{\sum_{n=-\infty}^{\infty} \frac{\an{n}}{\Gamma\left(-n-\nu-i\tau\right)\Gamma\left(1+n+\nu-i\tau\right)}}{\sum_{n=-\infty}^{\infty}\an{n}}
\nn,\\
\RinrefIntNorm=&
(2M\kappa)^{-2s}
e^{i\kappa\left(\epsilon-\tau+2\ln(\kappa)(\epsilon\kappa-\tau)/q^2\right)}
\Gamma\left(1-s-2i\epsilon_+\right)\Gamma\left(-s-2i\epsilon_-\right)\frac{\sum_{n=-\infty}^{\infty} \frac{\an{n}}{\Gamma\left(-n-\nu-s-i\epsilon\right)\Gamma\left(1+n+\nu-s-i\epsilon\right)}}{\sum_{n=-\infty}^{\infty}\an{n}}.
\end{align}

We now define a new radial function 
\begin{equation}\label{eq:psi}
{}_s\psi^{\text{in}}_{\omega \ell m}(r)\equiv 
\Delta^{s/2}
\sqrt{\frac{r^2+a^2}{r_+^2+a^2}}
\Rinhat(r,\omega).
\end{equation}
It follows from \eqref{eq:f,near hor} that its asymptotics are
\begin{align} \label{eq:psi,near hor}
{}_s\psi^{\text{in}}_{\omega \ell m}(r)& \sim
\left\{\begin{array}{l l}
\B{s}  \Delta^{-s/2}e^{-i \omega_-\rsST}+\A{s} \Delta^{s/2}e^{+i \omega_-\rsST}, & r\rightarrow r_-, \\
\Delta^{-s/2}e^{-i\omega_+ \rsST}
, &
r\to r_+, 
\end{array}
\right.
\end{align}
where
\begin{equation}\label{eq:ABST-Breftra}
\A{s}\equiv \sqrt{
\frac{r_-}{r_+}}
\RinrefIntNorm, 
\quad
\B{s}\equiv \sqrt{
\frac{r_-}{r_+}}
\RintraIntNorm. 
\end{equation}
For $s=0$, the new radial function
${}_s\psi^{\text{in}}_{\omega \ell m}$ in Eq.~\eqref{eq:psi} satisfies Eq.~\eqref{eq:radial1}, which is satisfied by $\psi_{\omega\ell m}^{\Lambda}$ (for generic integer $s$, ${}_s\psi^{\text{in}}_{\omega \ell m}$ satisfies Eq.~(5.2) in~\cite{Teukolsky:1973ha}).
The asymptotics for
${}_{s=0}\psi^{\text{in}}_{\omega \ell m}$ in Eq.~\eqref{eq:psi,near hor}  are similar to those for $\psi_{\omega\ell m}^{\text{in-II}}$ in \eqref{eq:inII-scattering}, except that the former are in terms of $\rsST$ whereas the latter are in terms of our $r_*$, with the difference between the tortoise coordinates is given in Eq.~\eqref{eq:tortoisediff}. Thus, finally, our interior coefficients defined in  \eqref{eq:inII-scattering} are given by
\begin{align}
A_{\omega\ell m}&
=e^{i(\omega_++\omega_-)\delta\tort}\A{0}
=\kappa^{2i(\epsilon-m/q)}\A{0},\nn \\
B_{\omega\ell m}&=e^{i(\omega_+-\omega_-)\delta\tort}\B{0}=\kappa^{2im\kappa/q}\B{0}.
\label{eq:AB ST-here}
\end{align}

We are finally in a position to obtain the small-$\epsilon$ expansions of $A_{\omega\ell m}$ and $B_{\omega\ell m}$ by combining\footnote{Note that the prefactor of $\B{0}$ in Eq.~\eqref{eq:AB ST-here} cancels out with the first factor in Eq.~\eqref{eq:hat Bref/tra} for $\RintraIntNorm$.} Eqs.~\eqref{eq:AB ST-here}, \eqref{eq:ABST-Breftra} and \eqref{eq:hat Bref/tra} and expanding them for small $\epsilon$ using the expansions for $a_n$ and $\nu$ provided in Appendix~B in~\cite{PhysRevD.94.124053} (and their higher-order extensions and generalizations for general $\ell$, including Eqs.~(3.6) and (3.7) therein).
We give here the expansion for the case of interest in this paper, namely $\ell=m=0$:
\begin{align}
A_{\omega 0 0}&=-\frac{\kappa }{q}-\frac{5 i \kappa 
   \epsilon }{6 q}+\frac{\left(\pi ^2-2
   \kappa ^2\right) \epsilon ^2}{6
   \kappa  q}-\frac{i \epsilon ^3
   \left(\kappa ^2 \left(72
   q^2+1489\right)-150 \pi ^2+1080 \psi
   ^{(2)}(1)\right)}{1080 \kappa 
   q}+O\left(\epsilon ^4\right),
   \\
B_{\omega 0 0}&=
   \frac{1}{q}+\frac{5 i \kappa  \epsilon
   }{6 q}-\frac{\left(2 q^2+\pi
   ^2-2\right) \epsilon ^2}{6
   q}+\frac{i \epsilon ^3 \left(18
   \kappa ^4+\left(1471-150 \pi
   ^2\right) \kappa ^2+1080 \psi
   ^{(2)}(1)\right)}{1080 \kappa 
   q}+O\left(\epsilon ^4\right),
   \end{align}
   where $\psi^{(n)}(z)$ is the $n$-th derivative of the digamma function $\psi(z)$.

For completeness, we also give the expansions for generic $\ell>0$ and $m=0$:
\begin{align}
(-1)^{\ell}\frac{q}{\kappa}A_{\omega \ell 0}&=
   -1+
   \frac{i   \epsilon}{2 }  \left(\frac{16 \ell^3+9 \ell^2-19 \ell+5}{8 \ell^3+12 \ell^2-2
   \ell-3}-4 H_\ell\right)+\epsilon ^2 
   \Bigg(
   \frac{8 \ell^3-3 \ell^2-17 \ell+8}{16 \ell^3+24 \ell^2-4
   \ell-6}-
   \\ & \left.
   \frac{\left(16 \ell^3+9 \ell^2-19 \ell+5\right) (\psi ^{(0)}(\ell+1)+\gamma )}{(2 \ell-1) (2 \ell+1) (2 \ell+3)}+2 (\psi
   ^{(0)}(\ell+1)+\gamma )^2+\frac{\pi ^2}{6-6 q^2}\right)+O\left(\epsilon ^3\right),\nn\\
(-1)^{\ell}qB_{\omega \ell 0}&=
  1-\frac{i  \epsilon}{2 \kappa  }  \left(\frac{\kappa ^2 \left(16 \ell^3+9 \ell^2-19 \ell+5\right)}{(2 \ell-1) (2
   \ell+1) (2 \ell+3)}-4 H_\ell\right)+\frac{ \epsilon ^2}{\kappa ^2 } \left(-\frac{\left(8 \ell^3-3 \ell^2-17
   \ell+8\right) \kappa^4}{2 (2 \ell-1) (2 \ell+1) (2 \ell+3)}+
   \right.\\ &\left.
   \frac{\left(16 \ell^3+9 \ell^2-19 \ell+5\right)
  \kappa^2 (\psi ^{(0)}(\ell+1)+\gamma )}{(2 \ell-1) (2 \ell+1) (2 \ell+3)}-2 (\psi ^{(0)}(\ell+1)+\gamma
   )^2-\frac{1}{6} \pi ^2 \kappa^2\right)+O\left(\epsilon ^3\right),\nn
     \end{align}
     where $H_{\ell}$ is $\ell$-th harmonic number. We note, in particular, that the expansion coefficients are real when expressed in a series in $i\epsilon$ (or equivalently $i\omega$; see the discussion below Eq.~\eqref{eq:AB-small-w}).

\subsection{Upgoing radial solution}

The second solution to the general-spin radial Teukolsky equation that we consider is only introduced in the exterior of the black hole and so is  contained in  Ref.~\cite{Sasaki:2003xr}. Its MST series representation is given in Eqs.~(159) and (153) in~\cite{Sasaki:2003xr}
and it satisfies the boundary conditions
\begin{equation}\label{eq:bc up}
\Rup(r,\omega)\sim
\left\{\begin{array}{l l}
\Cref \Delta^{-s}e^{-i\omega \rsST}+\Cinc e^{i\omega \rsST}, & r\to r_+,
\\
\Ctra r^{-1-2s}e^{i\omega \rsST}, & r\to \infty,
\end{array}
\right.
\end{equation}
for some upgoing, exterior scattering coefficients $C^{\text{inc/ref/tra}}$.
Similarly to the steps \eqref{eq:psi} and \eqref{eq:hatted slns} for the ingoing solution, we define 
\begin{equation}\label{eq:psi-up}
\psiup{s}(r)\equiv 
\Delta^{s/2}\sqrt{\frac{r^2+a^2}{r_+^2+a^2}}
\frac{\Rup(r,\omega)}{\Cinc},
\end{equation}
which readily satisfies   the following boundary conditions:
\begin{equation}\label{eq:psi up,bc}
\psiup{s}(r)\sim
\left\{\begin{array}{l l}
\rhoup{s}\Delta^{-s/2} e^{-i\omega \rsST}+ \Delta^{s/2}e^{i\omega \rsST}, & r\to r_+,
\\
\tauup{s} r^{-s}e^{i\omega \rsST}, & r\to \infty,
\end{array}
\right.
\end{equation}
where 
\begin{equation}\label{eq:Rup coeffs}
\rhoup{s}\equiv \frac{\Cref }{\Cinc},
\quad
\tauup{s}\equiv \frac{1}{\sqrt{r_+^2+a^2}} \frac{\Ctra }{\Cinc}.
\end{equation}
It is useful to relate the scattering coefficients of the upgoing solution here to those of the ingoing solution in the previous subsection via the following Wronskian relation:
\begin{equation}\label{eq:Wronsk-rhoupST}
\rhoup{s}=-\frac{\Btra \Brefcc}{\Btracc \Binc}.
\end{equation}
For $s=0$, 
$\psiup{s}$ in Eq.~\eqref{eq:psi-up} also satisfies Eq.~\eqref{eq:radial1} 
(and Eq.~(5.2) in~\cite{Teukolsky:1973ha} for generic integer $s$)
and its asymptotics  in Eq.~\eqref{eq:psi up,bc}  are similar to those for $\psi_{\omega\ell m}^{\text{up-I}}$ in \eqref{eq:upI-scattering}, except that the former are in terms of $\rsST$ while the latter are in terms of our $r_*$. Thus, our exterior  coefficients defined in \eqref{eq:upI-scattering} are given by
\begin{equation}\label{eq:rhoup-rhoupST}
\rho^\text{up}_{\omega \ell m}= e^{-2i\omega_+\delta \tort}\rhoup{s},\quad
\tau^\text{up}_{\omega \ell m}= e^{im\Omega_+\delta \tort}\tauup{s}.
\end{equation}

An expression for the upgoing reflection coefficient $\rho^\text{up}_{\omega \ell m}$ may be obtained by  combining Eqs.~\eqref{eq:rhoup-rhoupST} and \eqref{eq:Wronsk-rhoupST} 
together with Eqs.~(167), (168) and (169) in~\cite{Sasaki:2003xr} for, respectively, $\Btra$,  $\Binc$ and  $\Bref$. 
The resulting expression for $\rho^\text{up}_{\omega \ell m}$ can then be expanded for small $\epsilon$  using the expansions for $a_n$ and $\nu$ provided in Appendix~B in~\cite{PhysRevD.94.124053}.
In fact, in the case of $\ell=m=0$, the expansions for $\Btra$,  $\Binc$ and  $\Bref$ are already provided in Eqs.~(6.7)--(6.10) in~\cite{PhysRevD.94.124053}. Inserting these directly into Eqs.~\eqref{eq:rhoup-rhoupST} and \eqref{eq:Wronsk-rhoupST} yields, for $\text{Re}(\omega)\geq 0$,
\begin{align}
\rho^\text{up}_{\omega 00}&=-1-i (\kappa +1) \epsilon +\frac{1}{2} (\kappa +1) (\kappa +3) \epsilon ^2+
\\
& \frac{\epsilon ^3}{18 \kappa ^2} (\kappa +1) \left(3 i \kappa ^4+27 i \kappa ^3+18 \pi  \kappa ^2+36 i \gamma  \kappa
   ^2-33 i \kappa ^2+36 i \kappa ^2 \log (2 \kappa )-11 i \pi ^2 \kappa -36 i \zeta (3)+36 i \kappa ^2 \log (\epsilon )\right)+o\left(\epsilon ^3\right),\nn
\end{align}
where $\zeta(n)$ is the Riemann zeta function.



\bibliography{references}


\end{document}

%% file: Kerr-Penrose.tex
\begin{tikzpicture}
\node (I)    at ( 0,-3)   {I};
\node (II)   at (-3,0)   {II};
\node (III)  at (-1,3)   {III};

\path  
  (I) +(90:3)  coordinate[label=-90:$i^+$]  (Itop)
       +(-90:3) coordinate[label=-90:$i^-$] (Ibot)
       +(0:3)   coordinate                  (Iright)
       +(180:3) coordinate                  (Ileft);
       
\path[fill=white] (Itop) circle (0.1);

\draw (Ileft) -- 
          node[yshift= 0.1cm, midway, above, sloped, font=\large]    {$\cal{H}^R$}
      (Itop) --
          node[midway, above, sloped, font=\large] {$\cal{I}^+$}
      (Iright) -- 
          node[midway, below, sloped, font=\large] {$\cal{I}^-$}
      (Ibot) --
          node[yshift=-.1cm, midway, below, sloped, font=\large]    {$\cal{H}^-$} 
      (Ileft) -- cycle;

\path 
   (II) +(0,3)  coordinate (IItop)
        +(0,-3) coordinate (IIbot)
        +(-3,0) coordinate (IIleft)
        +(3,0)   coordinate (IIright);

\draw  (IIleft) --
        node[yshift=0.1cm, midway, above, sloped, font=\large]    {$\cal{CH}^L$}
           (IItop) -- 
        node[yshift=0.1cm, midway, above, sloped, font=\large]    {$\cal{CH}^R$}
         (IIright) --  (IIbot) -- 
       node[yshift=-0.1cm, midway, below, sloped, font=\large]    {$\cal{H}^L$} 
       (IIleft) -- cycle;

\path  
  (III) +(1,3)    coordinate    (IIItop)
       +(-2,0)    coordinate    (IIIleft)
       +(1,-3)     coordinate   (IIIbot)
       +(4,0)      coordinate   (IIIright);

\path 
    (Iright)   +(-45:0.1)   coordinate (IIleftV)
    (Ibot)     +(-45:.1)    coordinate (IItopV)
    (IIIright)   +(-45:0.1)   coordinate (IIIrightV)
    (IIIbot)     +(-45:.1)    coordinate (IIIbotV);
    
\draw (IIIbot) -- (IIIleft) --node[yshift=0.1cm, midway, above, sloped, font=\large]    {$\cal{CH}^+$} (IIItop) -- node[yshift=0.1cm, midway, above, sloped, font=\large]    {$\overline{\cal{I}}^+$} (IIIright) -- node[yshift=0.05cm, midway, below, sloped, font=\large]    {$\overline{\cal{I}}^-$}(IIIbot);
        
\draw[decorate,decoration=snake] (IIItop) -- (IIIbot)
      node[midway, right, inner sep=2mm] {$\rho^2=0$};

\path  
  (IIleft) +(-90:0.12)    coordinate    (IIleftu)
       +(-90:0.12)    coordinate    (IIleftUkruskal);
\path  
  (Ileft) +(-90:0.12)    coordinate    (Ileftu)
       +(-90:0.12)    coordinate    (IleftUkruskal);
\path  
  (Ibot) +(-90:0.12)    coordinate    (Ibotu)
       +(-135:0.12)    coordinate    (IbotUkruskal);

\draw[->,>=latex, draw=blue, line width=1pt, scale=3]     
(IItopV)--  node[yshift=-0cm, xshift=0.75cm, below, color=red, text=blue, sloped, font=\large]   {$v$} (IIleftV);

\draw[->,>=latex, draw=blue, line width=1pt, scale=3]     
(IIIbotV)--  node[yshift=-0cm, xshift=0.75cm, below, color=red, text=blue, sloped, font=\large]   {$v$} (IIIrightV);

\draw[-,>=latex, draw=green, line width=1.5pt, scale=3]($(IIleft)$) to[out=-40, in=-160, looseness=1] 
node[below, color=green, text=green, font=\large]   {$\Sigma$} (Iright);

\fill[green!20, opacity=0.3]($(IIleft)$) to[out=-40, in=-160, looseness=1] (Iright) -- (IItop) -- (IIleft) -- cycle;

\path (III) +(110:0.7) coordinate (p);

\draw[<-,>=latex, draw=red, line width=1pt, scale=3] (IIleftu) -- node[xshift=0.25cm, yshift=-0.35cm, below, color=red, text=red, font=\large]   {$u$}(Ileftu);

\draw[->,>=latex, draw=red, line width=1pt, scale=3] (Ibotu) -- node[xshift=0.25cm, yshift=-0.35cm, below, color=red, text=red, font=\large]   {$u$}(Ileftu);

\end{tikzpicture}

%% file: U-underbar.tex
\begin{tikzpicture}

\node (II)   at (-3,0) { };
\node at (-3,0.3) {II};
\node (III)  at (-1,3)   {III};

\path  
  (II) +(0:3)  coordinate[label=-4:$i^+$]  (Itop)
       +(-90:3) coordinate                  (Ileft);
       
\path[fill=white] (Itop) circle (0.1);

\draw (Ileft) -- 
          node[yshift= -0.05cm, midway, below, sloped, font=\large]    {$\cal{H}^R$}
      (Itop) -- cycle;

\path 
   (II) +(0,3)  coordinate (IItop)
        +(0,-3) coordinate (IIbot)
        +(-3,0) coordinate (IIleft)
        +(3,0)   coordinate (IIright);

\draw  (IIleft) --
        node[yshift=0.1cm, midway, above, sloped, font=\large]    {$\cal{CH}^L$}
           (IItop) -- 
        node[yshift=0.0cm, pos=0.65, below, sloped, font=\large]    {$\cal{CH}^R$}
         (IIright) --  (IIbot) -- 
       node[yshift=-0.1cm, midway, below, sloped, font=\large]    {$\cal{H}^L$} 
       (IIleft) -- cycle;

\path  
  (III) +(1,3)    coordinate    (IIItop)
       +(-2,0)    coordinate    (IIIleft)
       +(1,-3)     coordinate   (IIIbot)
       +(4,0)      coordinate   (IIIright);

\path 
    (IIbot)   +(-45:0.06)   coordinate (Ileftv)
    (IIright)     +(-45:0.06)    coordinate (Itopv)
    (IIIright)     +(-45:0.06)    coordinate (IIIrightv);
    
\draw (IIIbot) -- (IIIleft) --node[yshift=0.1cm, midway, above, sloped, font=\large]    {$\cal{CH}^+$} (IIItop) -- node[yshift=0cm, above, sloped, font=\large]    {$\overline{\cal{I}}^+$} (IIIright) -- node[yshift=0cm, midway, below, sloped, font=\large]    {$\overline{\cal{I}}^-$}(IIIbot);
        
\draw[decorate,decoration=snake] (IIItop) -- (IIIbot);

\path  
  (IIleft) +(-90:0.12)    coordinate    (IIleftu)
       +(-90:0.12)    coordinate    (IIleftUkruskal);
\path  
  (Ileft) +(-90:0.12)    coordinate    (Ileftu)
       +(-90:0.12)    coordinate    (IleftUkruskal);

\draw[->,>=latex,, line width=0.8pt, scale=3]     (Ileftv) --  node[yshift=-0cm, xshift=0.75cm, below, sloped, font=\large]   {$v$} (Itopv);
\draw[->,>=latex, line width=0.8pt, scale=3]     (Itopv) --  node[yshift=-0cm, xshift=0.75cm, below, sloped, font=\large]   {$v$} (IIIrightv);



\path (III) +(110:0.7) coordinate (p);

\draw[<-,>=latex, line width=0.8pt, scale=3] (IIleftu) -- node[xshift=0.25cm, yshift=-0.35cm, below, font=\large]   {$u$}(Ileftu);


\draw[->,>=latex, draw=red, line width=1.5pt, scale=3]($(IIItop)!.52!(IIIright)$) --  node[yshift=-0.7cm, xshift=-1.45cm, above, color=red, text=red, sloped, font=\large, rotate=-0 ]    {$\underline{\tau}_{\ \ \omega\ell m}^\text{out}$}  ($(IIright)!.48!(IItop)$); 

\draw[->,>=latex, draw=red, line width=1.5pt, scale=3]  ($(IIItop)!.52!(IIIright)$) to[out=-135, in=135, looseness=1.5] node[yshift=-0.8cm, xshift=-0.7cm, color=red, text=red, sloped, font=\large, rotate=-130]   {$\underline{\rho}_{\ \ \omega\ell m}^\text{out}$} ($(IIIbot)!.5!(IIIright)$);

\draw[->,>=latex, draw=blue, line width=1.5pt, scale=3]($(IIIleft)!.47!(IIItop)$) --  node[yshift=-0.7cm, xshift=1.45cm, above, color=red, text=blue, sloped, font=\large, rotate=-0 ]    {$\underline{\tau}_{\ \ \omega\ell m}^\text{down}$}  ($(IIright)!.47!(IIIright)$);

\draw[->,>=latex, draw=blue, line width=1.5pt, scale=3]  ($(IIItop)!.53!(IIIleft)$) to[out=-45, in=45, looseness=1.5] node[yshift=-0.7cm, xshift=0.6cm, color=red, text=blue, sloped, font=\large, rotate=135]   {$\underline{\rho}_{\ \ \omega\ell m}^\text{down}$} ($(IIIbot)!.52!(IIIleft)$);

\draw[->,>=latex, draw=orange, line width=1.5pt, scale=3]  ($(IItop)!.52!(IIleft)$) to[out=-45, in=45, looseness=1.5] ($(IIbot)!.52!(IIleft)$);
\draw[->,>=latex, draw=orange, line width=1.5pt, scale=3]  ($(IItop)!.52!(IIleft)$) -- ($(IIbot)!.48!(IIright)$);

\draw[->,>=latex, draw=green, line width=1.5pt, scale=3]  ($(IItop)!.52!(IIright)$) to[out=-135, in=135, looseness=1.5] ($(IIbot)!.52!(IIright)$);
\draw[->,>=latex, draw=green, line width=1.5pt, scale=3]  ($(IItop)!.52!(IIright)$) -- ($(IIbot)!.48!(IIleft)$);

\node[green, font=\large] at ($(IItop)!.5!(IIright)+(-0.6,0)$) {$f^\mathcal{R}_{\omega\ell m}$};
\node[orange, font=\large] at ($(IItop)!.5!(IIleft)+(0.65,0)$) {$f^\mathcal{L}_{\omega\ell m}$};

\node[red, font=\large] at ($(IIItop)!.5!(IIIright)+(-0.6,0)$) {$f^{\text{out}}_{\omega\ell m}$};

\node[blue, font=\large] at ($(IIItop)!.5!(IIIleft)+(0.65,0)$) {$f^{\text{down}}_{\omega\ell m}$};

\end{tikzpicture}